\documentclass[11pt]{article}

\usepackage{float}
\usepackage[T1]{fontenc}
\usepackage[utf8]{inputenc}
\usepackage{lmodern}
\usepackage{amsmath,amssymb,amsfonts}
\usepackage{bm}
\usepackage{siunitx}
\usepackage{graphicx}
\usepackage{booktabs}
\usepackage{multirow}
\usepackage{caption}
\usepackage{subcaption}
\usepackage{hyperref}
\usepackage{geometry}
\usepackage{xcolor}
\usepackage{bookmark}
\usepackage{natbib}
\hypersetup{colorlinks=true, linkcolor=blue, citecolor=blue, urlcolor=blue}

\title{Physics-Informed Neural Networks for Depth-Averaged Avalanche Dynamics}

\author{
Pradyumn Singh Sikarwar \quad
Vishal Sharma \quad
Gaurav Bhutani\thanks{Corresponding author: \texttt{gaurav@iitmandi.ac.in}}
\\[0.5em]
\normalsize School of Mechanical and Materials Engineering,\\
\normalsize Indian Institute of Technology Mandi, India
}

\date{}

\begin{document}
\maketitle

\begin{abstract}
Accurate prediction of avalanche motion is essential for hazard assessment in mountainous terrain. This study developed and evaluated a physics-informed neural network (PINN) framework for the Savage--Hutter model of depth-averaged granular flow, progressing systematically from one-dimensional analytical verification to two-dimensional experimental validation.
In the first stage, three one-dimensional problems of increasing complexity were verified against the analytical solution: height prediction with prescribed velocity, velocity prediction with prescribed height, and coupled prediction of both fields using the conservative formulation. The decoupled tests demonstrated that PINNs accurately reconstructed the spatio-temporal evolution of each field when the other was prescribed. The coupled conservative formulation learned both fields simultaneously without any prescribed data, achieving mean height and velocity RMSEs of $4.3 \times 10^{-2}$ and $7.9 \times 10^{-2}$ in non-dimensional units, respectively, compared with the analytical solution.
A systematic hyperparameter sensitivity study evaluated the influence of network depth, width, collocation density, learning rate, and number of epochs on the conservative PINN formulation.
In the second stage, the framework was extended to two dimensions and validated against laboratory experiments on the collapse of a cylindrical granular pile on an inclined plane, with \texttt{TITAN2D} providing numerical comparisons. Training with purely physics-based losses led to a collapse to the trivial zero solution; this was resolved by augmenting the loss with 10 sparse training data points, drawn from the final deposit profiles, thereby resulting in a physics-informed, data-assisted hybrid framework. Peak flow depth, depth-averaged velocity, RMSE, and wetted-area IoU were used to evaluate agreement at both global and local levels. The framework achieved global height RMSE values ranging from $2.7$ to $6.7\,\text{mm}$ across the four experimental cases, with mean wetted-area IoU values ranging from $69$ to $81\,\%$, demonstrating consistent performance across variations in pile mass and slope angle.
\end{abstract}


\section{Introduction}
\label{sec:introduction}
Rapid granular flows driven by gravity, such as snow, rock, and debris avalanches, pose severe hazards to mountain communities and infrastructure worldwide. Since full-scale experiments with such flows are dangerous, expensive, and rarely repeatable, physics-based modelling has become the primary means of understanding their dynamics and assessing the associated risks. The behaviour of these flows is governed by strong gravitational forcing and frictional resistance, giving rise to nonlinear, unsteady dynamics that are difficult to describe analytically and demanding to resolve numerically.

When the flow thickness is much smaller than the downslope length scale, the vertical dependence can be integrated out, reducing the three-dimensional governing equations to a depth-averaged two-dimensional system for the flow thickness and depth-averaged velocity components. The model of \citet{savage1989}, arguably the most influential formulation in this class, captures the essential physics of rapid dry granular avalanches, including inertia, basal Coulomb friction, internal shear stress, and lateral earth pressure, within a hyperbolic system of conservation laws. Extensive work has since built on this framework, both in extending the theoretical analysis of the governing equations \citep{gray2014depth} and in developing robust numerical techniques for their solution \citep{tai2002shock, mangeney2003numerical}. In particular, \citet{pitman2003computing} and \citet{patra2005} developed \texttt{TITAN2D}, a parallel adaptive-mesh Godunov solver for depth-averaged granular flows over natural terrain, which has become a widely used tool for hazard assessment. Despite their accuracy, mesh-based solvers can be difficult to set up and apply in practice. Discretisation choices, mesh generation over complex terrain, and the engineering effort required to develop general-purpose codes all impose non-trivial overhead before any simulation can be run. These considerations motivate the search for alternative solution strategies that reduce implementation complexity while retaining physical fidelity.

In recent years, physics-informed neural networks (PINNs) have emerged as a mesh-free alternative for solving differential equations \citep{raissi2019physics, karniadakis2021physics}. Rather than discretising the domain, PINNs encode the governing equations, initial conditions, boundary conditions, and, where available, observed data, directly into a neural network loss function, with derivatives computed via automatic differentiation. This framework has since been applied to advection-dominated flows \citep{mao2020physics,jagtap2020conservative} and systems with complex source terms \citep{karniadakis2021physics}, with substantial effort directed toward improving PINN trainability through gradient balancing~\citep{wang2021gradient}, spectral-bias analysis~\citep{wang2022ntk}, and domain-decomposition strategies~\citep{jagtap2020xpinns}. 

Within the context of depth-averaged flow equations, PINNs have recently been successfully applied to the shallow-water equations (SWEs) in hydraulic settings. Applications include spherical-domain geophysical flows \citep{bihlo2022pinn}, inverse PINNs that infer bed elevation from surface observations~\citep{dazzi2024pinn}, benchmarking against finite-volume solvers for free-surface flows~\citep{qi2024pinn}, river-model downscaling~\citep{feng2023pinn}, and, more recently, unsteady riverine and complex-terrain domains~\citep{yin2025pinn_swe, tian2025pinn_swe}.

Meanwhile, PINNs have also begun to appear in the broader granular-flow literature, though exclusively through continuum-level rheological models rather than depth-averaged formulations. \citet{baldoni2025rheological} used PINNs to identify the parameters of the $\mu(I)$-rheology from synthetic granular column collapse data, treating it as an inverse problem on the full Navier--Stokes-like equations. \citet{pai2025adaptive} proposed a Lagrangian PINN framework incorporating the $\mu(I)$-rheology for granular flows, demonstrating its ability to capture free-surface evolution and particle trajectories in the column-collapse problem. To the best of the authors' knowledge, this is the first study that applies PINNs to the depth-averaged granular avalanche equations. Although these equations share the structure of hydraulic shallow-water equations, they differ substantially in their constitutive physics, incorporating Coulomb basal friction and slope-dependent source terms arising from the inclined-plane geometry.

The Savage--Hutter system presents several features that complicate PINN training. First, basal resistance follows a direction-dependent Coulomb friction law rather than the smooth drag formulations typical of hydraulic models, introducing a non-smooth operator into the governing equations. Second, the Mohr--Coulomb earth pressure closure introduces sign-dependent source terms that can obstruct gradient-based optimisation.  Third, the solution is compactly supported: the flow thickness vanishes outside a finite, time-dependent wetted region.
Near the moving wet--dry interface, the sharp transition in flow depth, together with non-smooth Coulomb friction and velocity-normalisation terms, can produce poorly conditioned residuals at collocation points and destabilise PINN training.  

The present study addresses these questions through a systematic two-stage investigation. Section~\ref{sec:formulation} presents the mathematical formulation, including the one-dimensional governing equations, non-dimensionalisation, the analytical similarity solution, and the extension to two-dimensional depth-averaged equations. Section~\ref{sec:pinn_framework} describes the PINN methodology, covering the network architecture, residual formulation, and composite loss function. In the first stage (Section~\ref{sec:1d_benchmarks}), the one-dimensional Savage--Hutter equations are solved and verified against the analytical solution for a parabolic-cap initial profile, through two decoupled single-field tests --- height prediction with prescribed velocity, and velocity prediction with prescribed height --- followed by the fully coupled prediction of both fields under the conservative formulation. In the second stage (Section~\ref{sec:2d}), the framework is extended to two dimensions, addressing the trivial-solution pathology through sparse experimental observations, and benchmarked against the finite-volume solver \texttt{TITAN2D}~\citep{patra2005}.
Results are also validated against the laboratory experiments of \citet{maeno2013unconfined} for the collapse of a cylindrical granular pile on an inclined plane, using root-mean-square error (RMSE), intersection-over-union (IoU) of wet-cell footprints, and time-resolved peak-height and velocity tracking. Section~\ref{sec:conclusions} summarises the findings and outlines future directions.

\section{Mathematical Formulation}
\label{sec:formulation}

Granular avalanches were modelled as a single-phase, incompressible continuum, implying that the bulk density $\rho$ remains constant in both space and time, and that volume changes due to granular dilation are negligible. This assumption has been widely adopted for dense flows where particles remain in contact and packing variations are minimal \citep{savage1989}. Under these conditions, the equations of mass and momentum conservation governing the motion of the continuum are expressed as:

\begin{equation}
    \nabla \cdot \mathbf{u} = 0,
    \label{eq:2D_continuity}
\end{equation}

\begin{equation}
    \rho\left(
\frac{\partial \mathbf{u}}{\partial t}
+ \nabla \cdot (\mathbf{u} \mathbf{u})
\right)
= -\nabla \cdot \boldsymbol{\sigma} + \rho \mathbf{g},
\label{eq:2D_momentum}
\end{equation}
where $\mathbf{u}$ is the velocity vector, $\boldsymbol{\sigma}$ is the negative of the Cauchy shear stress tensor (so that compressive stresses are positive), and $\mathbf{g}$ denotes the acceleration due to gravity. In the one-dimensional depth-averaged formulation, the $x$- and $z$-directions denote the downslope and bed-normal directions, respectively. The $y$-direction represents the cross-slope direction, introduced in the two-dimensional formulation in Section~\ref{sec:2d_equations}.

\subsection{One-dimensional depth-averaged equations}
\label{sec:1d_equations}
The depth-averaged approach is valid when the flow thickness is much smaller than the downslope length scale, allowing the flow properties to be averaged over the depth. At the free surface, kinematic and stress-free boundary conditions were applied. At the bed, a no-penetration condition and a Coulomb friction law were imposed, the latter relating basal shear stress to the normal stress and the bed friction angle. In the depth-averaged formulation, vertical velocity and accelerations were neglected, and the vertical normal stress $\sigma_{zz}$ was taken to be hydrostatic:

\begin{equation}
	\sigma_{zz}(z) = \rho g_z (h-z).
    \label{eq:vertical_stress}
\end{equation}
The normal stress in the $x$-direction was related to the vertical normal stress through the earth pressure coefficient, $k_{ap}$, such that:
\begin{equation}
    \sigma_{xx} = k_{ap}\,\sigma_{zz}, 
\end{equation}
where $k_{ap}$ represents the ratio of horizontal to vertical normal stress. In this study, $k_{ap} = 1$ was adopted, which is standard for unconfined granular flows on inclined planes \citep{mangeney2005saint_venant, mangeney2007levee}. The same value was used in the two-dimensional formulation.

Integrating Equations~\eqref{eq:2D_continuity} and \eqref{eq:2D_momentum} from the bed to the free surface, applying Leibniz's rule, and imposing the boundary conditions yields the one-dimensional depth-averaged continuity and momentum equations:
\begin{equation}
	\frac{\partial h}{\partial t}
	+ \frac{\partial (h u)}{\partial x}
	= 0,
	\label{eq:1d_mass}
\end{equation}
and
\begin{equation}
	\frac{\partial (h u)}{\partial t}
	+ \frac{\partial}{\partial x}
	\left(
	h u^{2} + \frac{\,g_z\,h^{2}}{2}
	\right)
	= S_x,
	\label{eq:1d_mom_cons}
\end{equation}
where $h$ denotes flow depth along the $z$-direction, $u$ is the depth-averaged velocity in the $x$-direction, $g_z = g \cos{\zeta}$ is the bed-normal component of gravitational acceleration, $\zeta$ is the basal inclination angle, and $S_x$ is the source term containing the gravitational and basal friction contributions. The $\frac{1}{2}g_z h^2$ represents the depth-averaged hydrostatic pressure arising from the horizontal normal stress. The source term is:
\begin{equation}
	S_x = g_x\,h - h\,g_z\,\mathrm{sgn}({u})\,\tan\delta,
	\label{eq:1d_source}
\end{equation}
Here, $g_x = g\sin\zeta$ is the downslope component of gravitational acceleration and $\delta$ is the bed friction angle. The sign function ensures that basal friction always opposes the direction of motion; this conservative form is particularly suited to PINN-based solvers, which propagate conservation-law solutions more accurately~\citep{jagtap2020conservative}.

\subsubsection{Non-dimensionalization}
\label{sec:nondim}
\citet{savage1989} derived their analytical solutions in non-dimensional form by separating the characteristic flow thickness $H$ from the downslope length scale $L$, with aspect ratio $\epsilon = H/L \ll 1$. The same non-dimensionalisation was adopted here for the one-dimensional verification cases. The scaling is:
\begin{equation}
    \hat{x} = \frac{x}{L}, \qquad
    \hat{h} = \frac{h}{H}, \qquad
    \hat{u} = \frac{u}{\sqrt{gL}}, \qquad
    \hat{t} = t\sqrt{\frac{g}{L}}.
    \label{eq:scaling}
\end{equation}
Substituting into Equations~\eqref{eq:1d_mass}--\eqref{eq:1d_mom_cons}, the non-dimensional continuity equation retains its form:
\begin{equation}
    \frac{\partial \hat{h}}{\partial \hat{t}}
    + \frac{\partial (\hat{h}\hat{u})}{\partial \hat{x}}
    = 0,
    \label{eq:1d_mass_nd}
\end{equation}
and the non-dimensional momentum equation in conservative form is:
\begin{equation}
    \frac{\partial (\hat{h}\hat{u})}{\partial \hat{t}}
    + \frac{\partial}{\partial \hat{x}}
    \left(
    \hat{h}\hat{u}^{2}
    +
    \frac{\epsilon\,\cos\zeta}{2}\,\hat{h}^{2}
    \right)
    = \hat{h}
    \left(
    \sin\zeta - \mathrm{sgn}(\hat{u})\,\tan\delta\cos\zeta
    \right).
    \label{eq:1d_mom_nd}
\end{equation}
The aspect ratio $\epsilon$ appears in the horizontal normal stress flux coefficient $\frac{\epsilon\,\cos{\zeta}}{2}$ as a direct consequence of the two-scale non-dimensionalisation: $h$ is scaled by  $H = \epsilon L$ while $x$ is scaled by $L$. The non-dimensional system is governed by the slope angle $\zeta$, 
the bed friction angle $\delta$, and the aspect ratio $\epsilon$.

\subsubsection{Analytical similarity solution}
\label{sec:analytical}
 
A classical self-similar analytical solution of the one-dimensional Savage--Hutter model~\citep{savage1989} was employed to verify the PINN predictions. This solution describes a granular mass undergoing rigid-body translation combined with self-similar lateral spreading. All quantities are expressed in non-dimensional form introduced in Section~\ref{sec:nondim}.

\begin{figure}[H]
    \centering
    \includegraphics[width=0.55\linewidth]{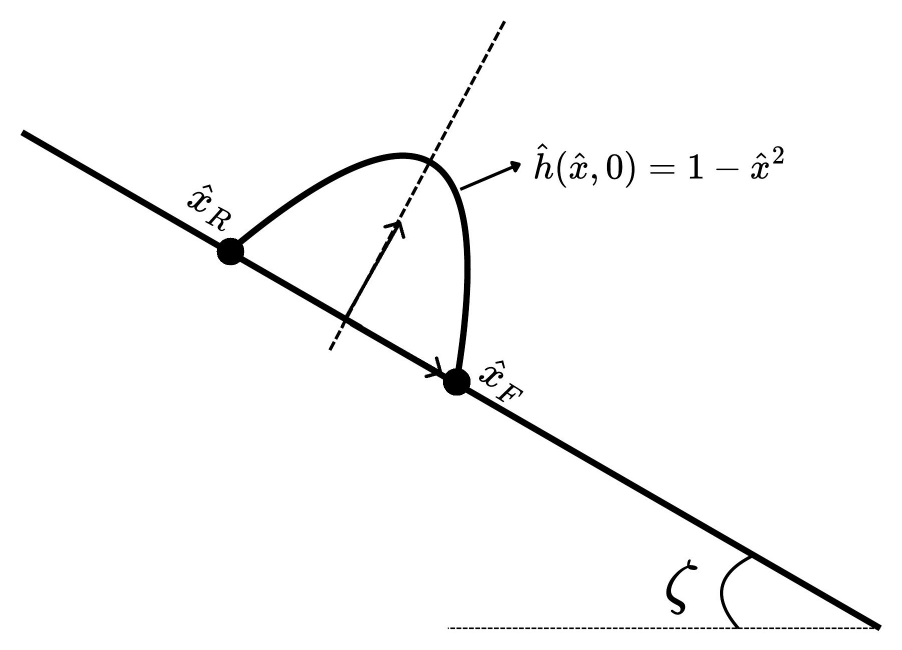}
    \caption{Geometric configuration of the one-dimensional benchmark problem. The initial pile follows the parabolic non-dimensional profile on an inclined plane of angle $\zeta$. The front and rear positions are denoted by $\hat{x}_F$ and $\hat{x}_R$.}
    \label{fig:schematic}
\end{figure}

A similarity coordinate centred on the accelerating pile is:
\begin{equation}
	\hat{\eta}(\hat{x},\hat{t}) = \frac{\hat{x} - \tfrac{1}{2}\hat{a}\,\hat{t}^{2}}{\hat{g}(\hat{t})},
\end{equation}
where $\hat{g}(\hat{t})$ denotes the non-dimensional half-width of the flowing mass and $\hat{a} = \sin\zeta - \cos\zeta\tan\delta$ is the non-dimensional effective downslope acceleration (equal to the dimensional acceleration $a$ normalised by $g$). The analytical flow height takes the parabolic similarity form:
\begin{equation}
	\hat{h}_{\mathrm{ana}}(\hat{x},\hat{t})
	= \frac{K}{\hat{g}(\hat{t})}\bigl(1 - \hat{\eta}^{2}\bigr),
	\label{eq:h_ana}
\end{equation}
which is taken non-zero only for $|\hat{\eta}|\le 1$, with $K = 2\epsilon\,\cos\zeta$. The initial parabolic pile $\hat{h}(\hat{x},0) = \max(1 - \hat{x}^2, 0)$ is shown in Figure~\ref{fig:schematic}.

The temporal evolution of $\hat{g}(\hat{t})$ follows from the implicit relation:
\begin{equation}
	\sqrt{\hat{g}(\hat{t})\bigl(\hat{g}(\hat{t})-1\bigr)}
	+ \ln\!\left[\sqrt{\hat{g}(\hat{t})} + \sqrt{\hat{g}(\hat{t})-1}\,\right]
	= \sqrt{K}\,\hat{t}.
	\label{eq:g_implicit}
\end{equation}
Since Equation~\eqref{eq:g_implicit} has no closed-form inversion, $\hat{g}(\hat{t})$ was reconstructed numerically using 200 uniformly distributed time instances over $0 \le \hat{t} \le 4$. The quadratic fit approximated the resulting profile well over the full time interval:
\begin{equation}
	\hat{g}(\hat{t}) \approx 0.11613\,\hat{t}^{2} + 0.74497\,\hat{t} + 1,
	\label{eq:g_fit}
\end{equation}
The analytical downslope velocity is:
\begin{equation}
	\hat{u}_{\mathrm{ana}}(\hat{x},\hat{t})
	= \sqrt{\frac{2K}{\,\hat{g}(\hat{t})\,\bigl(\hat{g}(\hat{t})-1\bigr)}}\;\hat{\eta}.
	\label{eq:u_ana}
\end{equation}

\subsection{Two-dimensional depth-averaged equations}
\label{sec:2d_equations}    

The one-dimensional depth-averaged equations were extended to include a cross-slope direction denoted as $y$, following the same derivation approach with additional cross-slope terms in the continuity and momentum equations; full details are given in \citet{sharma2026computational}.

The depth-averaged continuity equation in two dimensions is given as:
\begin{equation}
	\frac{\partial h}{\partial t}
	+ \frac{\partial (h u)}{\partial x}
	+ \frac{\partial (h v)}{\partial y}
	= 0,
	\label{eq:2d_mass}
\end{equation}
where $h(x,y,t)$ denotes the flow depth in the $z$-direction in a bed-fitted (local) orthogonal coordinate system, with the $x$ aligned downslope and the $y$ in the cross-slope direction. The variables $u$ and $v$ are the depth-averaged velocity components in the $x$- and $y$-directions, respectively.

The normal stresses in the $x$- and $y$-directions follow the same $k_{ap} = 1$ relation:
\begin{equation}
	{\sigma}_{xx} = k_{ap}\, {\sigma}_{zz}, \qquad {\sigma}_{yy} = k_{ap}\, {\sigma}_{zz}.
\end{equation}

Basal resistance followed a dry Coulomb friction law, $\tau^b = \sigma_{zz}^{b} \tan\delta$, with the basal normal stress from Equation~\eqref{eq:vertical_stress}, given by:
\begin{equation}
	\sigma_{zz}^b
	=
	\rho h g_z.
	\label{eq:total_stress}
\end{equation}
The internal shear stress $\bar{\sigma}_{xy}$, obtained from the Coulomb yielding and Mohr-circle considerations, is:
\begin{equation}
	\bar{\sigma}_{xy}
	=
	\mathrm{sgn}\!\left(\frac{\partial {u}}{\partial y}\right)\,
	\bar{\sigma}_{xx}\,\sin\phi,
	\label{eq:internal_stress}
\end{equation}
where the overbar denotes the depth-averaging operator.

Substituting the body force, basal friction, and internal stress expressions into the depth-averaged momentum equations yields:
\begin{equation}
	\frac{\partial }{\partial t}(h {u}) + \frac{\partial }{\partial x}\left(h {u}^2 + \frac{ g_z h^2}{2}\right) + \frac{\partial }{\partial y}(h {u} {v}) = S_x, 
	\label{eq:da_x_mom}
	\end{equation}
	\begin{equation}
	\frac{\partial }{\partial t}(h {v}) + \frac{\partial }{\partial x}(h {u} {v}) + \frac{\partial }{\partial y}\left(h {v}^2 + \frac{ g_z h^2}{2}\right) = S_y, 
	\label{eq:da_y_mom}
\end{equation}
where the Mohr--Coulomb source terms are:
\begin{equation}
	S_x = g_x h - \frac{{u}}{|\mathbf{{u}}|} \left( h\, g_z \, \mu \right)
	- h \, \text{sgn} \left( \frac{\partial {u}}{\partial y} \right) \sin{\phi} \, \frac{\partial (g_z h)}{\partial y},
	\label{eq:source_MC_x}
\end{equation}
\begin{equation}
	S_y = g_y h - \frac{{v}}{|\mathbf{{u}}|} \left( h \,g_z \, \mu \right) 
	- h \, \text{sgn} \left( \frac{\partial {v}}{\partial x} \right) \sin{\phi} \, \frac{\partial (g_z h)}{\partial x},
	\label{eq:source_MC_y}
\end{equation}
where $\mathbf{u} = (u, v)$ is the depth-averaged velocity field, and $\mu=\tan\delta$ is the basal friction coefficient. The general formulation includes curvature-correction terms involving the basal radii of curvature (see \citet{sharma2026computational} for the complete derivation); these vanish for the planar inclined-plane geometry used here, yielding Equations~\eqref{eq:source_MC_x}--\eqref{eq:source_MC_y} as written. The same governing equations were solved by both \texttt{TITAN2D} and the PINN framework. In the two-dimensional setting, dimensional equations were used throughout, as \texttt{TITAN2D} provides its solution in dimensional form~\citep{patra2005}.

\section{Physics-Informed Neural Network Framework}
\label{sec:pinn_framework}

\begin{figure}[H]
\centering
\includegraphics[width=\linewidth]{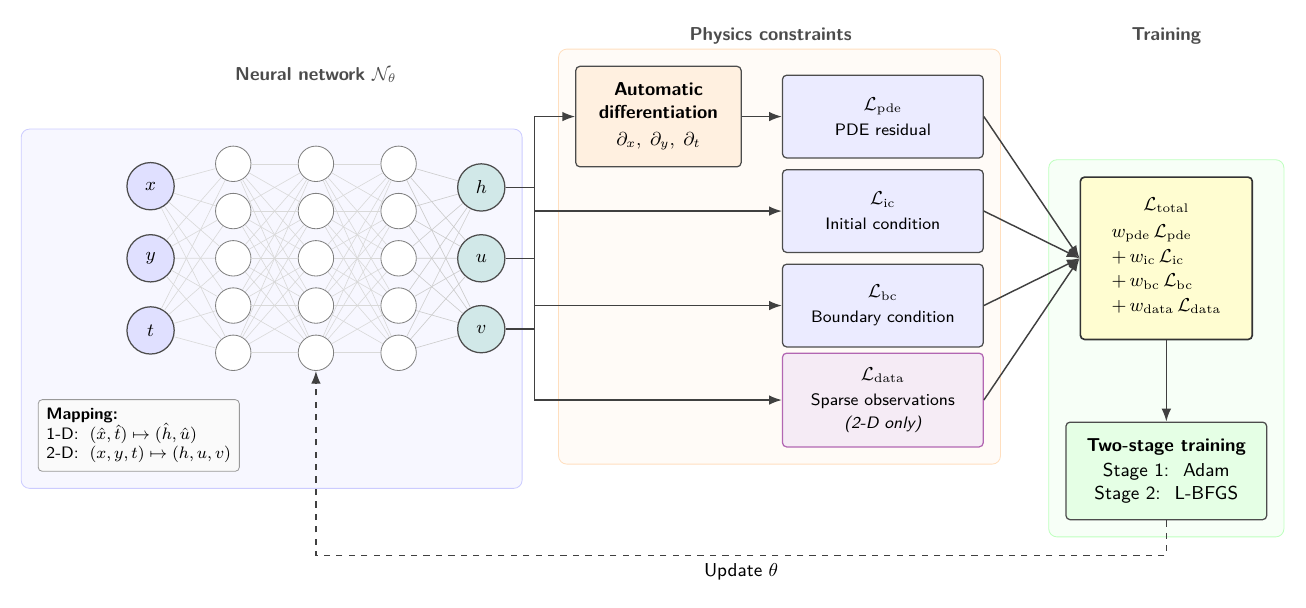}
\caption{Schematic of the PINN framework used in this study.}
\label{fig:pinn_architecture}
\end{figure}

A PINN framework was constructed and trained to solve the depth-averaged Savage--Hutter equations in both one and two dimensions. The framework consists of a fully connected feed-forward neural network whose parameters were optimised by minimising a composite loss function encoding the governing equations, initial conditions, boundary conditions, and, where available, sparse training data points. All spatial and temporal derivatives required to evaluate the governing equations were computed using automatic differentiation, thereby eliminating the need for a mesh or finite-difference stencil~\citep{baydin2018automatic}. A schematic of the framework is shown in Figure~\ref{fig:pinn_architecture}.

The solution fields were approximated by a fully connected feed-forward neural network,
\begin{equation}
    \mathbf{U}_{\theta}(\mathbf{x}) = \mathcal{N}_{\theta}(\mathbf{x}),
    \label{eq:network}
\end{equation}
where $\mathcal{N}_{\theta}$ denotes the network with trainable parameters $\theta$, $\mathbf{x}$ the input coordinates, and $\mathbf{U}_{\theta}$ the predicted solution fields: $(\hat{x}, \hat{t}) \mapsto (\hat{h}, \hat{u})$ for the one-dimensional cases and $(x, y, t) \mapsto (h, u, v)$ for the two-dimensional cases.

All input coordinates were linearly normalised to $[-1, 1]$ using the respective domain bounds. A hyperbolic tangent ($\tanh$) activation, a standard for PINNs applied to conservation laws~\citep{raissi2019physics, mao2020physics}, was used at each hidden layer. In the two-dimensional formulation, a softplus activation $\sigma(z) = \ln(1 + e^{z})$ was additionally applied to the $h$ output neuron to enforce $h \geq 0$; the velocity outputs $u$ and $v$ were left unconstrained. In the one-dimensional setting, all outputs were left unconstrained; $h$ remained non-negative throughout training without enforcement. All weights were initialised using the Xavier uniform scheme, and biases were initialised to zero.

Substituting the network output $\mathbf{U}_{\theta}$ into the governing equations yields PDE residuals that are minimised at collocation points distributed over the interior of the spatio-temporal domain. Automatic differentiation constructs a computational graph through which gradients propagate back to the network parameters, rather than merely evaluating derivatives at given points~\citep{baydin2018automatic}.

For the two-dimensional system, substituting $\mathbf{U}_{\theta} = (h, u, v)$
into Equations~\eqref{eq:2d_mass},~\eqref{eq:da_x_mom}, and~\eqref{eq:da_y_mom} yields three residuals:
\begin{align}
\mathcal{R}_{\mathrm{mass}}
&= \frac{\partial h}{\partial t}
   + \frac{\partial (hu)}{\partial x}
   + \frac{\partial (hv)}{\partial y},
\label{eq:res_mass}\\[4pt]
\mathcal{R}_{x}
&= \frac{\partial (hu)}{\partial t}
   + \frac{\partial}{\partial x}
     \!\left(hu^{2}+\tfrac{1}{2}g_{z}h^{2}\right)
   + \frac{\partial (huv)}{\partial y}
   - S_{x},
\label{eq:res_xmom}\\[4pt]
\mathcal{R}_{y}
&= \frac{\partial (hv)}{\partial t}
   + \frac{\partial (huv)}{\partial x}
   + \frac{\partial}{\partial y}
     \!\left(hv^{2}+\tfrac{1}{2}g_{z}h^{2}\right)
   - S_{y},
\label{eq:res_ymom}
\end{align}
where $S_{x}$ and $S_{y}$ are the Mohr--Coulomb source terms defined in Equations~\eqref{eq:source_MC_x}--\eqref{eq:source_MC_y}. The corresponding one-dimensional residuals follow directly from Equations~\eqref{eq:1d_mass_nd}--\eqref{eq:1d_mom_nd} and are not repeated here. If the network exactly satisfies the governing equations, all three residuals vanish pointwise; the training objective was therefore constructed to minimise them at collocation points sampled within the computational domain.

The conservative flux terms $hu$, $hv$, $hu^{2}$, $hv^{2}$, and $huv$ were assembled as composite tensor expressions and differentiated directly through the computational graph rather than expanded analytically via the product rule beforehand, which would dissolve the coupling between $h$ and $u$ (or $v$) into separate derivative branches before differentiation. \citet{tian2025pinn_swe} showed that this approach yields substantially higher accuracy than the product-rule expansion for shallow-water residuals; the same approach was adopted here.

The network was trained by minimising a composite loss function consisting of weighted physics-based terms. The PDE residual loss enforces the governing conservation laws at $N_{f}$ collocation points sampled within the interior of the spatio-temporal domain:
\begin{equation}
\mathcal{L}_{\mathrm{pde}}
=
\frac{1}{N_f}
\sum_{i=1}^{N_f}
\left(
\left|\mathcal{R}_{\mathrm{mass}}(x_i,y_i,t_i)\right|^2
+
\left|\mathcal{R}_{x}(x_i,y_i,t_i)\right|^2
+
\left|\mathcal{R}_{y}(x_i,y_i,t_i)\right|^2
\right),
\label{eq:loss_pde}
\end{equation}
The initial condition loss enforces agreement with the prescribed initial state at $N_{\mathrm{ic}}$ points sampled along the initial profile:
\begin{equation}
\mathcal{L}_{\mathrm{ic}}
=
\frac{1}{N_{\mathrm{ic}}}
\sum_{j=1}^{N_{\mathrm{ic}}}
\left|
\mathbf{U}_\theta(x_j,y_j,0)
-
\mathbf{U}_{\mathrm{ic}}(x_j,y_j)
\right|^2,
\label{eq:loss_ic}
\end{equation}
where $\mathbf{U}_{\mathrm{ic}}(x_j, y_j)$ denotes the prescribed initial profile evaluated at the $j$-th point. Boundary conditions were incorporated through
\begin{equation}
\mathcal{L}_{\mathrm{bc}}
=
\frac{1}{N_{\mathrm{bc}}}
\sum_{k=1}^{N_{\mathrm{bc}}}
\left|
\mathbf{U}_\theta(x_k,y_k,t_k)
-
\mathbf{U}_{\mathrm{bc}}(x_k,y_k,t_k)
\right|^2,
\label{eq:loss_bc}
\end{equation}
where $\mathbf{U}_{\mathrm{bc}}(x_k, y_k, t_k)$ denotes the prescribed Dirichlet boundary value at the $k$-th point and time $t_k$.

In the two-dimensional formulation, the Savage--Hutter system is homogeneous in the depth-averaged fields: $h \equiv 0$, $u \equiv 0$, $v \equiv 0$ satisfy Equations~\eqref{eq:2d_mass}--\eqref{eq:da_y_mom} identically and therefore minimise $\mathcal{L}_{\mathrm{pde}}$ from the outset of training (see Section~\ref{sec:trivial}). To break this trivial attractor, a sparse data loss was appended:
\begin{equation}
\mathcal{L}_{\mathrm{data}}
=
\frac{1}{N_{d}}
\sum_{m=1}^{N_{d}}
\left|
h_{\theta}(x_m, y_m, t_m) - h_{m}^{\mathrm{obs}}
\right|^2,
\label{eq:loss_data}
\end{equation}
where $N_{d}$ is the number of observed depth values, $h_{m}^{\mathrm{obs}}$ is the measured flow depth at location $(x_m, y_m, t_m)$, and $h_{\theta}$ is the network's depth prediction at that point. The complete training objective was therefore
\begin{equation}
\mathcal{L}_{\mathrm{total}}
=
w_{\mathrm{pde}}\,\mathcal{L}_{\mathrm{pde}}
+
w_{\mathrm{ic}}\,\mathcal{L}_{\mathrm{ic}}
+
w_{\mathrm{bc}}\,\mathcal{L}_{\mathrm{bc}}
+
w_{\mathrm{data}}\,\mathcal{L}_{\mathrm{data}},
\label{eq:total_loss}
\end{equation}
with fixed scalar weights $w_{\mathrm{pde}}$, $w_{\mathrm{ic}}$, $w_{\mathrm{bc}}$, and $w_{\mathrm{data}}$. For the one-dimensional verification cases, $w_{\mathrm{data}} = 0$ and the framework operated as a data-free solver. In the two-dimensional setting $w_{\mathrm{data}} > 0$, rendering the framework a physics-informed, data-assisted hybrid.

The optimisation strategy differed between the one- and two-dimensional settings. 
The one-dimensional verification cases were trained using Adam with a fixed learning 
rate of $\alpha = 10^{-3}$, with the number of epochs specified in  Table~\ref{tab:hp_settings}. For the two-dimensional cases, training was carried out 
in two sequential stages: Adam ($\alpha = 10^{-3}$) for early-stage optimisation, 
followed by up to 1000 steps of L-BFGS to refine the solution once the parameters 
were near a local minimum.
    
The collocation strategy differed between the two settings. For the one-dimensional cases, collocation points were drawn from a uniform distribution over $\hat{x} \in [-8, 10]$, $\hat{t} \in [0, 4]$. For the two-dimensional cases, the temporal coordinates of PDE collocation points were drawn from a Beta$(1, 3)$ distribution to concentrate sampling near $t = 0$, where depth gradients and the growth of the velocity field are steepest.  Collocation points were resampled every 500 epochs during the Adam stage to prevent overfitting, and a \texttt{ReduceLROnPlateau} scheduler halved $\alpha$ whenever $\mathcal{L}_{\mathrm{total}}$ did not decrease for 5000 consecutive epochs. No scheduler was applied in the one-dimensional setting, where a fixed learning rate produced stable convergence throughout.

\section{One-Dimensional Analytical Verification}
\label{sec:1d_benchmarks}
 
\subsection{Problem configurations}
\label{sec:1d_setup}

Three one-dimensional test cases were set up to verify the PINN formulation against the analytical solution of \citet{savage1989}: a decoupled height test, a decoupled velocity test, and a fully coupled test. In all cases, the spatial domain spanned $\hat{x} \in [-8,\,10]$ and the simulation ran over $\hat{t} \in [0,4]$. These bounds were chosen to keep the granular mass within the incline plane throughout, with the asymmetry in $\hat{x}$ reflecting the net downslope transport. The initial condition in all three cases was a parabolic granular pile at rest (Figure~\ref{fig:schematic}),

\begin{equation}
\hat{h}(\hat{x},0) = \max\!\left(1 - \hat{x}^{2},\,0\right),
\qquad
\hat{u}(\hat{x},0) = 0,
\end{equation}
with homogeneous Dirichlet conditions imposed at the spatial boundaries,
\[
\hat{h}(-8,\hat{t})=\hat{h}(10,\hat{t})=0,
\]
to ensure the mass remained confined within the domain. The three configurations are summarised in Table~\ref{tab:hp_settings}.

In the decoupled continuity test, the flow depth $\hat{h}(\hat{x},\hat{t})$ was predicted from the continuity equation alone, while the velocity field was prescribed from the analytical solution $\hat{u}_{\mathrm{ana}}(\hat{x},\hat{t})$ and substituted directly into the continuity residual during training. In the decoupled momentum test, the velocity field  $\hat{u}(\hat{x},\hat{t})$ was learned from the momentum equation, with the analytical height profile $\hat{h}_{\mathrm{ana}}(\hat{x},\hat{t})$ substituted into the momentum residual. In the coupled test, both $\hat{h}(\hat{x},\hat{t})$ and $\hat{u}(\hat{x},\hat{t})$ were learned simultaneously from the full conservative system (Equations~\eqref{eq:1d_mass_nd}--\eqref{eq:1d_mom_nd}), with no field prescribed. The network architectures and sampling parameters for all three cases are given in Table~\ref{tab:hp_settings}; the decoupled and coupled hyperparameter settings were both determined through the sensitivity study presented in Appendix~\ref{app:hp_sensitivity}.
 
\begin{table}[H]
\centering
\caption{Hyperparameter settings for the one-dimensional decoupled and
coupled formulations.}
\label{tab:hp_settings}
\begin{tabular}{llll}
\toprule
Hyperparameter & Decoupled-$h$ & Decoupled-$u$ & Coupled \\
\midrule
\multicolumn{4}{l}{\textit{Network architecture}} \\
Output variables          & $h(x,t)$  & $u(x,t)$ & $h(x,t)$, $u(x,t)$ \\
Width (Neurons per layer)              & 8         & 32        & 64 \\
Depth  (Hidden layers)              & 5          & 5         & 5 \\
Activation                & $\tanh$    & $\tanh$   & $\tanh$ \\
Weight initialisation     & Xavier uniform & Xavier uniform  & Xavier uniform \\
\midrule
\multicolumn{4}{l}{\textit{Training}} \\
Optimiser                  & Adam       & Adam      & Adam \\
Learning rate               & $10^{-3}$  & $10^{-3}$ & $10^{-3}$ \\
Epochs                     & 5,000     & 10,000     & 20,000 \\
\midrule
\multicolumn{4}{l}{\textit{Sampling}} \\
$N_{\mathrm{col}}$ (PDE points)        & 2,000 & 10,000 & 20,000 \\
$N_{\mathrm{ic}}$ (IC points)          & 1,000  & 2,000 & 3,000 \\
$N_{\mathrm{bc}}$ (BC points)          & 1,000  & ---   & 3,000 \\
\midrule
\multicolumn{4}{l}{\textit{Domain}} \\
$x$ range                  & $[-8, 10]$  & $[-8, 10]$  & $[-8, 10]$ \\
$t$ range                  & $[0, 4]$    & $[0, 4]$    & $[0, 4]$ \\
Input normalisation        & $[-1, 1]$  &$[-1, 1]$ &  $[-1, 1]$\\
\midrule
\multicolumn{4}{l}{\textit{Loss weights}} \\
$w_{\mathrm{pde}}$         & 1.0          & 1.0       & 1.0 \\
$w_{\mathrm{ic}}$          & 1.0          & 1.0       & 1.0 \\
$w_{\mathrm{bc}}$          & 1.0          & ---       & 1.0 \\
\bottomrule
\end{tabular}
\end{table}
\subsection{Decoupled verification}
\label{sec:decoupled}

\begin{figure}[H]
    \centering
    \begin{subfigure}[b]{0.32\textwidth}
        \includegraphics[width=\textwidth]{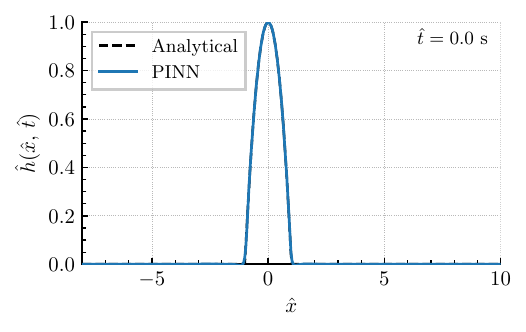}
        \caption{$\hat{t} = 0.0$}
    \end{subfigure}
    \hfill
    \begin{subfigure}[b]{0.32\textwidth}
        \includegraphics[width=\textwidth]{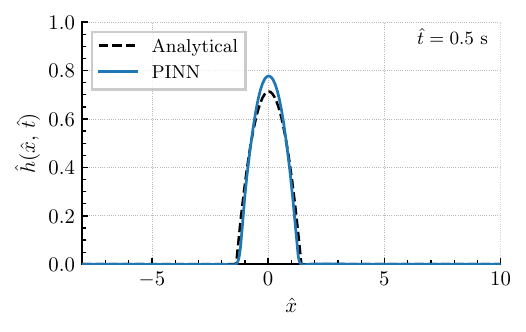}
        \caption{$\hat{t} = 0.5$}
    \end{subfigure}
    \hfill
    \begin{subfigure}[b]{0.32\textwidth}
        \includegraphics[width=\textwidth]{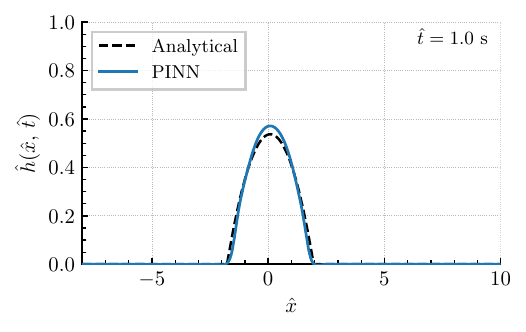}
        \caption{$\hat{t} = 1.0$}
    \end{subfigure}

    \vspace{0.5em}

    \begin{subfigure}[b]{0.32\textwidth}
        \includegraphics[width=\textwidth]{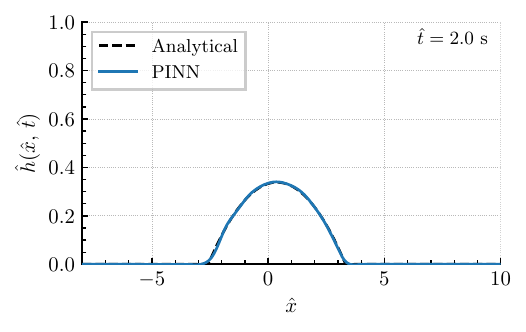}
        \caption{$\hat{t} = 2.0$}
    \end{subfigure}
    \hfill
    \begin{subfigure}[b]{0.32\textwidth}
        \includegraphics[width=\textwidth]{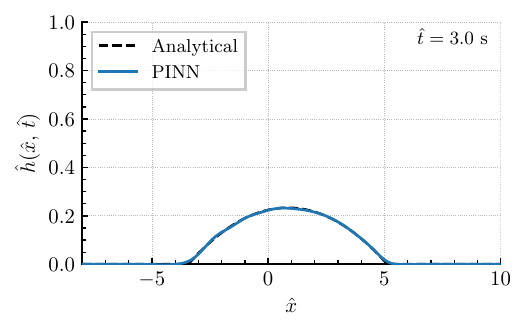}
        \caption{$\hat{t} = 3.0$}
    \end{subfigure}
    \hfill
    \begin{subfigure}[b]{0.32\textwidth}
        \includegraphics[width=\textwidth]{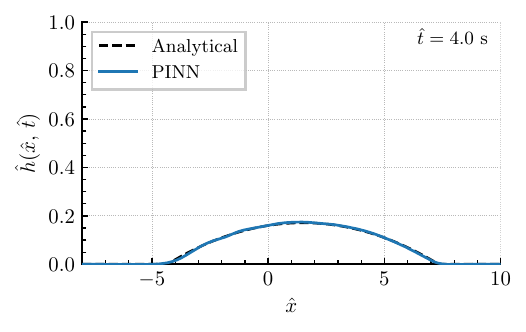}
        \caption{$\hat{t} = 4.0$}
    \end{subfigure}

    \caption{Decoupled-$h$: analytical and PINN-predicted height 
    profiles at six time instants.}
    \label{fig:decoupled_height}
\end{figure}

In both decoupled tests, the PINN reproduced the analytical solution well across the full simulation window. For the Decoupled-$h$ test (Figure~\ref{fig:decoupled_height}), the initial parabolic shape was predicted accurately, and the PINN tracked the analytical profiles closely as the pile spread.

\begin{figure}[H]
    \centering
    \begin{subfigure}[b]{0.32\textwidth}
        \includegraphics[width=\textwidth]{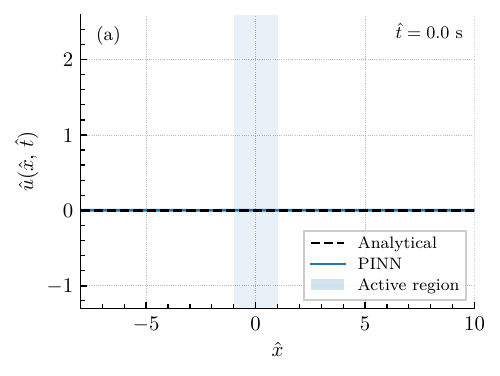}
        \caption{$\hat{t} = 0.0$}
    \end{subfigure}
    \hfill
    \begin{subfigure}[b]{0.32\textwidth}
        \includegraphics[width=\textwidth]{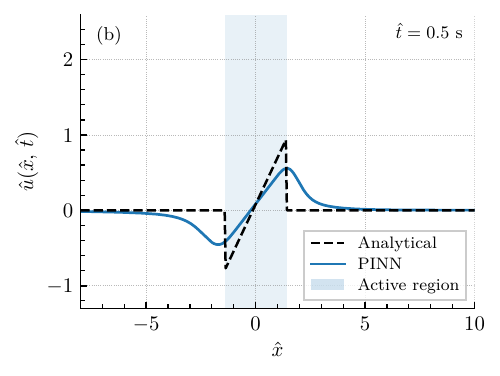}
        \caption{$\hat{t} = 0.5$}
    \end{subfigure}
    \hfill
    \begin{subfigure}[b]{0.32\textwidth}
        \includegraphics[width=\textwidth]{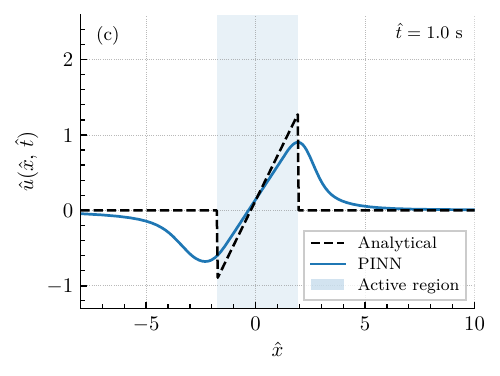}
        \caption{$\hat{t} = 1.0$}
    \end{subfigure}

    \vspace{0.5em}

    \begin{subfigure}[b]{0.32\textwidth}
        \includegraphics[width=\textwidth]{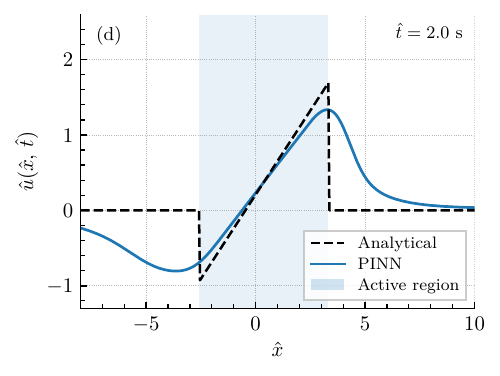}
        \caption{$\hat{t} = 2.0$}
    \end{subfigure}
    \hfill
    \begin{subfigure}[b]{0.32\textwidth}
        \includegraphics[width=\textwidth]{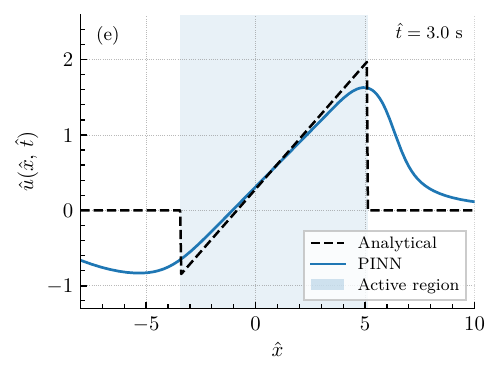}
        \caption{$\hat{t} = 3.0$}
    \end{subfigure}
    \hfill
    \begin{subfigure}[b]{0.32\textwidth}
        \includegraphics[width=\textwidth]{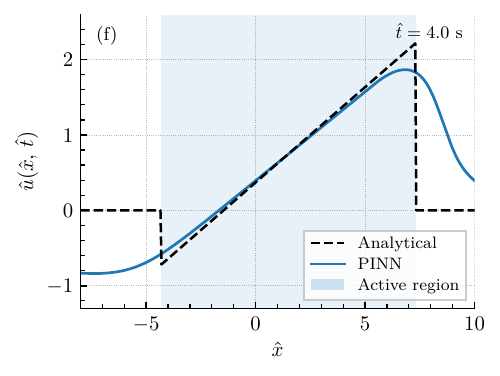}
        \caption{$\hat{t} = 4.0$}
    \end{subfigure}

    \caption{Decoupled-$u$: analytical and PINN-predicted velocity 
    profiles at six time instants.}
    \label{fig:decoupled_velocity}
\end{figure}

For the Decoupled-$u$ test, the predicted velocity profiles are shown in Figure~\ref{fig:decoupled_velocity}. At $\hat{t}=0$, the network returned a near-zero velocity field. As the flow developed, the PINN captured the increasing velocity and tracked the width of the active region throughout the process. Errors were concentrated at the front and rear edges of the pile, where the analytical velocity drops sharply, but the network prediction tapers smoothly beyond the support. This is a known limitation of smooth function approximators.

\begin{figure}[H]
    \centering
    \begin{subfigure}[b]{0.48\textwidth}
        \includegraphics[width=\textwidth]{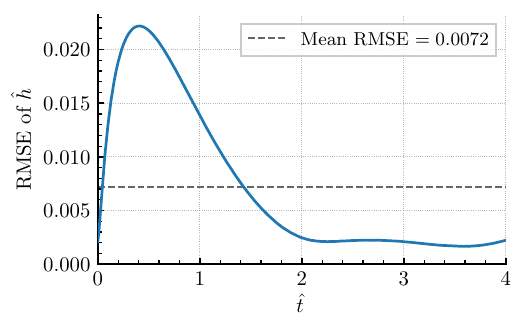}
        \caption{Height RMSE}
    \end{subfigure}
    \hfill
    \begin{subfigure}[b]{0.48\textwidth}
        \includegraphics[width=\textwidth]{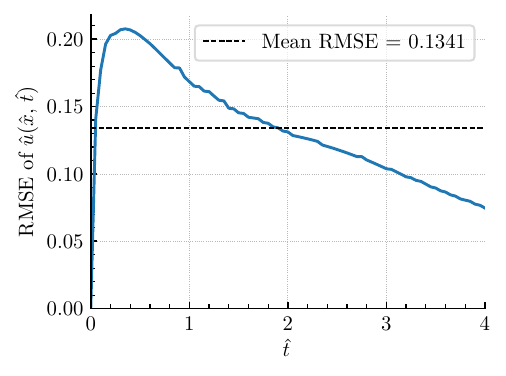}
        \caption{Velocity RMSE}
    \end{subfigure}

    \caption{Decoupled verification: temporal evolution of RMSE for the height field (a) and velocity field (b).}
    \label{fig:decoupled_rmse}
\end{figure}

To quantify agreement in both tests, the time-resolved RMSE was computed as
\begin{equation}
\mathrm{RMSE}_{\phi}(\hat{t})
=
\sqrt{
\frac{1}{N}
\sum_{i=1}^{N}
\left(
\hat{\phi}_{\mathrm{PINN}}(\hat{x}_i,\hat{t}) -
\hat{\phi}_{\mathrm{ana}}(\hat{x}_i,\hat{t})
\right)^2
},
\label{eq:rmse_decoupled}
\end{equation}
where $\hat{\phi} \in \{\hat{h},\,\hat{u}\}$. The temporal evolution of both errors is shown in Figure~\ref{fig:decoupled_rmse}. Both curves share the same qualitative shape: a sharp rise during the early transient, and a steady decline thereafter. The early peak has two causes: the initial profile contains a sharp discontinuity at the pile edges that a smooth $\tanh$ network cannot represent exactly, and the flux divergence is largest and most localised near the compact front at early times. Once the pile flattens and spreads, both difficulties ease.

The difference in magnitude between the two curves is notable. The height RMSE (Figure~\ref{fig:decoupled_rmse}a) peaked at roughly $0.020$ and averaged $7.2\times10^{-3}$ over the full interval, while the velocity RMSE (Figure~\ref{fig:decoupled_rmse}b) peaked near $0.20$ and settled at a higher level throughout. This gap reflects the inherent difference in difficulty between the two equations: the continuity equation is linear in $\hat{h}$ when $\hat{u}$ is prescribed, whereas the momentum equation requires balancing nonlinear flux gradients against Coulomb source terms. Consistently, the Decoupled-$u$ test required a wider network (32 vs.\ 8 neurons) and five times as many collocation points.

\subsection{Coupled verification}
\label{sec:coupled}

In the coupled formulation, a single neural network maps $(\hat{x}, \hat{t}) \mapsto (\hat{h}, \hat{u})$ and learns both fields simultaneously from the conservative Equations~\eqref{eq:1d_mass_nd}--\eqref{eq:1d_mom_nd}. The decoupled tests established that the momentum equation is the harder component; this difficulty is compounded here by the nonlinear coupling between the two fields, so a wider, more densely sampled network was required than in either decoupled test. 

\begin{figure}[H]
\centering
\includegraphics[width=0.78\linewidth]{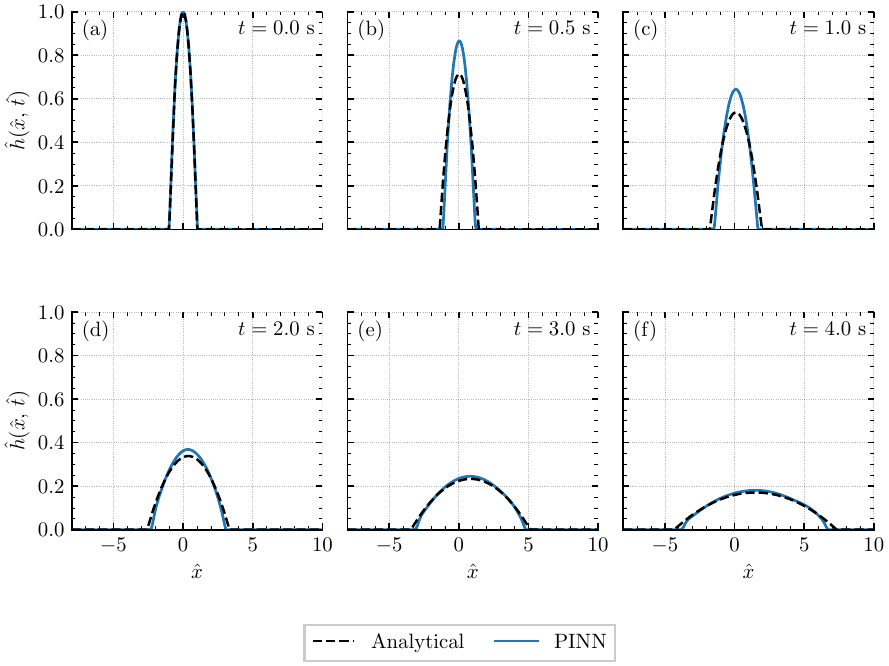}
\caption{Coupled formulation: analytical and PINN-predicted height profiles at
$\hat{t} = 0$, $0.5$, $1.0$, $2.0$, $3.0$, and $4.0$.}
\label{fig:C_height}
\end{figure}

\begin{figure}[H]
\centering
\includegraphics[width=0.78\linewidth]{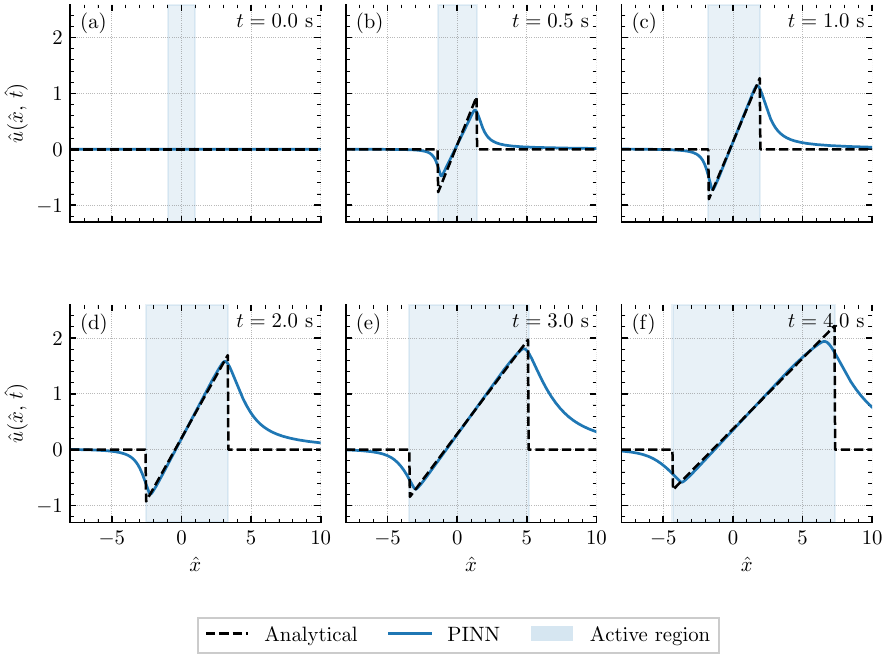}
\caption{Coupled formulation: analytical and PINN-predicted velocity profiles at $\hat{t} = 0$, $0.5$, $1.0$, $2.0$, $3.0$, and $4.0$. The shaded band marks the active region.}
\label{fig:C_velocity}
\end{figure}

\begin{figure}[H]
\centering

\begin{subfigure}[b]{0.48\textwidth}
    \includegraphics[width=\textwidth]{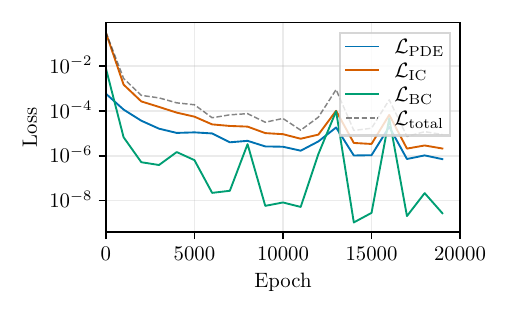}
    \caption{Training loss components}
    \label{fig:C_loss}
\end{subfigure}
\hfill
\begin{subfigure}[b]{0.48\textwidth}
    \includegraphics[width=\textwidth]{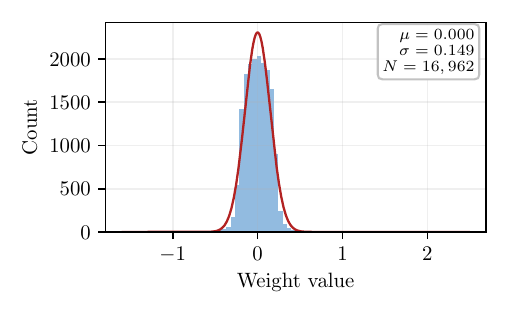}
    \caption{Network weight histogram at final epoch}
    \label{fig:C_weights}
\end{subfigure}

\vspace{0.5em}

\begin{subfigure}[b]{0.60\textwidth}
    \centering
    \includegraphics[width=\textwidth]{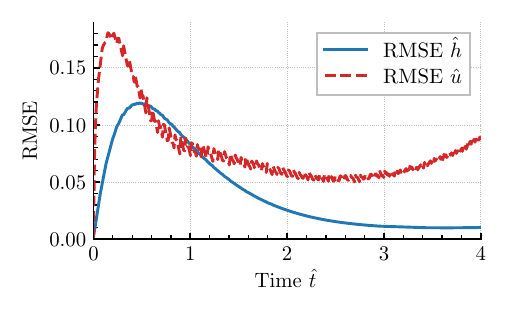}
    \caption{RMSE of height and velocity fields vs.\ time}
    \label{fig:C_rmse_h}
\end{subfigure}
\caption{Coupled PINN training diagnostics: (a)~component losses, (b)~final-epoch weight histogram ($\mu = 0.000$, $\sigma = 0.149$), and (c)~RMSE of height and velocity fields over time.}
\label{fig:C_diag}
\end{figure}

The predicted height and velocity profiles are compared against the analytical solution in Figures~\ref{fig:C_height} and ~\ref{fig:C_velocity}, and the training diagnostics are shown in Figure~\ref{fig:C_diag}. In both fields, the PINN reproduced the analytical solution accurately across the full simulation window, with errors concentrated at the pile peaks during the early transient phase. The height profiles (Figure~\ref{fig:C_height}) show that the initial parabolic pile was recovered accurately at $\hat{t} = 0$, with both the peak height and support width correctly reproduced. As the pile spread and flattened, the PINN closely tracked the analytical shape. At earlier times ($\hat{t} = 0.5$--$1.0$), small deviations appeared near the pile peak.

The velocity profiles (Figure~\ref{fig:C_velocity}) showed the same overall pattern. At $\hat{t} = 0$, the network returned the near-zero velocity field, consistent with the initial condition. As the flow developed, the PINN captured the increasing velocity and reproduced the linear velocity structure within the active region, with both slope and magnitude in good agreement with the analytical solution throughout. Discrepancies were confined to the front and rear edges of the pile, where the prediction tapered smoothly beyond the sharp cutoff of the analytical solution.

The training diagnostics (Figure~\ref{fig:C_diag}) confirmed that all three loss components decreased steadily over 20,000 epochs, with $\mathcal{L}_{\mathrm{PDE}}$ and $\mathcal{L}_{\mathrm{IC}}$ reaching $\mathcal{O}(10^{-5}\text{--}10^{-6})$ by the end of training. The final-epoch weight histogram was Gaussian ($\mu = 0.000$ and $\sigma = 0.149$), with no sign of exploding or vanishing weights. The RMSE for both fields peaked near $\hat{t} \approx 0.3$--$0.5$ and then decayed, with the height error settling below $0.025$ for the remainder of the simulation. Averaged over the full simulation window, the mean RMSE was $4.3 \times 10^{-2}$ for height and $7.9 \times 10^{-2}$ for velocity.

\section{Two-Dimensional Validation: Experimental Granular Avalanche}
\label{sec:2d}

\subsection{Experimental configuration}
\label{sec:2d_exp}
The framework was next validated against the laboratory experiments of \citet{maeno2013unconfined}, in which a hollow cylinder filled with dry sand was rapidly lifted from an inclined surface, allowing the pile to collapse freely under gravity. Four experimental cases were considered, spanning two pile heights and two slope angles, as summarised in Table~\ref{tab:maeno_cases}.

\begin{table}[H]
    \centering
    \caption{
        Experimental cases from \citet{maeno2013unconfined} were used to validate the two-dimensional PINN. All cases share $R_0 = \SI{0.05}{\metre}$, $\rho = \SI{1481}{\kilogram\per\metre\cubed}$, $\phi = 24^{\circ}$, and $\delta = 28^{\circ}$.}
    \label{tab:maeno_cases}
    \begin{tabular}{lS[table-format=1.1]S[table-format=2.0]
                      S[table-format=1.3]S[table-format=1.2]}
        \toprule
        {Case}
            & {$M$ (\si{\kilogram})}
            & {$\zeta$ (\si{\degree})}
            & {$H_0$ (\si{\metre})}
            & {$t_{\max}$ (\si{\second})} \\
        \midrule
        C1 & 1.0 & 10 & 0.085 & 0.70 \\
        C2 & 1.0 & 15 & 0.085 & 0.70 \\
        C3 & 2.5 & 10 & 0.215 & 0.95 \\
        C4 & 2.5 & 15 & 0.215 & 0.95 \\
        \bottomrule
    \end{tabular}
\end{table}

\begin{figure}[H]
    \centering
    \begin{subfigure}[b]{0.48\textwidth}
        \centering
        \includegraphics[width=\textwidth]{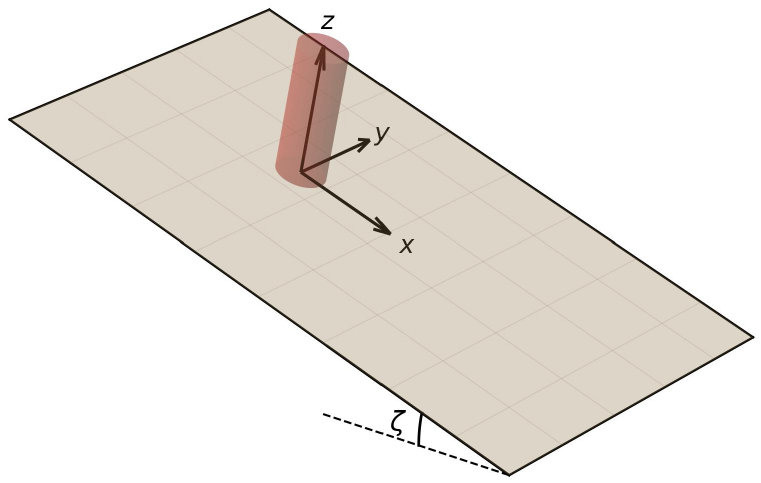}
        \caption{}
        \label{fig:setup_initial}
    \end{subfigure}
    \hfill
    \begin{subfigure}[b]{0.48\textwidth}
        \centering
        \includegraphics[width=\textwidth]{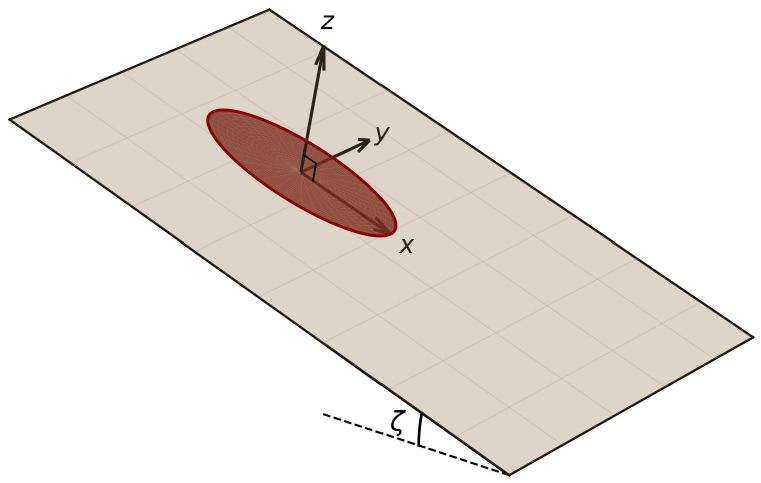}
        \caption{}
        \label{fig:setup_final}
    \end{subfigure}
    \caption{Three-dimensional schematic of the experimental configuration
        on a plane inclined at angle $\zeta$ to the horizontal, with $x$
        directed down-slope, $y$ cross-slope, and $z$ normal to the
        inclined surface. (a)~Initial condition: a cylindrical granular
        pile of radius $R_0$ and height $H_0$ centred at the origin.
        (b)~Final deposit: the collapsed pile spreads into an elongated
        elliptical deposit with greater runout in the downslope direction.}
    \label{fig:domain_setup}
\end{figure}

The domain for all cases spanned $x \in [-0.4,\,0.8]\,\si{\metre}$ in the downslope direction and $y \in [-0.4,\,0.4]\,\si{\metre}$ in the cross-slope direction, with the pile centroid initially at the origin. These bounds were chosen so the spreading mass remained within the domain throughout. The geometry is illustrated in Figure~\ref{fig:domain_setup}.

The primary analysis focuses on Case~C4 ($M = \SI{2.5}{\kilogram}$, $\zeta = 15^{\circ}$), which combines the highest aspect ratio with the steepest slope and therefore represents the most demanding test of the framework. Results for Cases~C1--C3 are also presented.

\subsection{Trivial-solution pathology in two-dimensional PINNs}
\label{sec:trivial}

A fundamental difficulty arises when the Savage--Hutter system is trained using only PDE, initial-condition and boundary-condition losses: the network converges to the trivial zero solution $h \equiv 0$, $u \equiv 0$, $v \equiv 0$, rather than the physically meaningful flow. Two mechanisms drive this collapse.  

First, the Savage--Hutter system is homogeneous in the depth-averaged fields, so $h = u = v = 0$ satisfies all governing equations exactly and yields a zero PDE residual, making the trivial solution a valid minimiser of the PDE loss from the start of training. Second, the physical deposit occupies only a small fraction of the domain; the optimiser is continuously pushed toward zero by a large proportion of collocation points lying outside the wetted region. The contribution of the second mechanism was investigated separately through a pseudo-2D strip-flow experiment, in which the cylindrical pile was replaced by a full-width strip to increase the wetted fraction of the domain; the results are presented in Appendix~\ref{app:obstacle}.

Several strategies were explored to prevent this collapse. These included hard enforcement of the initial condition, causal weighting to prioritise the early-time solution, learning-rate scheduling, an additional mass-conservation loss, and non-dimensionalisation of the governing equations. Hard enforcement of the initial condition prevented the solution from collapsing to zero, but the network instead reproduced an almost stationary pile with little spreading or downslope motion. Causal weighting improved the temporal evolution and allowed the pile to spread, but the predicted height still decreased excessively at later times. Learning-rate scheduling slowed the progression toward the trivial state but did not prevent the eventual collapse. Similarly, the mass-conservation loss reduced the rate of mass loss but was not sufficient to maintain a physically meaningful solution throughout the simulation. Non-dimensionalisation improved the overall flow behaviour and produced the most physically consistent result among the purely physics-based formulations, but the predicted height still collapsed at later times. Therefore, although these modifications improved various aspects of the training behaviour, none was sufficient on its own to yield a stable and physically meaningful solution over the entire simulation period.

Together, these two mechanisms create a loss landscape in which the trivial solution sits in a deep, wide basin that is difficult to escape through gradient-based optimisation alone. To break this, a sparse data loss was appended to the training objective, incorporating a small number of experimental observations that provide a signal strong enough to pull the network away from zero and anchor it toward a physically meaningful solution. 

\subsection{Minimum-data requirement study}
\label{sec:ablation}

To determine the minimum number of data points required to escape the trivial solution identified in Section~\ref{sec:trivial}, a systematic study of data requirements was conducted, in which the number of data points per profile, $N$, was varied from 1 to 20. At each value of $N$, points were sampled along the centerline profile ($y = 0$) and the cross-slope profile ($x = 0$) of the deposit at the final time, yielding $2N$ data points in total. Three independent training runs were performed at each $N$ to assess repeatability; details are provided in the Appendix.

The global height RMSE decreased sharply up to $N = 3$, showed only a small further improvement by $N = 5$, and changed little thereafter. On this basis, $N = 5$ points per profile was adopted for all subsequent cases, giving 10 training points per case in total.

\begin{figure}[H]
\centering
\includegraphics[width=0.48\textwidth]{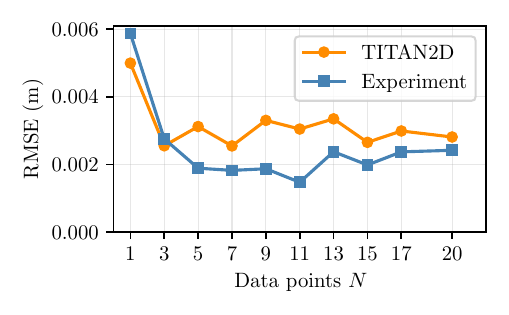}
\caption{Global height RMSE as a function of the number of observations per profile, $N$, for C4 ($2.5\,\text{kg}$, $15^\circ$), compared with \texttt{TITAN2D} (orange circles) and experimental measurements (blue squares).}
\label{fig:rmse_vs_n}
\end{figure}

The ten training data points were selected to cover the key features of the deposit geometry in both directions: the five centerline points sampled the upstream face, the deposit peak, and the downstream tail, while the five cross-slope points spanned the full lateral extent of the deposit about the centerline. Exact point coordinates are provided in Appendix~\ref{app:data_points}.

\subsection{Network architecture and two-stage training}
\label{sec:2d_arch}

The two-dimensional PINN took the input triple $(x, y, t)$ and produced the output triple $(h, u, v)$ via a fully connected network. All inputs were linearly scaled to $[-1, 1]$ using the domain bounds before entering the network. A softplus activation was applied to the $h$ output to enforce non-negativity of the flow depth, while $u$ and $v$ were left unconstrained. The architecture is summarised in Table~\ref{tab:arch_2d}.

\begin{table}[H]
\centering
\caption{2D PINN architecture and training configuration for the
\citet{maeno2013unconfined} cases.}
\label{tab:arch_2d}
\begin{tabular}{ll}
\toprule
Property & Value \\
\midrule
Input dimension                  & 3 ($x$, $y$, $t$) \\
Hidden layers                    & 5 \\
Neurons per layer                & 64 \\
Activation                       & $\tanh$ \\
Output activation on $h$         & softplus \\
Weight initialisation            & Xavier uniform \\
Total trainable parameters       & 17,091 \\
\midrule
Optimisers                       & Adam, L-BFGS \\
Adam epochs                      & 500,000 \\
Initial learning rate $\alpha$   & $10^{-3}$ \\
L-BFGS steps                     & 1,000 \\
PDE collocation points $N_f$     & 50,000 \\
Initial-condition points $N_{\mathrm{ic}}$ & 5,000 \\
Boundary-condition points $N_{\mathrm{bc}}$ & 4,000 \\
$w_{\mathrm{pde}},\,w_{\mathrm{ic}},\,w_{\mathrm{bc}}$ & 1 \\
$w_{\mathrm{data}}$              & 50 \\
\bottomrule
\end{tabular}
\end{table}

Training followed the two-stage protocol of Section~\ref{sec:pinn_framework}, using the configuration in Table~\ref{tab:arch_2d}: 500,000 Adam epochs with a scheduler and collocation resampling, followed by up to 1,000 L-BFGS steps. The physics loss weights were set to $w_{\mathrm{pde}} = w_{\mathrm{ic}} = w_{\mathrm{bc}} = 1$, while the data weight was elevated to $w_{\mathrm{data}} = 50$. This elevation was necessary to prevent data loss from vanishing before the PDE and IC losses drove the network away from the trivial zero solution. The value $w_{\mathrm{data}} = 50$ was selected empirically to consistently prevent collapse across repeated initialisations.

All two-dimensional experiments were run on an NVIDIA RTX~A5000 GPU, requiring approximately 6.5\,h per case, compared with 3--12\, min for the corresponding \texttt{TITAN2D} simulations (Table~\ref{tab:cost_comparison}). At present, the PINN framework is  computationally expensive to train than the conventional solver. 
This cost, however, corresponds to training each case independently from random initialisation. To investigate whether this overhead could be reduced, a preliminary transfer-learning study was conducted in which the network trained for Case~C4 was used to initialise Case~C2 rather than being reinitialised from scratch. Under this scheme, the training time was reduced by approximately $94\,\%$ relative to training from scratch, while the predicted deposit remained in close agreement with the corresponding \texttt{TITAN2D} reference.  Although demonstrated for only a single case pair, this result suggests that a pre-trained network can serve as an effective initialisation for related flow conditions, substantially reducing the computational cost of training new cases. 
Beyond computational performance, PINNs offer an additional practical advantage over conventional numerical methods. Once the governing equations are incorporated into the loss function, the same computational framework can be applied to different classes of partial differential equations without developing new discretisation schemes or implementing specialised numerical treatments for additional terms. Consequently, the effort required to derive, code, and validate discretisations for modified governing equations can be significantly reduced.
\begin{table}[H]
\centering
\caption{Wall-clock training and simulation times for all four experimental cases. PINN training was performed on an NVIDIA RTX A5000 GPU; \texttt{TITAN2D} simulations were run on 8 CPU cores. The PINN time is identical across cases because training duration depends solely on the machine-learning parameters, which are held fixed throughout.}
\label{tab:cost_comparison}
\begin{tabular}{llcc}
\toprule
Case
    & Configuration
    & PINN training time (h)
    & \texttt{TITAN2D} simulation time (min) \\
\midrule
C1 & $1.0\,\text{kg}$, $10^{\circ}$ & $\approx 6.5$ &  3 \\
C2 & $1.0\,\text{kg}$, $15^{\circ}$ & $\approx 6.5$ &  7 \\
C3 & $2.5\,\text{kg}$, $10^{\circ}$ & $\approx 6.5$ &  6 \\
C4 & $2.5\,\text{kg}$, $15^{\circ}$ & $\approx 6.5$ & 12 \\
\bottomrule
\end{tabular}
\end{table}

\subsection{Results: primary case C4}
\label{sec:2d_results}

The results for C4 are presented in three parts: training convergence, spatial and temporal evolution of the deposit, and quantitative error metrics. The repeatability of these results across different random initialisations is assessed in Appendix~\ref{app:repeatability}.

\begin{figure}[H]
\centering
\begin{subfigure}[b]{0.48\textwidth}
    \includegraphics[width=\textwidth]{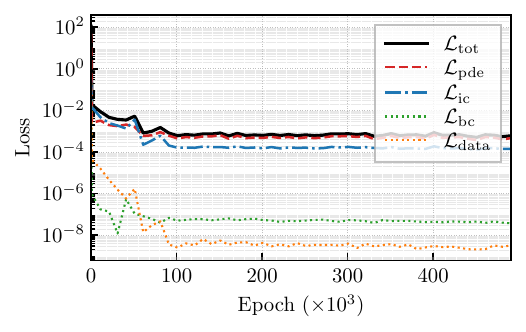}
    \caption{Adam stage (500\,000 epochs)}
\end{subfigure}
\hfill
\begin{subfigure}[b]{0.48\textwidth}
    \includegraphics[width=\textwidth]{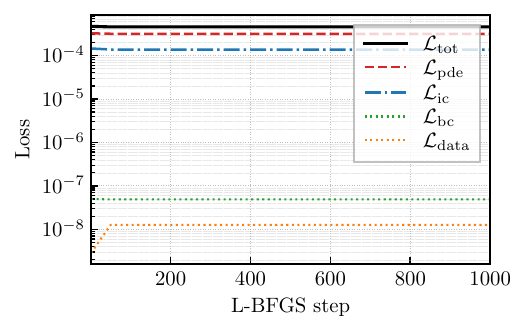}
    \caption{L-BFGS stage (1000 steps)}
\end{subfigure}
\caption{Training loss history for C4.  (a)~Adam stage: all loss
components decreased monotonically; the data loss dominated early training and guided the network away from the trivial solution.  (b)~L-BFGS stage:
rapid convergence within $\approx 50$ steps, reducing $\mathcal{L}_{\text{total}}$ from $6.0\times10^{-4}$ to $4.5\times10^{-4}$.}
\label{fig:2d_loss}
\end{figure}

Training convergence is shown in Figure~\ref{fig:2d_loss}. By the end of the Adam stage, the total loss has decreased to $\mathcal{L}_{\text{total}} \approx 6.0\times10^{-4}$, with $\mathcal{L}_{\text{pde}} \approx 4.6\times10^{-4}$ and $\mathcal{L}_{\text{ic}} \approx 1.4\times10^{-4}$. The data loss fell to near zero by mid-training, indicating the network had fitted the ten training points and was extrapolating via the PDE and IC losses. The subsequent L-BFGS stage converged within approximately 50 steps, reducing the total loss to $\mathcal{L}_{\text{total}} \approx 4.5\times10^{-4}$.

\begin{figure}[H]
\centering
\setlength{\abovecaptionskip}{2pt}
\setlength{\belowcaptionskip}{0pt}
\captionsetup[subfigure]{aboveskip=1pt, belowskip=-2pt}

\begin{subfigure}[b]{0.32\textwidth}
    \includegraphics[width=\textwidth]{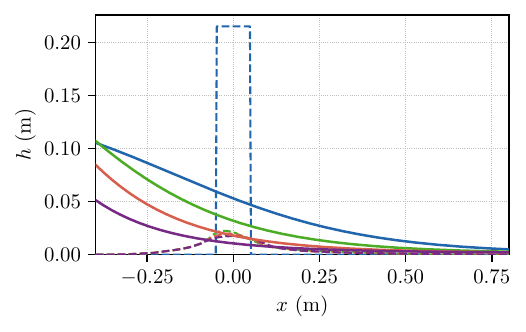}
    \caption{Epoch $10^{4}$}
\end{subfigure}
\hfill
\begin{subfigure}[b]{0.32\textwidth}
    \includegraphics[width=\textwidth]{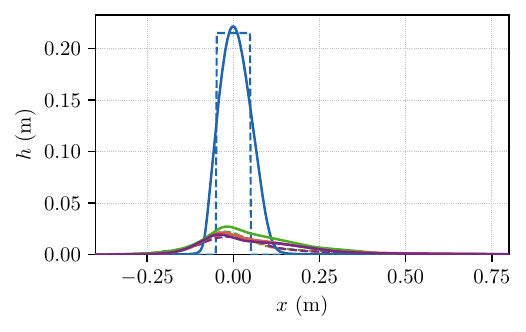}
    \caption{Epoch $10^{5}$}
\end{subfigure}
\hfill
\begin{subfigure}[b]{0.32\textwidth}
    \includegraphics[width=\textwidth]{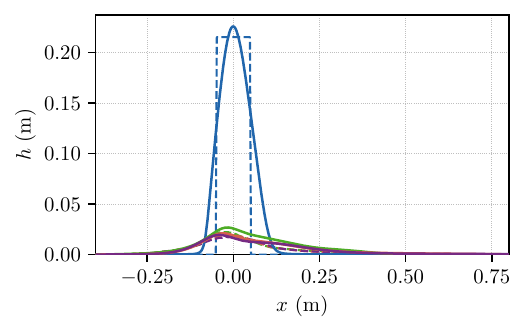}
    \caption{Epoch $2\times10^{5}$}
\end{subfigure}

\vspace{0.5em}

\begin{subfigure}[b]{0.32\textwidth}
    \includegraphics[width=\textwidth]{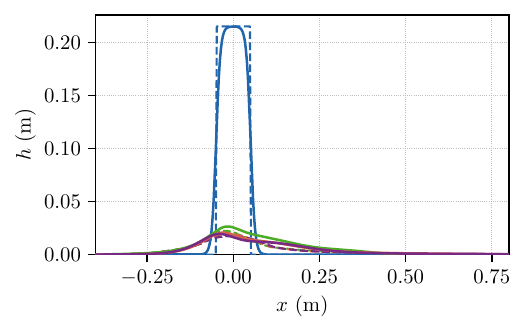}
    \caption{Epoch $3\times10^{5}$}
\end{subfigure}
\hfill
\begin{subfigure}[b]{0.32\textwidth}
    \includegraphics[width=\textwidth]{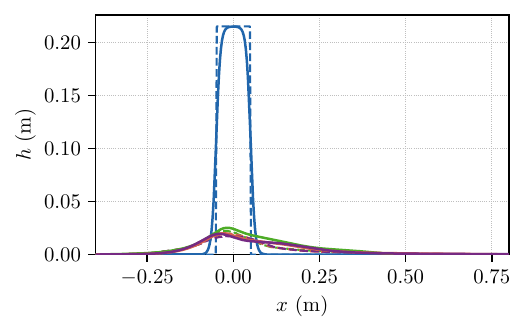}
    \caption{Epoch $5\times10^{5}$}
\end{subfigure}
\hfill
\begin{subfigure}[b]{0.32\textwidth}
    \includegraphics[width=\textwidth]{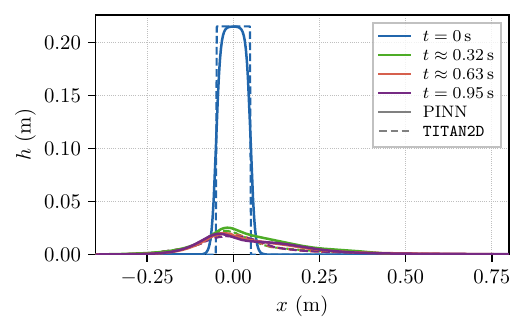}
    \caption{L-BFGS (final)}
\end{subfigure}

\caption{Centerline profile $h(x,\,y{=}0,\,t)$ for C4 at six
training checkpoints. PINN (solid), \texttt{TITAN2D} (dashed) at
$t = 0$, $0.32$, $0.63$, $0.95\,\text{s}$.}
\label{fig:2d_centerline_epochs}
\end{figure}

Figure~\ref{fig:2d_centerline_epochs} presents the predicted centerline height profile across six training checkpoints. At $10^4$ Adam epochs, the network had not yet recovered the initial condition or any later-time profiles. The initial condition was captured by $10^{5}$ epochs, and the bulk of the dynamic information was captured between $10^{5}$ and $3\times10^{5}$ epochs, as the profiles progressively aligned with \texttt{TITAN2D}. Beyond $3\times10^{5}$ epochs, additional Adam iterations yielded only marginal changes, whereas L-BFGS sharpened the wavefront and removed edge artefacts.

\begin{figure}[H]
\centering
\begin{subfigure}[b]{0.48\textwidth}
\includegraphics[width=\textwidth]{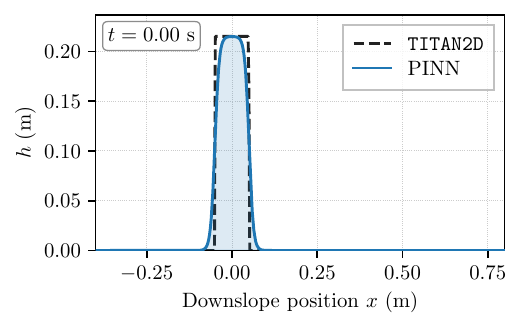}
\end{subfigure}
\hfill
\begin{subfigure}[b]{0.48\textwidth}
\includegraphics[width=\textwidth]{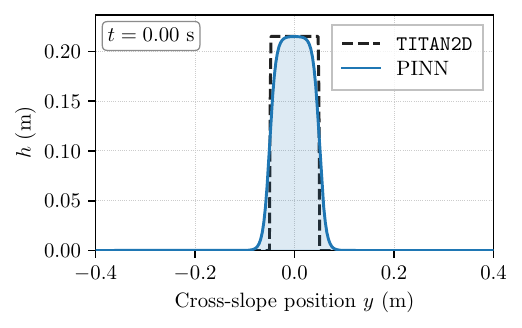}
\end{subfigure}
\vspace{-0.6em}
\begin{subfigure}[b]{0.48\textwidth}
\includegraphics[width=\textwidth]{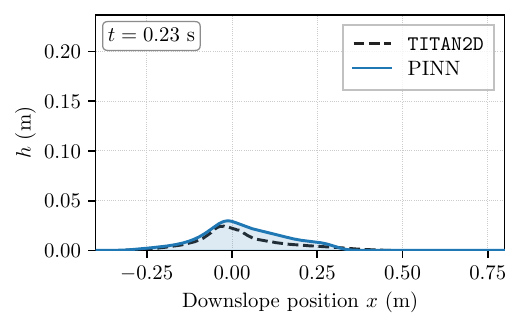}
\end{subfigure}
\hfill
\begin{subfigure}[b]{0.48\textwidth}
\includegraphics[width=\textwidth]{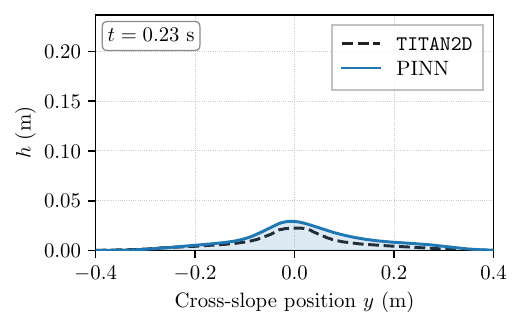}
\end{subfigure}
\vspace{-0.6em}
\begin{subfigure}[b]{0.48\textwidth}
\includegraphics[width=\textwidth]{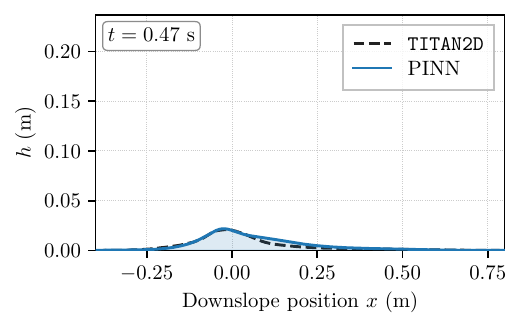}
\end{subfigure}
\hfill
\begin{subfigure}[b]{0.48\textwidth}
\includegraphics[width=\textwidth]{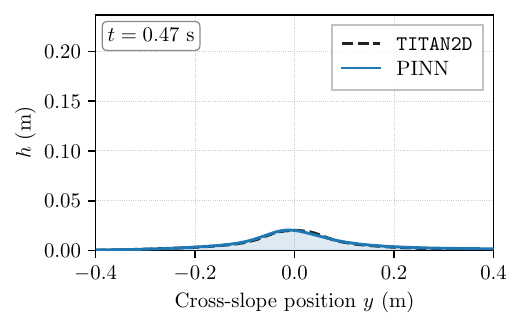}
\end{subfigure}
\vspace{-0.6em}
\begin{subfigure}[b]{0.48\textwidth}
\includegraphics[width=\textwidth]{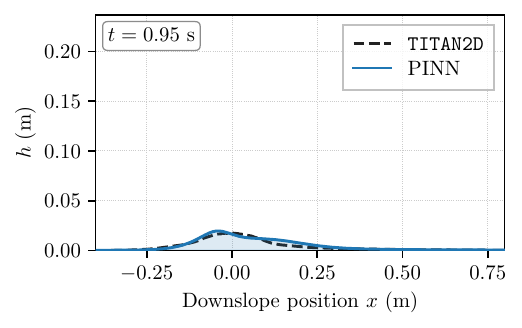}
\end{subfigure}
\hfill
\begin{subfigure}[b]{0.48\textwidth}
\includegraphics[width=\textwidth]{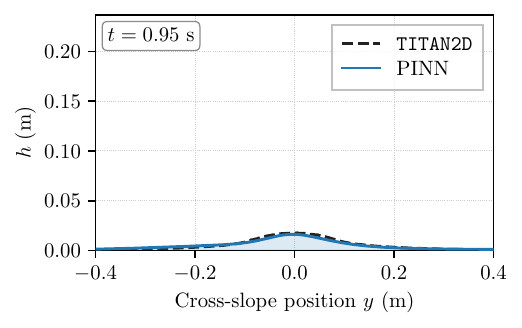}
\end{subfigure}
\caption{Centerline (left) and cross-slope (right) height profiles at four time snapshots for C4. PINN (solid blue) and ~\texttt{TITAN2D} (dashed orange).}
\label{fig:2d_profiles_time}
\end{figure}

The centerline and cross-slope height profiles at the four time snapshots are shown in Figure~\ref{fig:2d_profiles_time}. The cylindrical initial condition was recovered accurately at $t = 0\,\text{s}$ in both directions, and the collapse and lateral spreading were closely tracked across all intermediate snapshots. The front position and runout length agreed well with \texttt{TITAN2D} along the centerline, and the deposit width and peak height were well reproduced in the cross-slope direction throughout. 

\begin{figure}[H]
\centering
\begin{subfigure}[b]{0.48\textwidth}
\includegraphics[width=\textwidth]{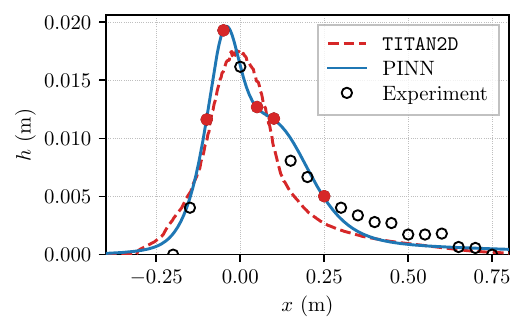}
\end{subfigure}
\hfill
\begin{subfigure}[b]{0.48\textwidth}
\includegraphics[width=\textwidth]{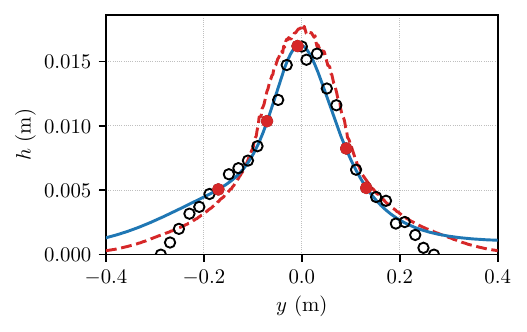}
\end{subfigure}
\caption{Final deposit profiles at $t = 0.95\,\text{s}$ for C4.
Left: centerline ($y = 0$); right: cross-slope ($x = 0$). PINN (solid
blue), \texttt{TITAN2D} (dashed red), and \citet{maeno2013unconfined} experimental data (circles). Filled red circles indicate the training data points.}
\label{fig:2d_final_deposit}
\end{figure}

The final deposit profiles at $t = 0.95\,\text{s}$ are compared 
with the experimental measurements of \citet{maeno2013unconfined} in Figure~\ref{fig:2d_final_deposit}, with the ten training data points indicated by filled red circles. Along the centerline, the PINN accurately reproduced the peak height and downstream tail out to the full runout, whereas \texttt{TITAN2D} slightly underpredicted both. Along the cross-slope direction, both solvers accurately reproduced the deposit width, though \texttt{TITAN2D} produced a sharper lateral cutoff, whereas the PINN yielded a smoother transition to the dry region. This closer agreement suggests that the combination of the physics constraint and sparse training data points steered the solution toward the measured deposit geometry.

\begin{figure}[H]
\centering
\begin{subfigure}[b]{0.24\textwidth}
\includegraphics[width=\textwidth]{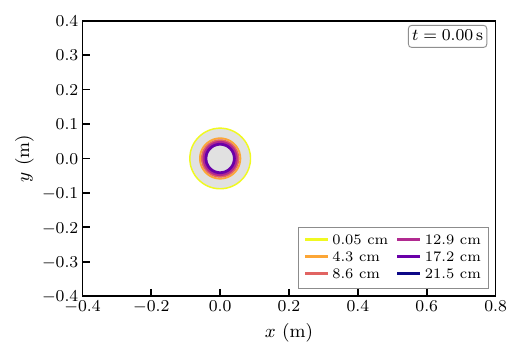}
\end{subfigure}
\hfill
\begin{subfigure}[b]{0.24\textwidth}
\includegraphics[width=\textwidth]{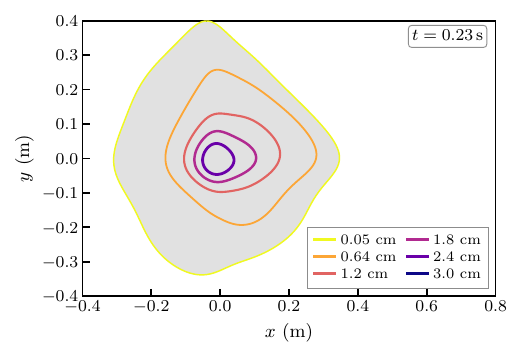}
\end{subfigure}
\hfill
\begin{subfigure}[b]{0.24\textwidth}
\includegraphics[width=\textwidth]{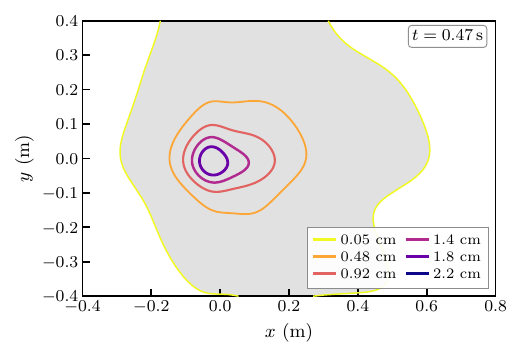}
\end{subfigure}
\hfill
\begin{subfigure}[b]{0.24\textwidth}
\includegraphics[width=\textwidth]{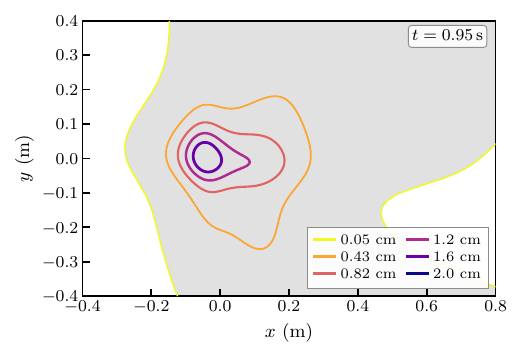}
\end{subfigure}
\vspace{-0.5em}
\begin{subfigure}[b]{0.24\textwidth}
\includegraphics[width=\textwidth]{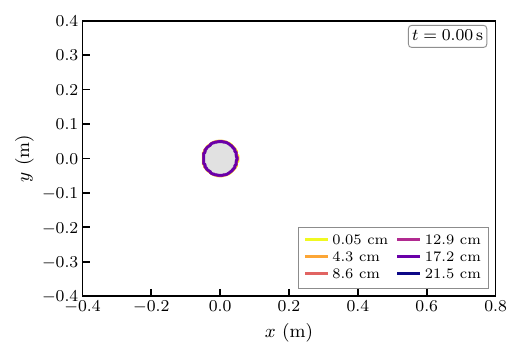}
\end{subfigure}
\hfill
\begin{subfigure}[b]{0.24\textwidth}
\includegraphics[width=\textwidth]{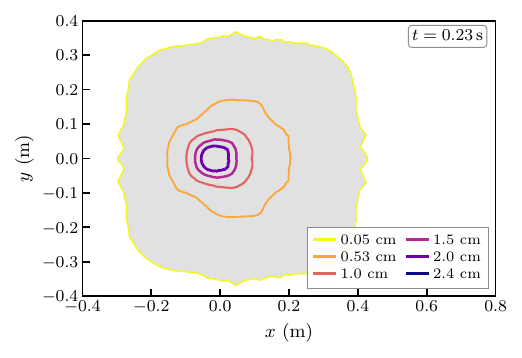}
\end{subfigure}
\hfill
\begin{subfigure}[b]{0.24\textwidth}
\includegraphics[width=\textwidth]{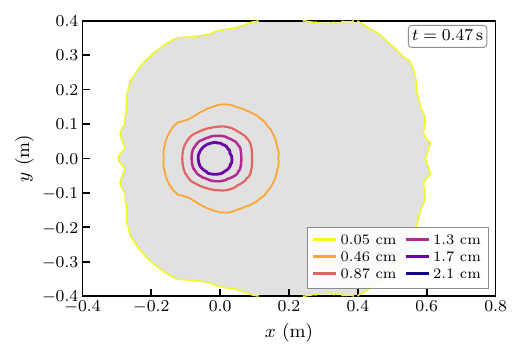}
\end{subfigure}
\hfill
\begin{subfigure}[b]{0.24\textwidth}
\includegraphics[width=\textwidth]{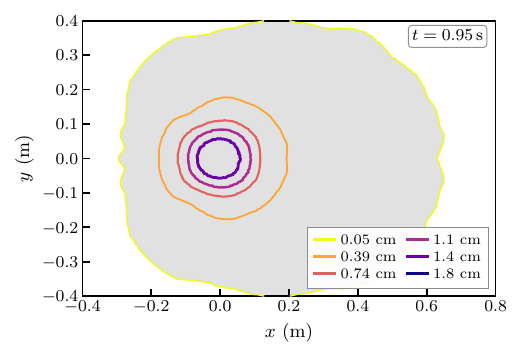}
\end{subfigure}
\caption{Depth contours for C4 at $t = 0.00$, $0.23$, $0.47$, and $0.95\,\text{s}$ (columns, left to right). Top row: PINN; bottom row: \texttt{TITAN2D}. Contour levels are identical across columns. The grey region marks the wetted footprint, defined as the area where the flow depth exceeds the minimum threshold.}
\label{fig:2d_contours}
\end{figure}

The depth contours at four time instants are compared in Figure~\ref{fig:2d_contours}. At $t = 0\,\text{s}$, both solutions reproduced the cylindrical initial pile with peak depths near $21.5\,\text{cm}$ and a compact circular footprint. As the avalanche developed, the deposit spread preferentially downslope, with PINN contour shapes and levels remaining in close agreement with \texttt{TITAN2D}. For visualisation, a depth cutoff of $0.05\,\text{cm}$ was used to display the low-depth extent of the flow and as the lowest contour level in each panel. This visualisation cutoff is distinct from the threshold $h_{\mathrm{thr}} = H_0/50$ used for the quantitative IoU and wetted-area calculations. By $t = 0.95\,\text{s}$, both solutions yielded a thin, elongated deposit with peak depths below $2\,\text{cm}$, but the PINN extended visibly farther downslope than \texttt{TITAN2D}.

\begin{figure}[H]
\centering
\begin{subfigure}[b]{0.48\textwidth}
\includegraphics[width=\textwidth]{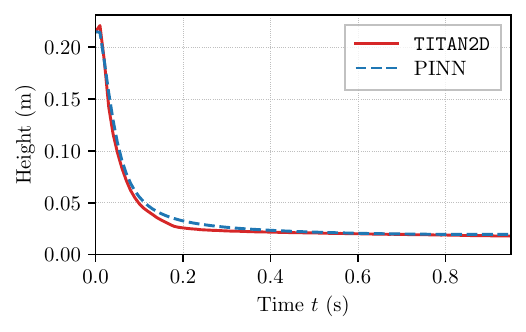}
\end{subfigure}
\hfill
\begin{subfigure}[b]{0.48\textwidth}
\includegraphics[width=\textwidth]{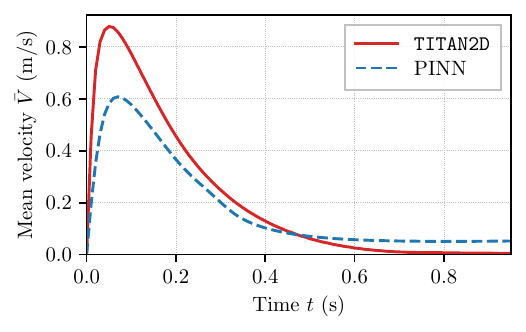}
\end{subfigure}
\vspace{-0.9em}
\begin{subfigure}[b]{0.48\textwidth}
\includegraphics[width=\textwidth]{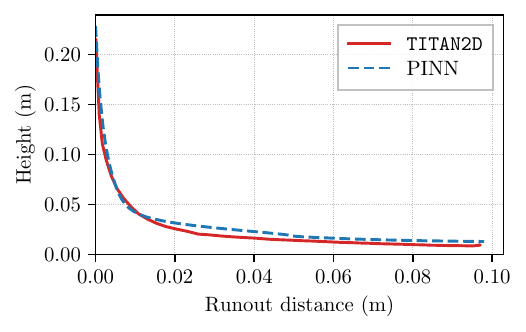}
\end{subfigure}
\hfill
\begin{subfigure}[b]{0.48\textwidth}
\includegraphics[width=\textwidth]{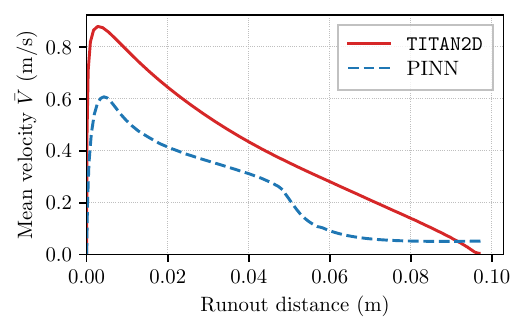}
\end{subfigure}
\caption{Integral flow diagnostics for C4: $h_{\max}$ and $\bar{V}$ versus time (top row) and runout distance (bottom row). \texttt{TITAN2D} (solid red); PINN (dashed blue).}
\label{fig:2d_diagnostics}
\end{figure}

The integral flow diagnostics are presented in Figure~\ref{fig:2d_diagnostics}. The maximum height $h_{\max}$ decayed monotonically from $21.5\,\text{cm}$ as material redistributed downslope, and the PINN tracked this decay closely throughout. The mean velocity $\bar{V}$ increased sharply over the first $0.1\,\text{s}$, peaked near $0.7\,\text{m/s}$, and then decreased steadily as basal friction dissipated momentum; both phases were well reproduced, with only a small offset near the peak. The same quantities, plotted against runout distance, show the height traces remaining closely aligned throughout, while the velocity trace shows a larger gap in the middle of the runout range before converging again near the end.

\begin{figure}[H]
\centering
\begin{subfigure}[b]{0.48\textwidth}
\includegraphics[width=\textwidth]{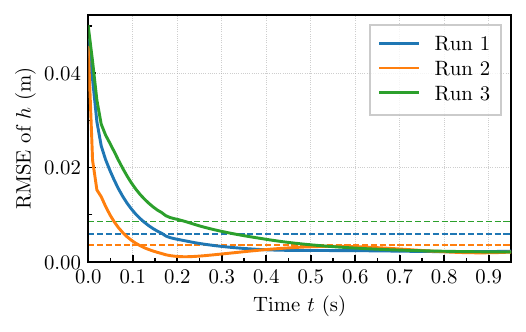}
\end{subfigure}
\hfill
\begin{subfigure}[b]{0.48\textwidth}
\includegraphics[width=\textwidth]{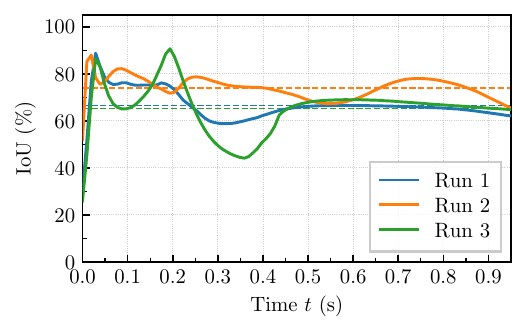}
\end{subfigure}
\caption{Time-resolved error metrics for C4. Left: RMSE of the depth
field $h$; the global value averaged over all time steps is
$6.7 \times 10^{-3}\,\text{m}$. Right: Intersection over Union (IoU) of
the wetted area; the time-averaged value is $69.0\,\%$.}
\label{fig:2d_metrics}
\end{figure}

The time-resolved RMSE and wetted-area IoU are shown in Figure~\ref{fig:2d_metrics}. The RMSE was largest during the early collapse phase and decreased steadily as the deposit thinned; the global value of $6.7\,\text{mm}$, averaged over the full simulation window, represents $3\,\%$ of the initial pile height. All quantitative wetted-region metrics reported in this section, including the IoU and wetted-area calculations, were evaluated using the relative depth threshold $h_{\mathrm{thr}} = H_0/50$. For Case~C4, this corresponds to $h_{\mathrm{thr}} = 4.3\,\text{mm}$.

\begin{figure}[H]
\centering
\includegraphics[width=0.65\textwidth]{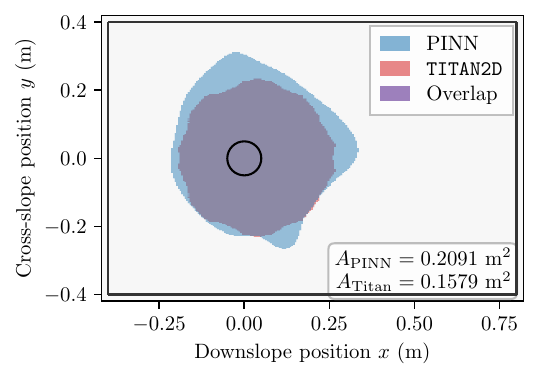}
\caption{Final wetted footprints at $t = 0.95\,\text{s}$ for C4. Blue: PINN ($A_{\text{PINN}} = 0.2091\,\text{m}^2$); red: \texttt{TITAN2D} ($A_{\text{Titan}} = 0.1579\,\text{m}^2$); purple: overlap region. The black circle marks the initial pile boundary. Wet cells were classified using the threshold $h_{\mathrm{thr}} = H_0/50 = 4.3\,\text{mm}$.}
\label{fig:2d_wetted}
\end{figure}

The final wetted footprints are compared in Figure~\ref{fig:2d_wetted}. The PINN wetted area of $0.2091\,\text{m}^2$ exceeded the \texttt{TITAN2D}
reference of $0.1579\,\text{m}^2$, with the excess confined to the deposit periphery.

\subsection{Multi-case validation}
\label{sec:multicase}

To assess whether the performance observed for C4 was maintained across different flow conditions, the framework was applied to three additional cases, C1 ($1\,\text{kg}$ at $10^\circ$), C2 ($1\,\text{kg}$ at $15^\circ$), and C3 ($2.5\,\text{kg}$ at $10^\circ$), using the identical network architecture, training procedure, and data points throughout. Each was evaluated from a single training run, as was C4. The repeatability associated with this choice is characterised in Appendix~\ref{app:repeatability}.

\begin{figure}[H]
\centering
\begin{subfigure}[b]{0.48\textwidth}
\includegraphics[width=\textwidth]{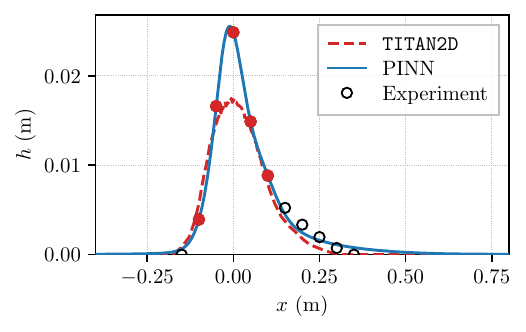}
\end{subfigure}
\hfill
\begin{subfigure}[b]{0.48\textwidth}
\includegraphics[width=\textwidth]{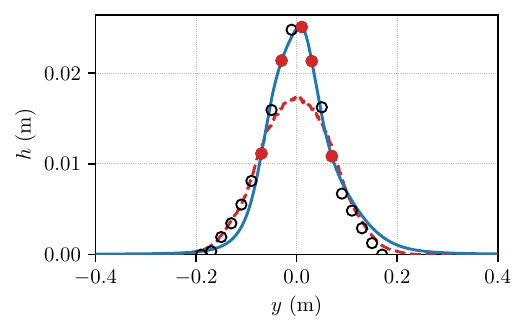}
\end{subfigure}
\vspace{-0.5em}
\begin{subfigure}[b]{0.48\textwidth}
\includegraphics[width=\textwidth]{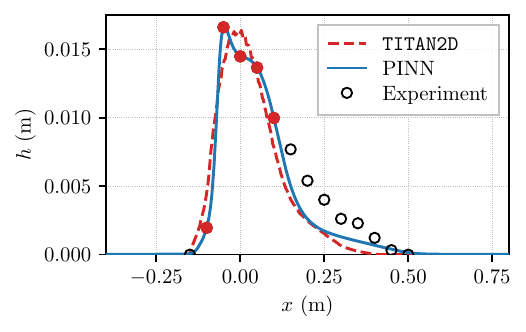}
\end{subfigure}
\hfill
\begin{subfigure}[b]{0.48\textwidth}
\includegraphics[width=\textwidth]{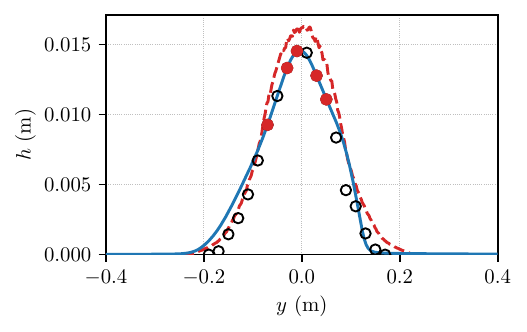}
\end{subfigure}
\vspace{-0.5em}
\begin{subfigure}[b]{0.48\textwidth}
\includegraphics[width=\textwidth]{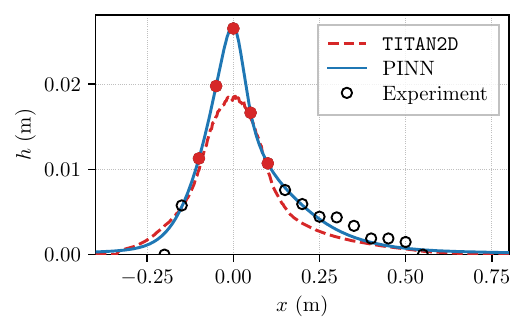}
\end{subfigure}
\hfill
\begin{subfigure}[b]{0.48\textwidth}
\includegraphics[width=\textwidth]{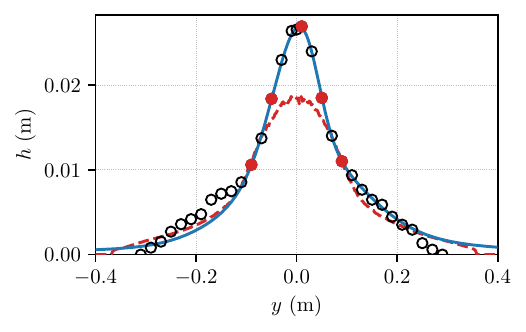}
\end{subfigure}
\caption{Final deposit profiles for Cases~C1--C3. Left column: centerline ($y = 0$); right column: cross-slope ($x = 0$). Rows from top to bottom: C1 ($1\,\text{kg}$~$10^\circ$),  C2 ($1\,\text{kg}$~$15^\circ$), and
C3 ($2.5\,\text{kg}$~$10^\circ$). PINN (solid blue), \texttt{TITAN2D} (dashed red), and experimental data from \citet{maeno2013unconfined} (circles). Filled red circles indicate training observations.}
\label{fig:multicase}
\end{figure}

The final deposit profiles for Cases~C1--C3 are shown in Figure~\ref{fig:multicase}. In all three cases, the PINN reproduced the experimental peak height and deposit width closely in both the centerline and cross-slope directions. Along the centerline, the PINN also tracked the downstream deposit tail more closely to the experiment than \texttt{TITAN2D}, which truncated this region earlier in all three cases. 

\begin{figure}[H]
\centering
\begin{subfigure}[b]{0.32\textwidth}
\includegraphics[width=\textwidth]{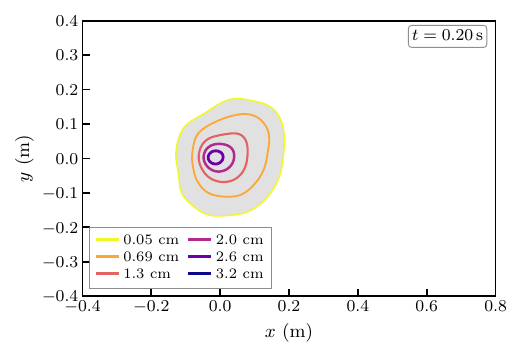}
\end{subfigure}
\hfill
\begin{subfigure}[b]{0.32\textwidth}
\includegraphics[width=\textwidth]{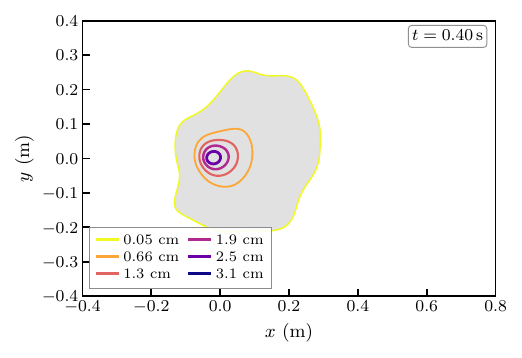}
\end{subfigure}
\hfill
\begin{subfigure}[b]{0.32\textwidth}
\includegraphics[width=\textwidth]{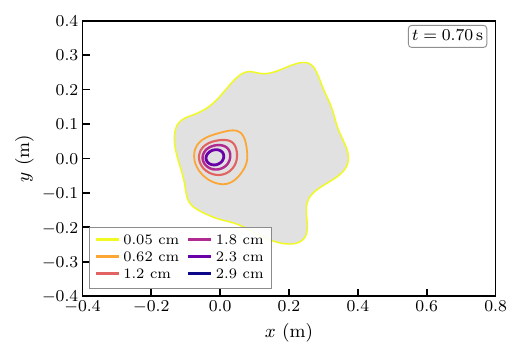}
\end{subfigure}
\vspace{-0.7em}
\begin{subfigure}[b]{0.32\textwidth}
\includegraphics[width=\textwidth]{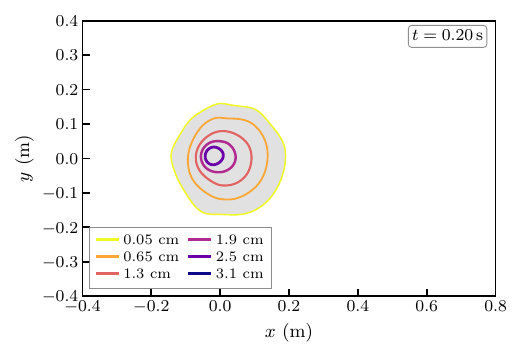}
\end{subfigure}
\hfill
\begin{subfigure}[b]{0.32\textwidth}
\includegraphics[width=\textwidth]{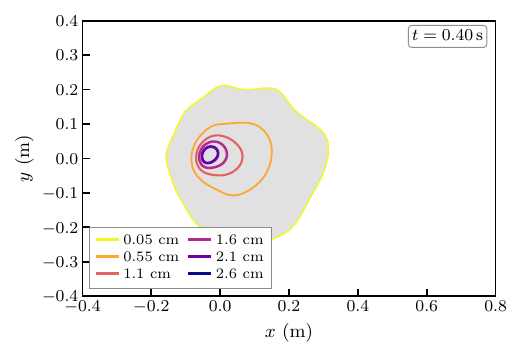}
\end{subfigure}
\hfill
\begin{subfigure}[b]{0.32\textwidth}
\includegraphics[width=\textwidth]{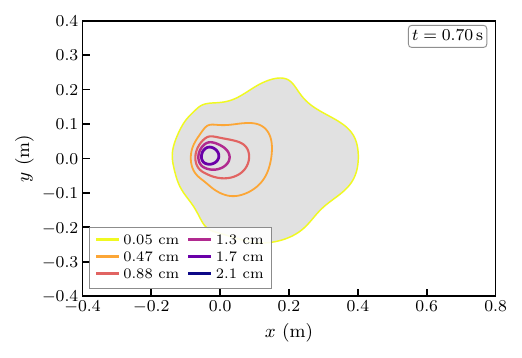}
\end{subfigure}
\vspace{-0.7em}
\begin{subfigure}[b]{0.32\textwidth}
\includegraphics[width=\textwidth]{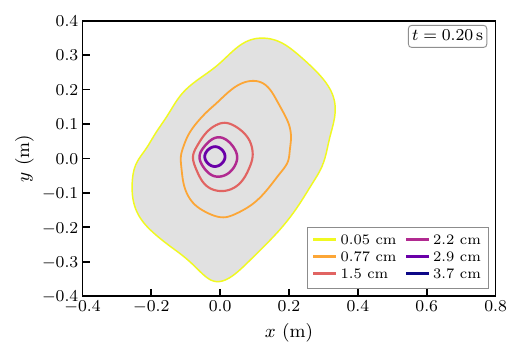}
\end{subfigure}
\hfill
\begin{subfigure}[b]{0.32\textwidth}
\includegraphics[width=\textwidth]{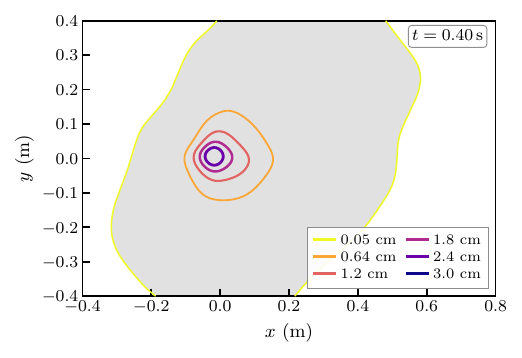}
\end{subfigure}
\hfill
\begin{subfigure}[b]{0.32\textwidth}
\includegraphics[width=\textwidth]{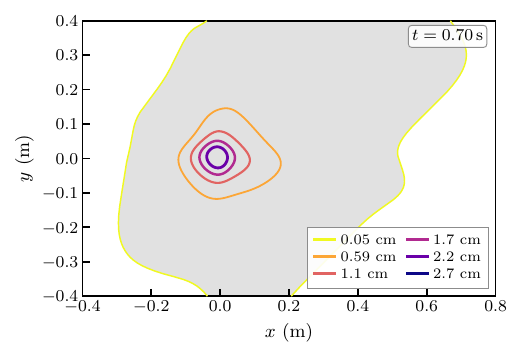}
\end{subfigure}
\caption{PINN depth contours at $t = 0.20$, $0.40$, and $0.70\,\text{s}$ (columns, left to right) for Cases~C1--C3 (rows, top to bottom): $1\,\text{kg}$~$10^\circ$, $1\,\text{kg}$~$15^\circ$, and $2.5\,\text{kg}$~$10^\circ$. The grey region marks the wetted footprint, defined using the same height cutoff of $0.05\,\text{cm}$ used for C4.}
\label{fig:multicase_contours}
\end{figure}

The depth contours at $t = 0.20$, $0.40$, and $0.70\,\text{s}$ for Cases~C1--C3 are presented in Figure~\ref{fig:multicase_contours}. In all three cases, the footprint grew, and the peak depth decreased over time, with the initially compact boundary developing local asymmetries
and lobed extensions that persisted from $t = 0.40$ to $0.70\,\text{s}$.

The global RMSE and mean IoU for all four cases are collected in Table~\ref{tab:metrics_summary}. RMSE increased with pile mass ($1\,\text{kg}$: $2.7$--$3.1\,\text{mm}$; $2.5\,\text{kg}$: $4.7$--$6.7\,\text{mm}$), reflecting the steeper depth gradients and more energetic early collapse of larger piles under a fixed collocation budget. Expressed relative to the initial pile height, the global RMSE ranged from $2.2\,\%$ (C3) to $3.6\,\%$ (C1) across the four cases. The IoU was highest for Case~C2 ($80.7\,\%$) and lowest for C4 ($69.0\,\%$). In all cases, RMSE peaked during the early collapse phase and decreased as the deposit settled, consistent with the trend observed for C4.

\begin{table}[H]
\centering
\caption{Global RMSE and mean IoU for all four experimental cases.}
\label{tab:metrics_summary}
\begin{tabular}{llcc}
\hline
Case & Configuration & Global RMSE (m) & Mean IoU (\%) \\
\hline
C1 & $1.0\,\text{kg}$, $10^\circ$ & $3.1 \times 10^{-3}$ & 70.9 \\
C2 & $1.0\,\text{kg}$, $15^\circ$ & $2.7 \times 10^{-3}$ & 80.7 \\
C3 & $2.5\,\text{kg}$, $10^\circ$ & $4.7 \times 10^{-3}$ & 73.1 \\
C4 & $2.5\,\text{kg}$, $15^\circ$ & $6.7 \times 10^{-3}$ & 69.0 \\
\hline
\end{tabular}
\end{table}

\begin{figure}[H]
\centering
\begin{subfigure}[b]{0.32\textwidth}
\includegraphics[width=\textwidth]{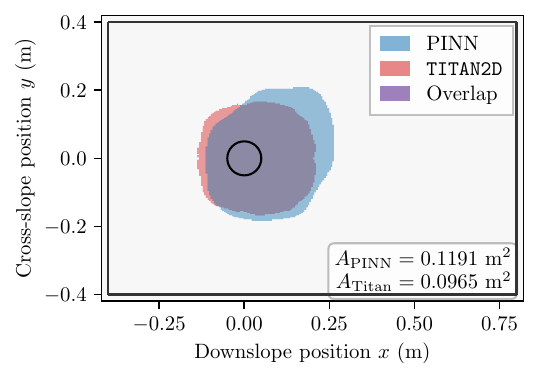}
\end{subfigure}
\hfill
\begin{subfigure}[b]{0.32\textwidth}
\includegraphics[width=\textwidth]{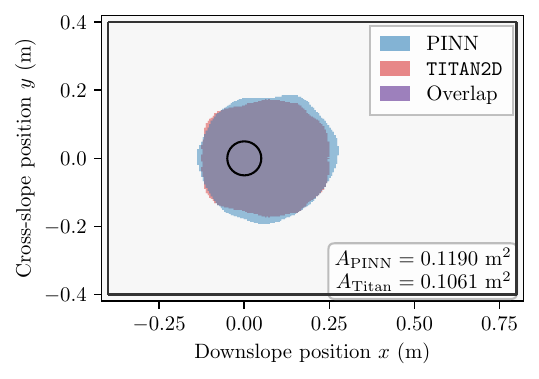}
\end{subfigure}
\hfill
\begin{subfigure}[b]{0.32\textwidth}
\includegraphics[width=\textwidth]{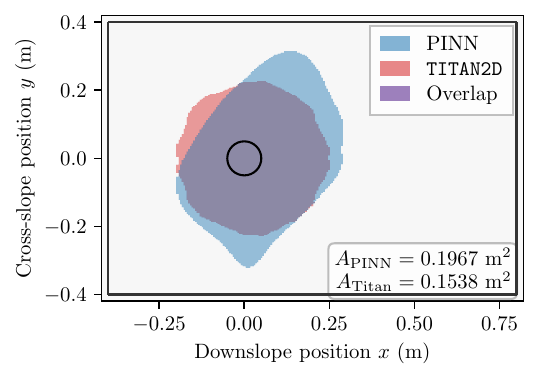}
\end{subfigure}
\caption{Final wetted footprints for Cases~C1--C3 (left to right): $1\,\text{kg}$~$10^\circ$, $1\,\text{kg}$~$15^\circ$, and $2.5\,\text{kg}$~$10^\circ$. PINN (blue), \texttt{TITAN2D} (red), and overlap (purple). The black circle marks the initial pile boundary. Wetted areas are annotated on each panel. Wet cells were classified using $h_{\mathrm{thr}} = H_0/50$, giving $1.7\,\text{mm}$ for C1--C2 and $4.3\,\text{mm}$ for C3.}
\label{fig:multicase_wetted}
\end{figure}

The final wetted footprints for Cases~C1--C3 are shown in Figure~\ref{fig:multicase_wetted}, and the corresponding wetted areas for all four cases are given in Table~\ref{tab:wetted_area}. The PINN wetted area exceeded the \texttt{TITAN2D} reference in every case, with the excess ranging from $12.1\,\%$ in Case~C2 to $32.4\,\%$ in Case~C4. The excess increased in proportion to mass at a fixed slope angle. In all cases, the overlap region covered the bulk of the \texttt{TITAN2D} footprint, indicating that the deposit interior was well reproduced and that the discrepancies were confined to the wet--dry boundary.

\begin{table}[H]
\centering
\caption{Final wetted areas for all four experimental cases.}
\label{tab:wetted_area}
\begin{tabular}{llccc}
\toprule
Case & Configuration
     & $A_{\text{PINN}}$ (m$^2$)
     & $A_{\texttt{TITAN2D}}$ (m$^2$)
     & Excess (\%) \\
\midrule
C1 & $1.0\,\text{kg}$, $10^\circ$ & 0.1191 & 0.0965 & 23.4 \\
C2 & $1.0\,\text{kg}$, $15^\circ$ & 0.1190 & 0.1061 & 12.1 \\
C3 & $2.5\,\text{kg}$, $10^\circ$ & 0.1967 & 0.1538 & 27.9 \\
C4 & $2.5\,\text{kg}$, $15^\circ$ & 0.2091 & 0.1579 & 32.4 \\
\bottomrule
\end{tabular}
\end{table}

\section{Conclusions}
\label{sec:conclusions}

This study demonstrated, for the first time, that physics-informed neural networks can solve the depth-averaged Savage--Hutter granular avalanche equations, progressing systematically from one-dimensional analytical verification to two-dimensional experimental validation against the laboratory measurements of \citet{maeno2013unconfined}.

In the one-dimensional setting, decoupled single-field tests showed that the continuity equation was readily satisfied by a compact network, whereas the momentum equation required greater architectural capacity and collocation density, reflecting the nonlinear coupling. The fully coupled conservative formulation resolved both fields simultaneously without any prescribed data, achieving mean height and velocity RMSEs of $4.3 \times 10^{-2}$ and $7.9 \times 10^{-2}$ in non-dimensional units, respectively, compared with the analytical parabolic-cap solution.

In the two-dimensional setting, training with purely physics-based losses led to collapse onto the trivial zero solution. Augmenting the loss with ten sparse training data points per case, drawn from the final deposit profiles along the centerline and cross-slope directions, provided sufficient signal to anchor the optimisation away from the zero attractor while preserving the physical consistency enforced by the PDE constraint. In post-event forensic scenarios, final deposit measurements are often the only data available at an avalanche site; the present results indicate that such measurements, combined with the governing equations, are sufficient to reconstruct the full spatio-temporal flow history with accuracy consistent across variations in pile mass and slope angle, as reported above.

Three limitations apply to the present results. The smooth output of a fully connected network cannot represent the sharp wet--dry interface characteristic of granular avalanche fronts, leading to a systematic overestimation of the wetted area. The data-assistance requirement means the framework is not a fully data-free solver, and its applicability depends on the availability of at least sparse final-deposit measurements. The training time of approximately $6.5\,\text{h}$ per case substantially exceeds the wall-clock time required for a single \texttt{TITAN2D} simulation.

Several directions for future work follow from these limitations. Extension to irregular terrain, via bed-fitted coordinate transformations or terrain-aware architectures, is required for real geophysical hazard applications. The integration of physics-informed approaches with operator learning architectures, such as neural operators, holds promise for producing generalisable solvers that retain physical consistency across diverse terrain and flow configurations. Building on these directions, future work will focus on making the framework more robust, more generalisable and more computationally efficient, broadening its applicability to a wider range of avalanche scenarios.

\section*{Acknowledgments}
P.S.S. gratefully acknowledges the Ministry of Education (MoE), Government of India, for funding his MTech (Research) studies. The authors thank the Data Science (DS) Laboratory at the Indian Institute of Technology Mandi for providing the computational resources used in this work. The authors would also like to acknowledge the developers of the open-source \texttt{TITAN2D} code.

\section*{Computer Code Availability}
Hardware requirements: GPU. Programming language: Python. The source code is available for download at \url{https://github.com/Prasik3182/PINNs_Avalanche.git}. The TITAN2D reference data used for benchmarking is archived on Zenodo at \url{https://doi.org/10.5281/zenodo.21393644}.

\bibliographystyle{unsrtnat}
\bibliography{ref}
 
\appendix

\section{Hyperparameter Sensitivity: Decoupled and Coupled Case}
\label{app:hp_sensitivity}

\begin{figure}[H]
\centering
\setlength{\tabcolsep}{2pt}
\begin{tabular}{cc}
\includegraphics[width=0.32\textwidth]{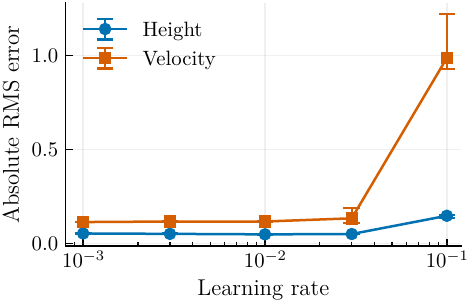} &
\includegraphics[width=0.32\textwidth]{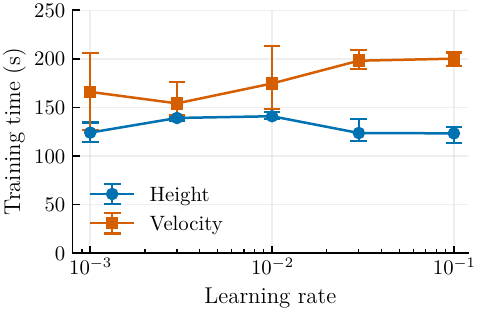} \\
\includegraphics[width=0.32\textwidth]{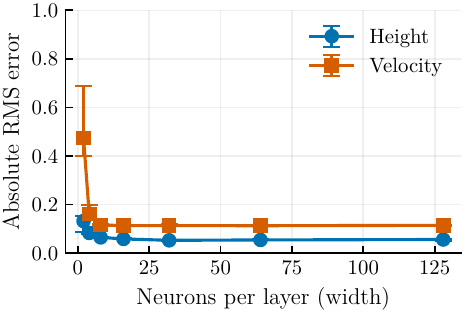} &
\includegraphics[width=0.32\textwidth]{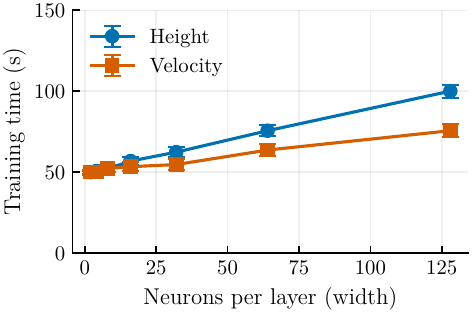} \\
\includegraphics[width=0.32\textwidth]{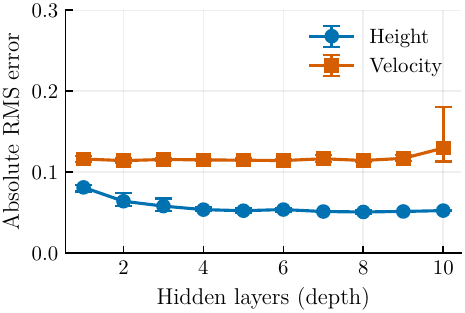} &
\includegraphics[width=0.32\textwidth]{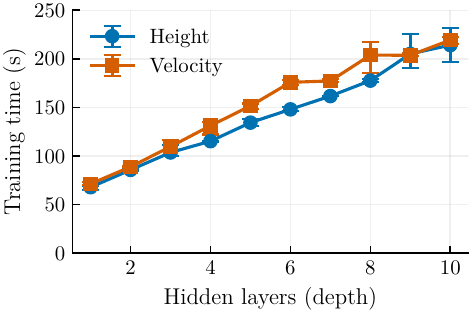} \\
\includegraphics[width=0.32\textwidth]{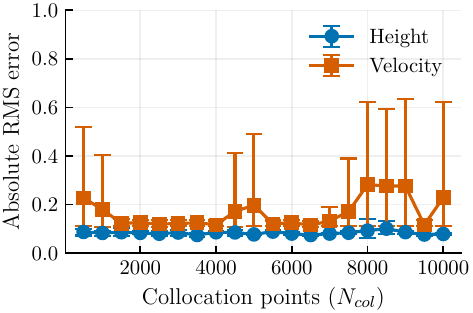} &
\includegraphics[width=0.32\textwidth]{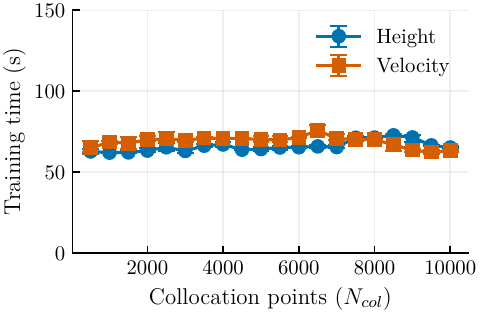} \\
\includegraphics[width=0.32\textwidth]{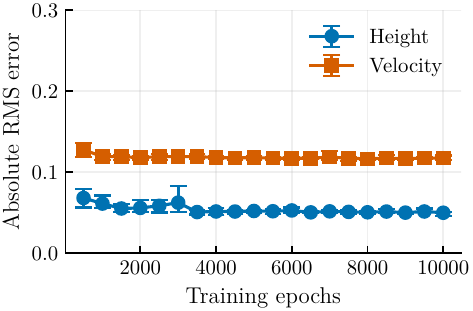} &
\includegraphics[width=0.32\textwidth]{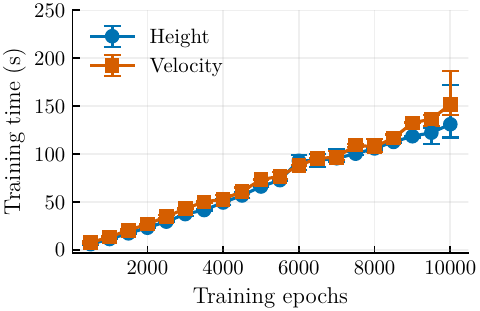} \\
\end{tabular}
\caption{Hyperparameter sensitivity of the decoupled one-dimensional PINN:
prediction RMSE (left) and wall-clock training time (right). Rows~(a)--(e):
learning rate $\alpha$, width $N_w$, depth $N_d$, collocation points
$N_{\mathrm{col}}$, and epochs $N_e$. Error bars show mean $\pm$ standard deviation over three independent runs. Height (blue circles); velocity
(orange squares).}
\label{fig:app_hp_sensitivity}
\end{figure}

The hyperparameter sensitivity results for the decoupled one-dimensional
tests are presented in Figure~\ref{fig:app_hp_sensitivity}. Each
hyperparameter was varied independently while all others were held fixed; error bars show mean $\pm$ standard deviation. The selected values, summarised in Table~\ref{tab:hp_settings}, are consistent with the trends observed here.

The height RMSE was insensitive to all five hyperparameters; the velocity
RMSE accounted for most of the variation, reflecting the greater difficulty
of the momentum equation established in the main text. The
learning rate $\alpha = 10^{-3}$ sat in the flat region before the sharp
velocity error increase at $\alpha = 10^{-1}$. Depth 5 fell within the flat region of the depth sweep, beyond which no further gain in accuracy was observed, and training cost increased linearly. Collocation density had no systematic effect on accuracy, confirming that the chosen values of 2000 and 10000 for both cases were not limiting factors.

\begin{figure}[H]
\centering
\begin{subfigure}[b]{0.42\textwidth}
    \includegraphics[width=\textwidth]{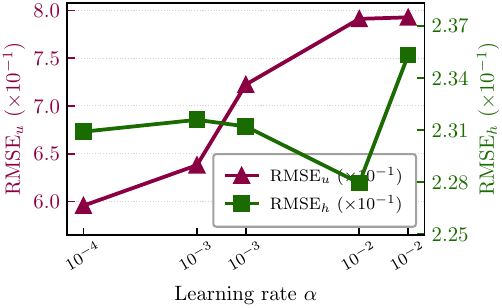}
\end{subfigure}
\hspace{0.02\textwidth}
\begin{subfigure}[b]{0.42\textwidth}
    \includegraphics[width=\textwidth]{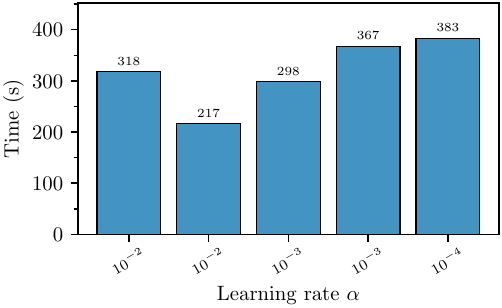}
\end{subfigure}
 
\vspace{0.2em}
 
\begin{subfigure}[b]{0.42\textwidth}
    \includegraphics[width=\textwidth]{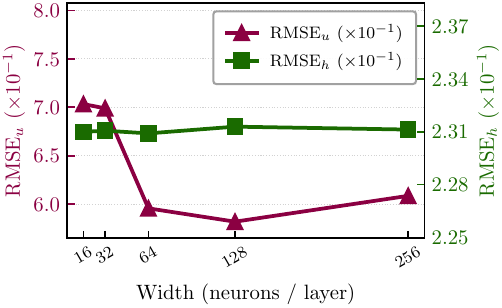}
\end{subfigure}
\hspace{0.02\textwidth}
\begin{subfigure}[b]{0.42\textwidth}
    \includegraphics[width=\textwidth]{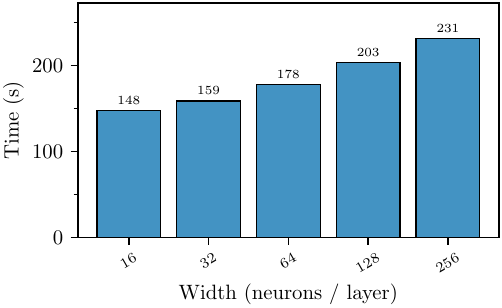}
\end{subfigure}
 
\vspace{0.2em}
 
\begin{subfigure}[b]{0.42\textwidth}
    \includegraphics[width=\textwidth]{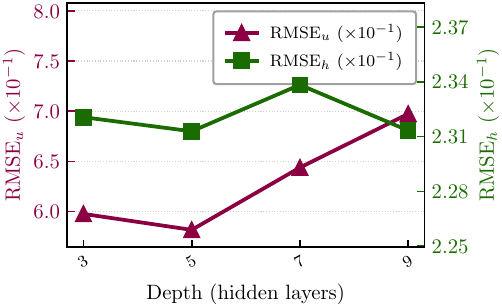}
\end{subfigure}
\hspace{0.02\textwidth}
\begin{subfigure}[b]{0.42\textwidth}
    \includegraphics[width=\textwidth]{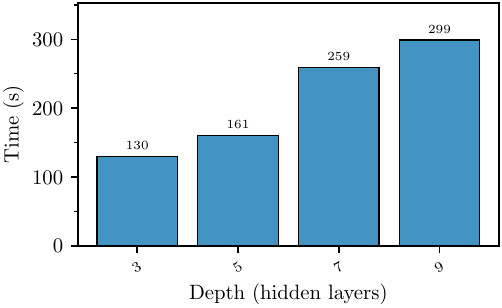}
\end{subfigure}
 
\vspace{0.1em}
 
\begin{subfigure}[b]{0.42\textwidth}
    \includegraphics[width=\textwidth]{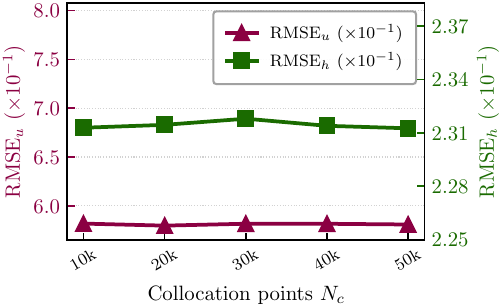}
\end{subfigure}
\hspace{0.02\textwidth}
\begin{subfigure}[b]{0.42\textwidth}
    \includegraphics[width=\textwidth]{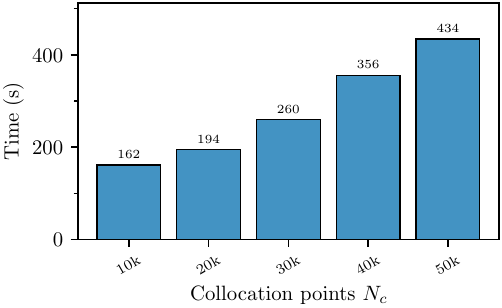}
\end{subfigure}
 
\vspace{0.2em}
 
\begin{subfigure}[b]{0.42\textwidth}
    \includegraphics[width=\textwidth]{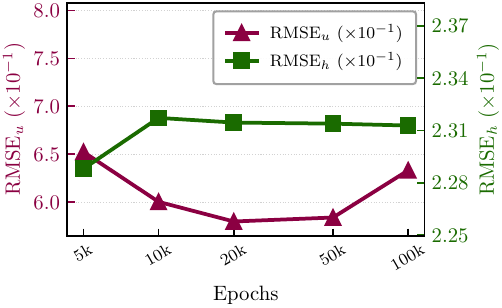}
\end{subfigure}
\hspace{0.02\textwidth}
\begin{subfigure}[b]{0.42\textwidth}
    \includegraphics[width=\textwidth]{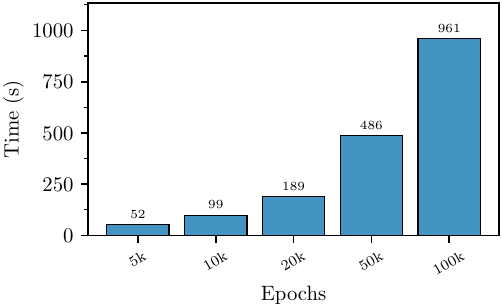}
\end{subfigure}
 
\caption{Hyperparameter sensitivity of the coupled one-dimensional PINN:
prediction accuracy (left) and wall-clock training time (right).
Rows~(a)--(e): learning rate $\alpha$, width $N_w$, depth $N_d$,
collocation points $N_c$, and epochs $N_e$.}
\label{fig:app_hp_coupled}
\end{figure}

The sensitivity study for the coupled formulation is presented in Figure~\ref{fig:app_hp_coupled}. $\mathrm{RMSE}_h$ remained nearly flat across all settings while the velocity error accounted for almost all the variation. Training cost scaled linearly with network size and collocation density. Based on these sweeps, a five-layer network with 64 neurons per layer was selected, as summarised in Table~\ref{tab:hp_settings}.

\section{Pseudo-2D Strip Flow: Diagnosing Dry-Region Dominance}
\label{app:obstacle}

Section~\ref{sec:trivial} identified two structural causes of the trivial-solution collapse in the two-dimensional setting: the mathematical homogeneity of the Savage--Hutter system, and the dominance of the dry region in the computational domain. The cylindrical initial condition used in Cases~C1--C4 occupies approximately $0.82\,\%$ of the $1.2\,\text{m} \times 0.8\,\text{m}$ domain, leaving approximately $99.18\,\%$ of the spatial domain initially dry. Although the wetted footprint increases as the granular mass spreads, the dry region remains substantially larger than the active flow region. To isolate the contribution of this dry-region dominance, the cylindrical pile was replaced by a full-width parabolic ridge uniform in the cross-slope ($y$) direction. This configuration is referred to as the pseudo-2D case: the initial condition and the resulting flow are uniform in $y$, so the problem is effectively one-dimensional in its dynamics while remaining formally two-dimensional in its spatial inputs and governing equations. The strip occupies a substantially larger fraction of the domain, directly increasing the proportion of collocation points that sample the active flow region.

\subsection{Problem configuration}
\label{app:obs_config}

The initial condition was a parabolic ridge uniform in $y$, centred at
$x = 0.40\,\text{m}$ with peak height $H_0 = 0.085\,\text{m}$. The
granular material properties matched those of Cases~C1--C4: bulk density
$\rho = \SI{1481}{\kilogram\per\metre\cubed}$, internal friction angle
$\phi = 24^{\circ}$, and bed friction angle $\delta = 28^{\circ}$. The
same five-layer, 64-neuron network architecture, loss formulation, and
training configuration described in the main text were used,
with one deliberate omission: no sparse data augmentation was included.
If dry-region dominance were the sole cause of the trivial collapse, the
wider wetted footprint should allow the physics loss alone to sustain a
physical solution. A volume ratio, defined as the ratio of the predicted
total volume at time $t$ to the initial volume $V_0$, was tracked
throughout training as the primary diagnostic of mass retention. The
experiment was run at slope angles of $15^{\circ}$, $20^{\circ}$, and
$30^{\circ}$, and the simulation window ran to $t_{\max} = 0.55\,\text{s}$
in each case.

\subsection{Results}
\label{app:obs_results}

At all three slope angles, training exhibited a brief collapse toward zero
near epochs 12000--14000, followed by recovery to a physical solution.
This recovery did not occur in the equivalent cylindrical-pile experiments
without data augmentation, confirming that the wider wetted footprint
provided sufficient gradient signal to escape the trivial attractor. The
final volume ratios are collected in Table~\ref{tab:pseudo2d_vr}: higher
slope angles produced higher retained volume, with the $30^{\circ}$ case
retaining $96.7\,\%$ of the initial volume while the $15^{\circ}$ case
settled near $0.675$ and was still slowly increasing at the end of
training. The centreline profiles show the flow spreading downslope with
increasing runout at steeper angles, and the cross-slope profiles remained
uniform in $y$ throughout, which is physically consistent with a strip
initial condition in the absence of a cross-slope gravity component.
Centreline profiles and depth contours for each slope angle are presented
in Sections~\ref{app:obs_15deg}--\ref{app:obs_30deg}.

\begin{table}[H]
\centering
\caption{Final volume ratios for the pseudo-2D experiment. Values are
reported at the end of training and averaged over the final
10000 epochs.}
\label{tab:pseudo2d_vr}
\begin{tabular}{ccc}
\toprule
$\zeta$ & Final volume ratio & Mean (last 10000 epochs) \\
\midrule
$15^{\circ}$ & 0.675 & 0.645 \\
$20^{\circ}$ & 0.745 & 0.723 \\
$30^{\circ}$ & 0.967 & 0.969 \\
\bottomrule
\end{tabular}
\end{table}

These results indicate that the trivial-solution collapse is driven by the dominance of the dry region within the computational domain: because the physical deposit occupies only a small fraction of the domain, the majority of collocation points lie outside the wetted footprint and continuously pull the PDE loss toward $h = u = v = 0$, overwhelming the comparatively weak gradient signal from the active flow region. The pseudo-2D experiment tested this mechanism directly by enlarging the wetted footprint through a full-width strip geometry. With a larger fraction of the domain occupied by flow, a greater proportion of collocation points contributed a non-trivial gradient signal, and recovery from the trivial solution became possible. This confirms that the imbalance between wetted and dry collocation points is a primary driver of the collapse, and motivates the data-augmentation strategy adopted in the main text, which supplies a gradient signal independent of the wetted-area fraction.

\subsubsection{\texorpdfstring{$\zeta = 15^{\circ}$}{zeta = 15 degrees}}
\label{app:obs_15deg}

At $15^{\circ}$, the flow decelerates throughout the simulation window
under the net frictional resistance. The centreline peak advances
progressively downslope and flattens as the deposit spreads, while the
wavefront position and peak height remain in close agreement with
\texttt{TITAN2D} at every snapshot. The depth contours show the expected
elongation of the deposit in the downslope direction.

\begin{figure}[H]
\centering
\begin{subfigure}[b]{0.32\textwidth}
\includegraphics[width=\textwidth]{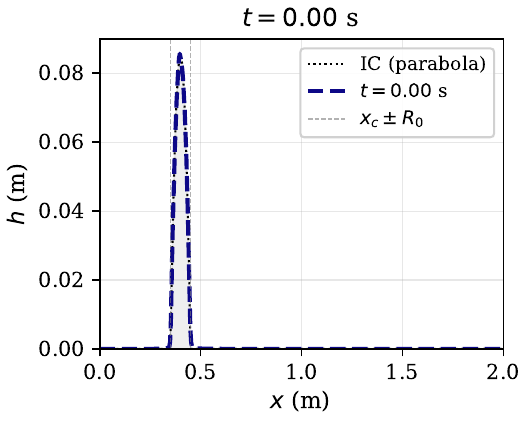}
\caption{$t = 0.00\,\text{s}$}
\end{subfigure}
\hfill
\begin{subfigure}[b]{0.32\textwidth}
\includegraphics[width=\textwidth]{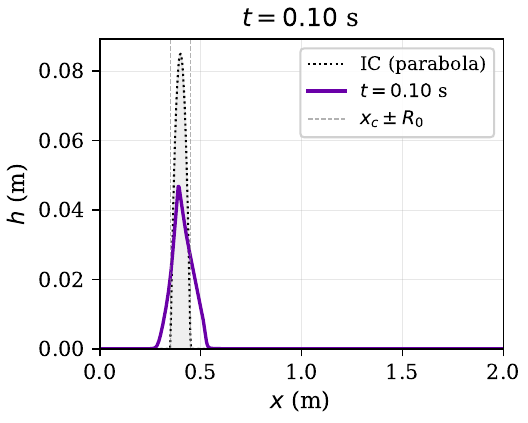}
\caption{$t = 0.10\,\text{s}$}
\end{subfigure}
\hfill
\begin{subfigure}[b]{0.32\textwidth}
\includegraphics[width=\textwidth]{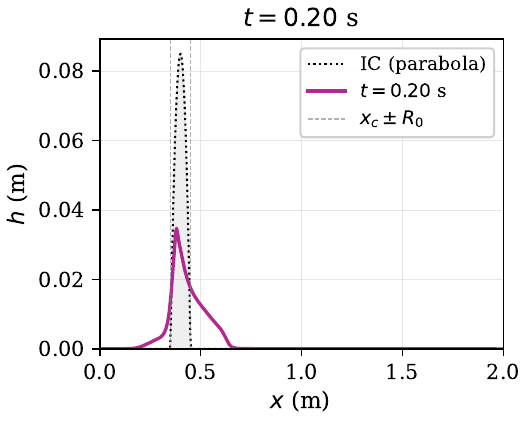}
\caption{$t = 0.20\,\text{s}$}
\end{subfigure}
\vspace{-0.3em}
\begin{subfigure}[b]{0.32\textwidth}
\includegraphics[width=\textwidth]{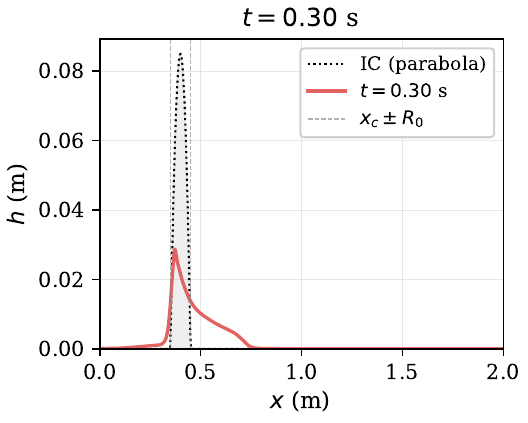}
\caption{$t = 0.30\,\text{s}$}
\end{subfigure}
\hfill
\begin{subfigure}[b]{0.32\textwidth}
\includegraphics[width=\textwidth]{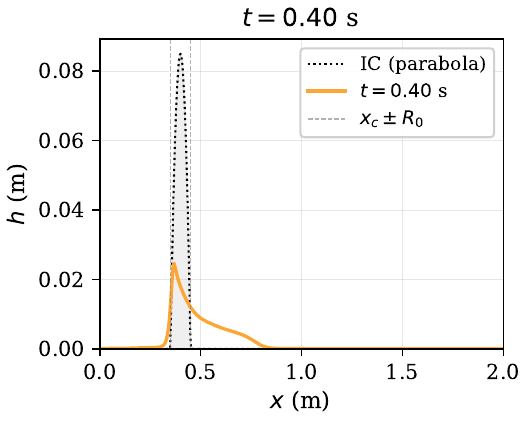}
\caption{$t = 0.40\,\text{s}$}
\end{subfigure}
\hfill
\begin{subfigure}[b]{0.32\textwidth}
\includegraphics[width=\textwidth]{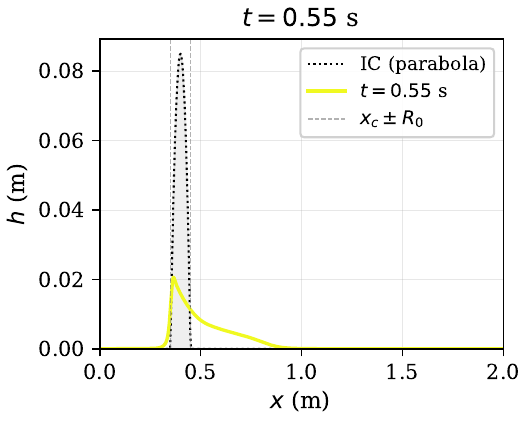}
\caption{$t = 0.55\,\text{s}$}
\end{subfigure}
\caption{Pseudo-2D case, $\zeta = 15^{\circ}$: centreline height profiles
at six time instants. PINN (solid blue) and \texttt{TITAN2D} (dashed red).}
\label{fig:app_obs_cl_15deg}
\end{figure}

\begin{figure}[H]
\centering
\begin{subfigure}[b]{0.32\textwidth}
\includegraphics[width=\textwidth]{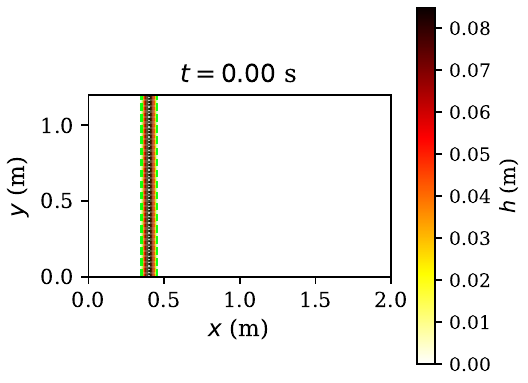}
\caption{$t = 0.00\,\text{s}$}
\end{subfigure}
\hfill
\begin{subfigure}[b]{0.32\textwidth}
\includegraphics[width=\textwidth]{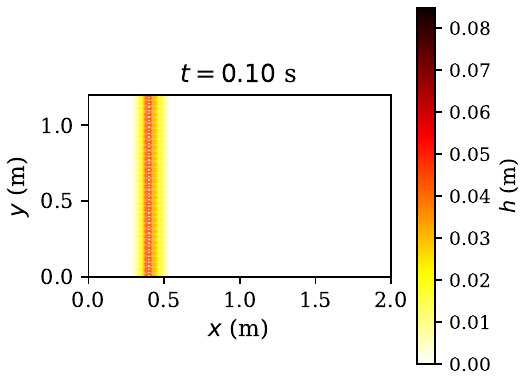}
\caption{$t = 0.10\,\text{s}$}
\end{subfigure}
\hfill
\begin{subfigure}[b]{0.32\textwidth}
\includegraphics[width=\textwidth]{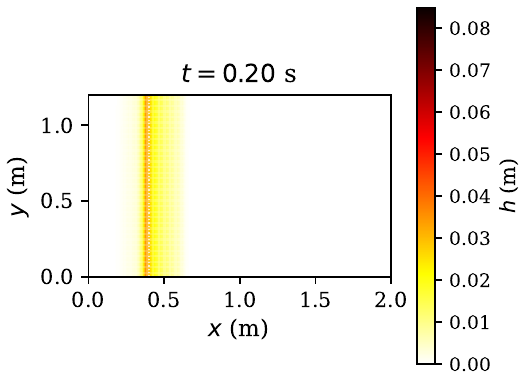}
\caption{$t = 0.20\,\text{s}$}
\end{subfigure}
\vspace{-0.3em}
\begin{subfigure}[b]{0.32\textwidth}
\includegraphics[width=\textwidth]{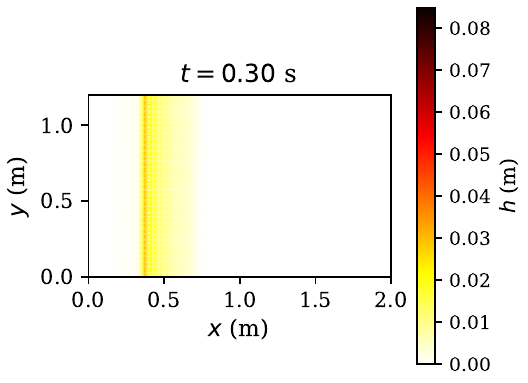}
\caption{$t = 0.30\,\text{s}$}
\end{subfigure}
\hfill
\begin{subfigure}[b]{0.32\textwidth}
\includegraphics[width=\textwidth]{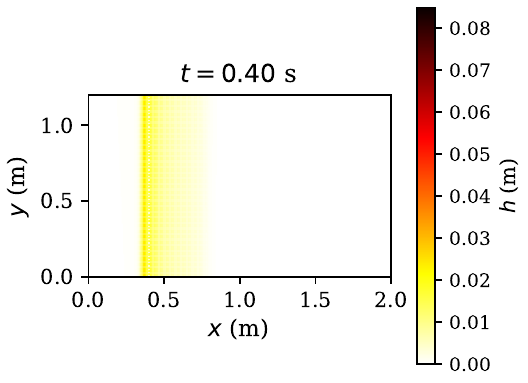}
\caption{$t = 0.40\,\text{s}$}
\end{subfigure}
\hfill
\begin{subfigure}[b]{0.32\textwidth}
\includegraphics[width=\textwidth]{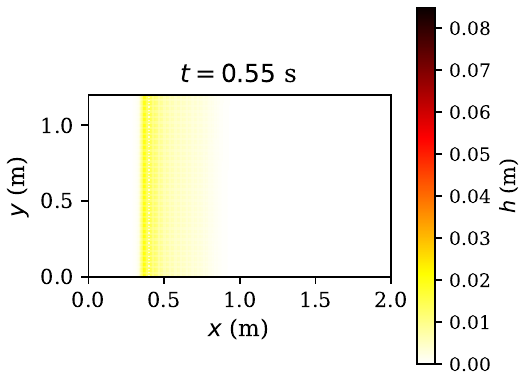}
\caption{$t = 0.55\,\text{s}$}
\end{subfigure}
\caption{Pseudo-2D case, $\zeta = 15^{\circ}$: top-view depth contours at
six time instants. Contour levels are identical across all panels.}
\label{fig:app_obs_top_15deg}
\end{figure}

\subsubsection{\texorpdfstring{$\zeta = 20^{\circ}$}{zeta = 20 degrees}}
\label{app:obs_20deg}

At $20^{\circ}$, the reduced frictional resistance produced a faster
collapse and greater downslope runout within the same simulation window.
The centreline peak decayed more rapidly, and the downstream tail extended
further than at $15^{\circ}$, while the wavefront position remained in
close agreement with \texttt{TITAN2D} throughout. The depth contours show
a more elongated deposit in the downslope direction relative to the
$15^{\circ}$ case.

\begin{figure}[H]
\centering
\begin{subfigure}[b]{0.32\textwidth}
\includegraphics[width=\textwidth]{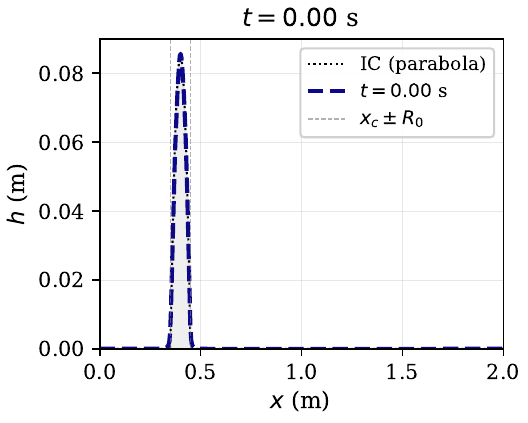}
\caption{$t = 0.00\,\text{s}$}
\end{subfigure}
\hfill
\begin{subfigure}[b]{0.32\textwidth}
\includegraphics[width=\textwidth]{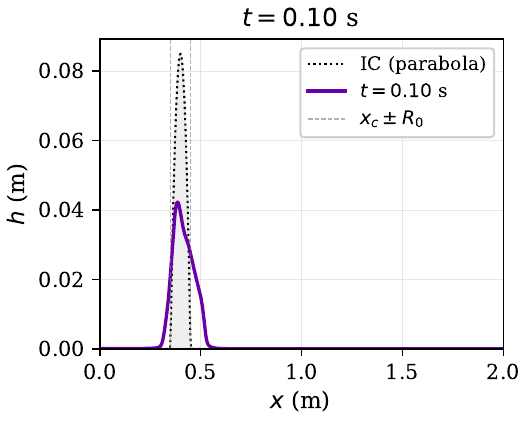}
\caption{$t = 0.10\,\text{s}$}
\end{subfigure}
\hfill
\begin{subfigure}[b]{0.32\textwidth}
\includegraphics[width=\textwidth]{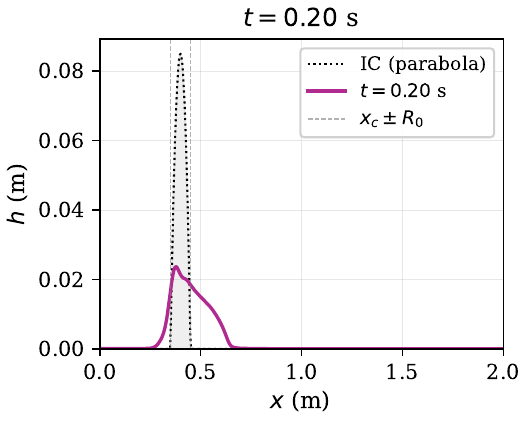}
\caption{$t = 0.20\,\text{s}$}
\end{subfigure}
\vspace{-0.3em}
\begin{subfigure}[b]{0.32\textwidth}
\includegraphics[width=\textwidth]{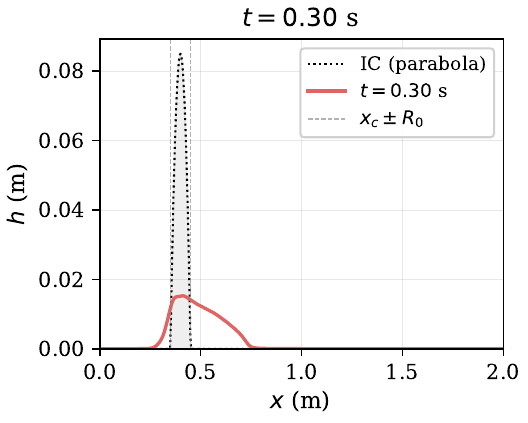}
\caption{$t = 0.30\,\text{s}$}
\end{subfigure}
\hfill
\begin{subfigure}[b]{0.32\textwidth}
\includegraphics[width=\textwidth]{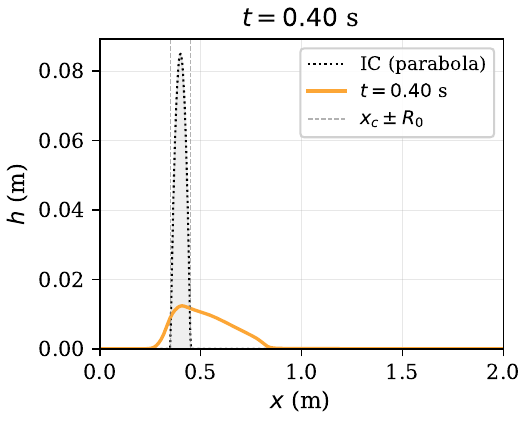}
\caption{$t = 0.40\,\text{s}$}
\end{subfigure}
\hfill
\begin{subfigure}[b]{0.32\textwidth}
\includegraphics[width=\textwidth]{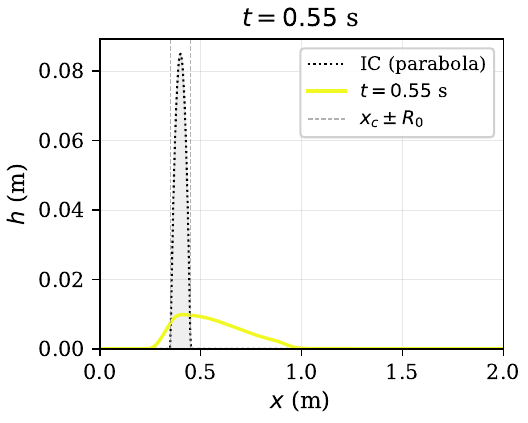}
\caption{$t = 0.55\,\text{s}$}
\end{subfigure}
\caption{Pseudo-2D case, $\zeta = 20^{\circ}$: centreline height profiles
at six time instants. PINN (solid blue) and \texttt{TITAN2D} (dashed red).}
\label{fig:app_obs_cl_20deg}
\end{figure}

\begin{figure}[H]
\centering
\begin{subfigure}[b]{0.32\textwidth}
\includegraphics[width=\textwidth]{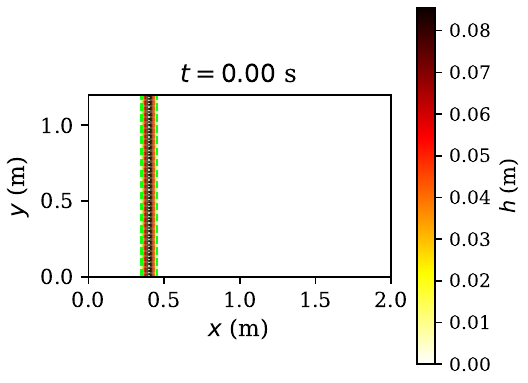}
\caption{$t = 0.00\,\text{s}$}
\end{subfigure}
\hfill
\begin{subfigure}[b]{0.32\textwidth}
\includegraphics[width=\textwidth]{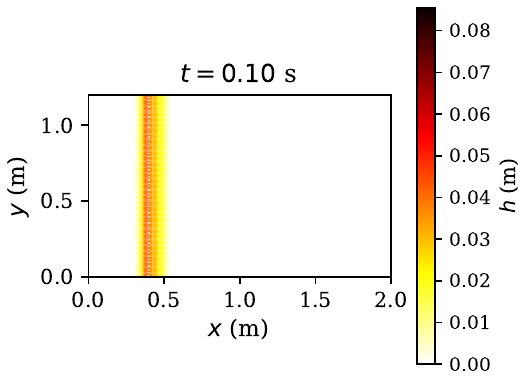}
\caption{$t = 0.10\,\text{s}$}
\end{subfigure}
\hfill
\begin{subfigure}[b]{0.32\textwidth}
\includegraphics[width=\textwidth]{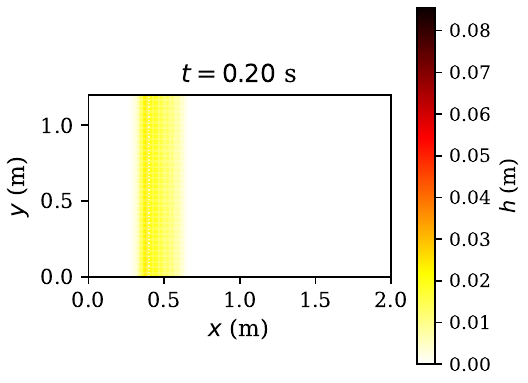}
\caption{$t = 0.20\,\text{s}$}
\end{subfigure}
\vspace{-0.3em}
\begin{subfigure}[b]{0.32\textwidth}
\includegraphics[width=\textwidth]{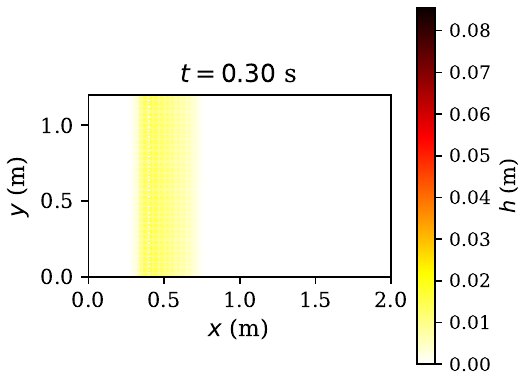}
\caption{$t = 0.30\,\text{s}$}
\end{subfigure}
\hfill
\begin{subfigure}[b]{0.32\textwidth}
\includegraphics[width=\textwidth]{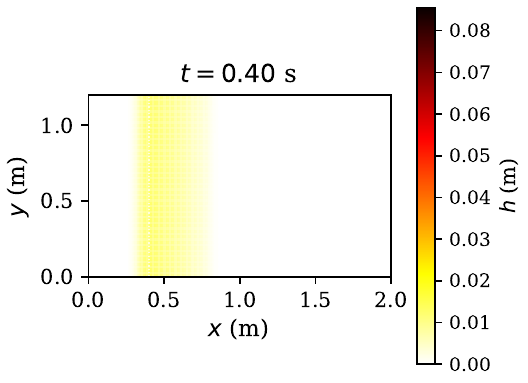}
\caption{$t = 0.40\,\text{s}$}
\end{subfigure}
\hfill
\begin{subfigure}[b]{0.32\textwidth}
\includegraphics[width=\textwidth]{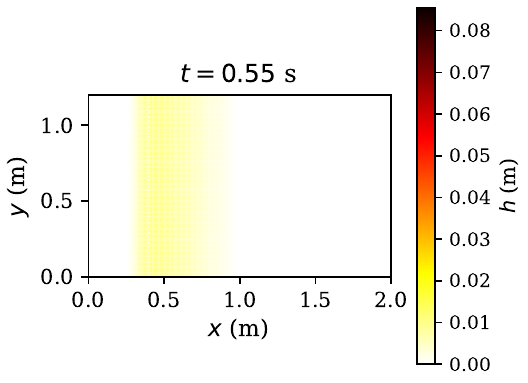}
\caption{$t = 0.55\,\text{s}$}
\end{subfigure}
\caption{Pseudo-2D case, $\zeta = 20^{\circ}$: top-view depth contours at
six time instants. Contour levels are identical across all panels.}
\label{fig:app_obs_top_20deg}
\end{figure}

\subsubsection{\texorpdfstring{$\zeta = 30^{\circ}$}{zeta = 30 degrees}}
\label{app:obs_30deg}

At $30^{\circ}$, the slope exceeds the bed friction angle, so the net gravitational driving force is positive, and the flow accelerates throughout the simulation window. The wavefront advanced rapidly, and the downstream tail extended well beyond the initial pile location by $t = 0.55\,\text{s}$, with only modest decay in peak height. The depth contours show the deposit elongated downslope, with the major axis considerably longer than in either of the decelerating cases, and the PINN remained in close agreement with \texttt{TITAN2D} throughout.

\begin{figure}[H]
\centering
\begin{subfigure}[b]{0.32\textwidth}
\includegraphics[width=\textwidth]{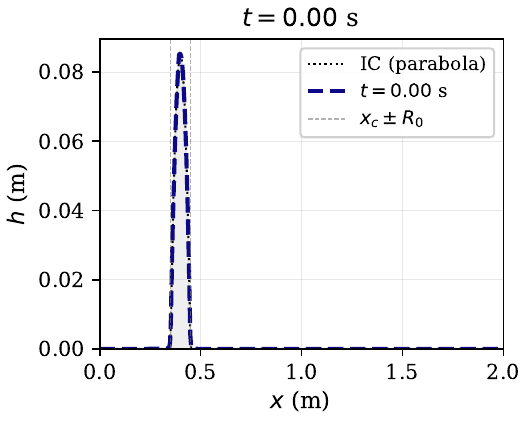}
\caption{$t = 0.00\,\text{s}$}
\end{subfigure}
\hfill
\begin{subfigure}[b]{0.32\textwidth}
\includegraphics[width=\textwidth]{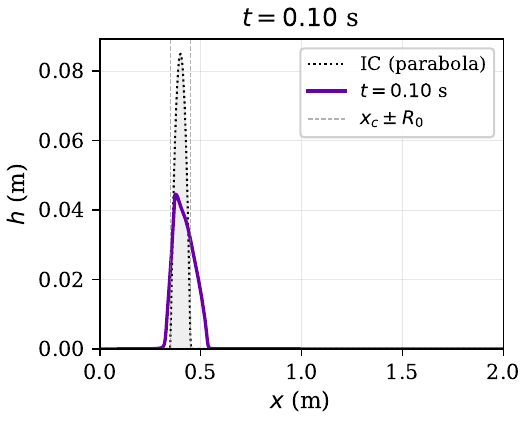}
\caption{$t = 0.10\,\text{s}$}
\end{subfigure}
\hfill
\begin{subfigure}[b]{0.32\textwidth}
\includegraphics[width=\textwidth]{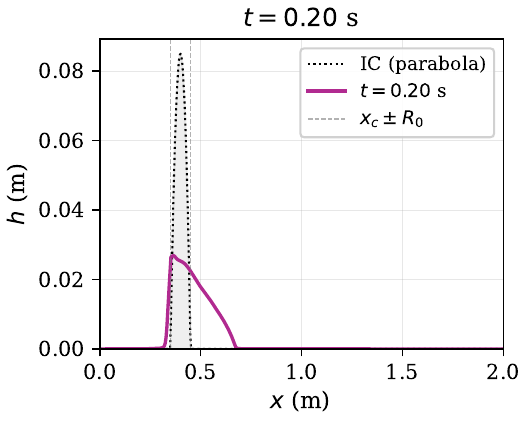}
\caption{$t = 0.20\,\text{s}$}
\end{subfigure}
\vspace{-0.3em}
\begin{subfigure}[b]{0.32\textwidth}
\includegraphics[width=\textwidth]{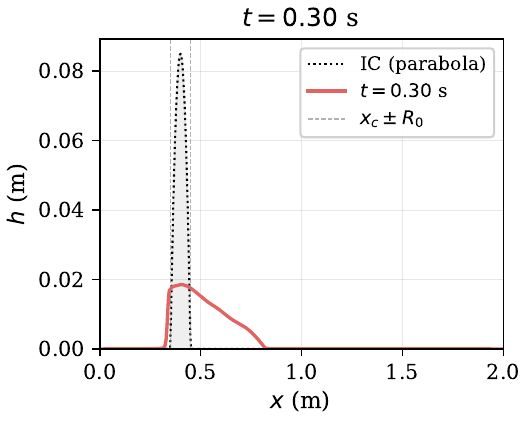}
\caption{$t = 0.30\,\text{s}$}
\end{subfigure}
\hfill
\begin{subfigure}[b]{0.32\textwidth}
\includegraphics[width=\textwidth]{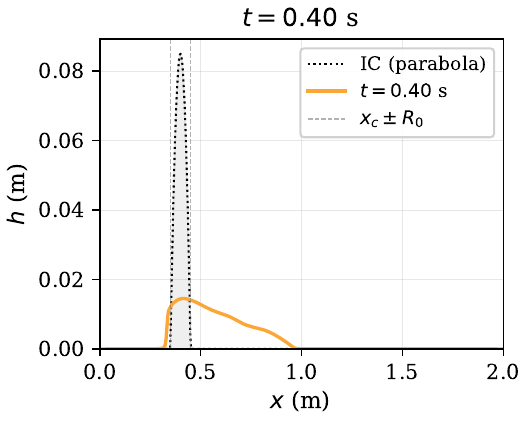}
\caption{$t = 0.40\,\text{s}$}
\end{subfigure}
\hfill
\begin{subfigure}[b]{0.32\textwidth}
\includegraphics[width=\textwidth]{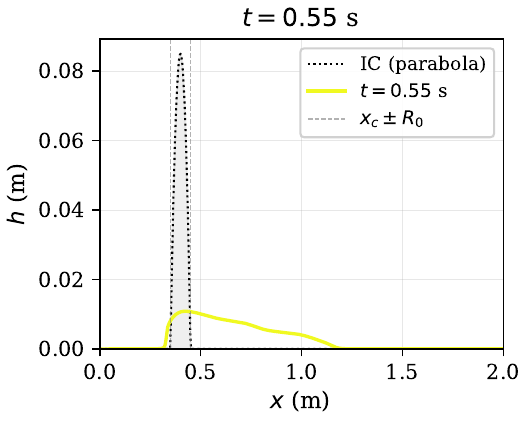}
\caption{$t = 0.55\,\text{s}$}
\end{subfigure}
\caption{Pseudo-2D case, $\zeta = 30^{\circ}$: centreline height profiles
at six time instants. PINN (solid blue) and \texttt{TITAN2D} (dashed red).}
\label{fig:app_obs_cl_30deg}
\end{figure}

\begin{figure}[H]
\centering
\begin{subfigure}[b]{0.32\textwidth}
\includegraphics[width=\textwidth]{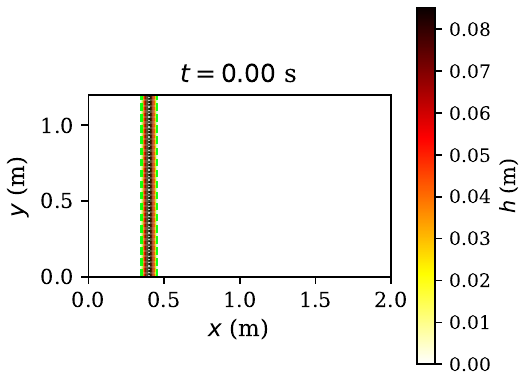}
\caption{$t = 0.00\,\text{s}$}
\end{subfigure}
\hfill
\begin{subfigure}[b]{0.32\textwidth}
\includegraphics[width=\textwidth]{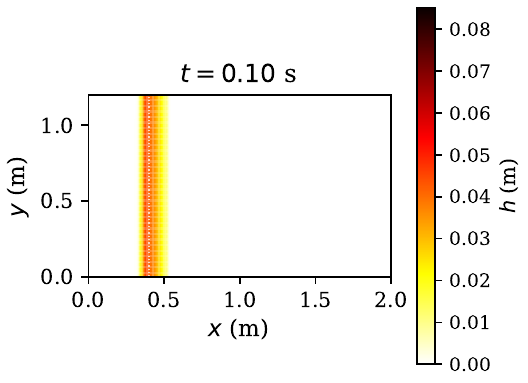}
\caption{$t = 0.10\,\text{s}$}
\end{subfigure}
\hfill
\begin{subfigure}[b]{0.32\textwidth}
\includegraphics[width=\textwidth]{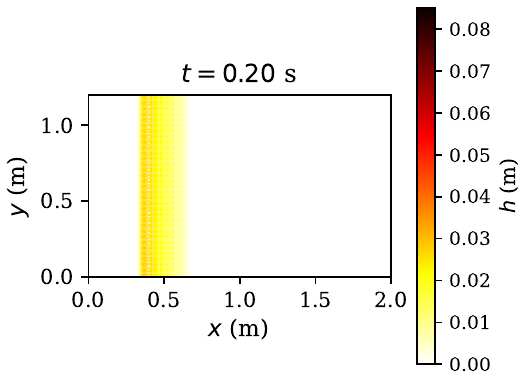}
\caption{$t = 0.20\,\text{s}$}
\end{subfigure}
\vspace{-0.3em}
\begin{subfigure}[b]{0.32\textwidth}
\includegraphics[width=\textwidth]{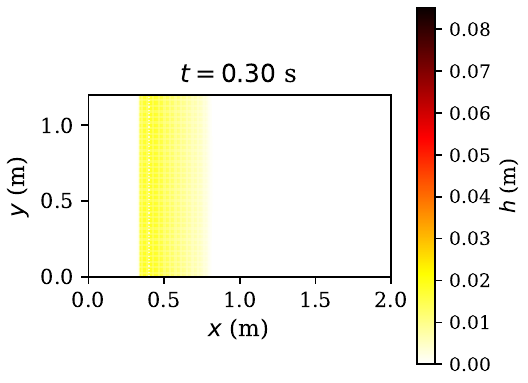}
\caption{$t = 0.30\,\text{s}$}
\end{subfigure}
\hfill
\begin{subfigure}[b]{0.32\textwidth}
\includegraphics[width=\textwidth]{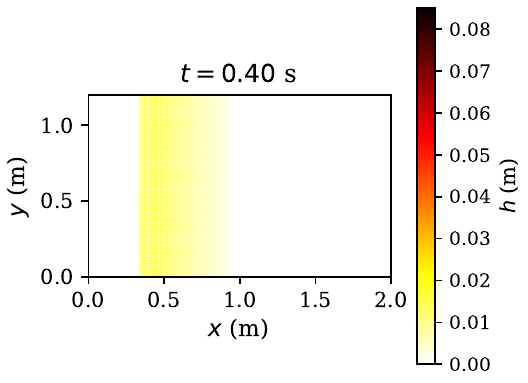}
\caption{$t = 0.40\,\text{s}$}
\end{subfigure}
\hfill
\begin{subfigure}[b]{0.32\textwidth}
\includegraphics[width=\textwidth]{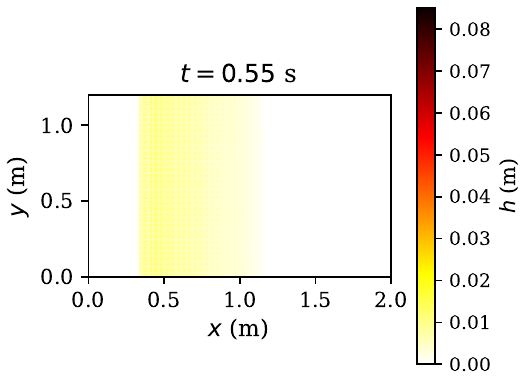}
\caption{$t = 0.55\,\text{s}$}
\end{subfigure}
\caption{Pseudo-2D case, $\zeta = 30^{\circ}$: top-view depth contours at
six time instants. Contour levels are identical across all panels.}
\label{fig:app_obs_top_30deg}
\end{figure}

\section{Extended Results for Secondary Cases C1--C3}
\label{app:extended_c1c3}

Time-resolved profile comparisons and integral diagnostics for Cases~C1,
C2 and C3 are presented here, complementing the final-deposit profiles in
the main text. For each case, the centreline and cross-slope
profiles at successive snapshots are followed by integral flow diagnostics
and the IoU evolution, mirroring the analysis for C4 in
the main text.

\subsection{\texorpdfstring{Case C1: 1\,kg, $10^\circ$}
{Case C1: 1 kg, 10 degrees}}
\label{app:c1}

\begin{figure}[H]
\centering
\begin{subfigure}[b]{0.48\textwidth}
\includegraphics[width=\textwidth]{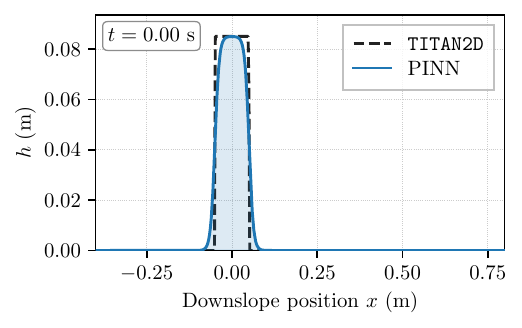}
\end{subfigure}
\hfill
\begin{subfigure}[b]{0.48\textwidth}
\includegraphics[width=\textwidth]{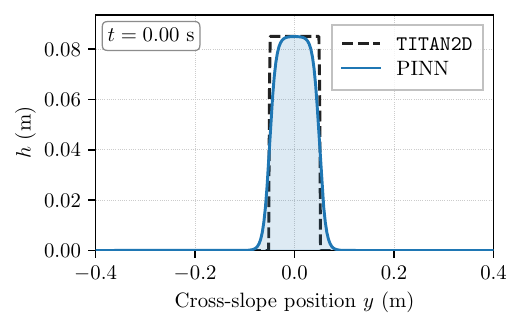}
\end{subfigure}
\vspace{-0.5em}
\begin{subfigure}[b]{0.48\textwidth}
\includegraphics[width=\textwidth]{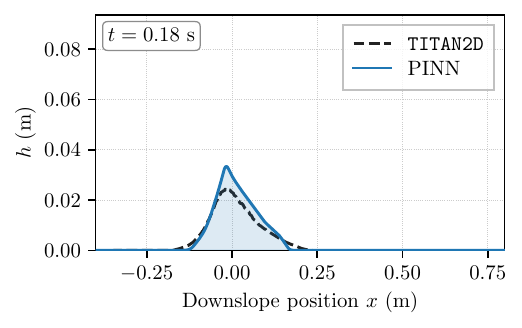}
\end{subfigure}
\hfill
\begin{subfigure}[b]{0.48\textwidth}
\includegraphics[width=\textwidth]{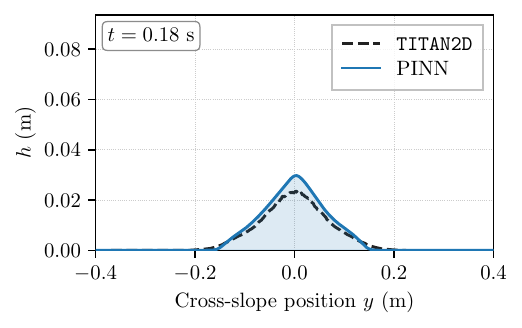}
\end{subfigure}
\vspace{-0.5em}
\begin{subfigure}[b]{0.48\textwidth}
\includegraphics[width=\textwidth]{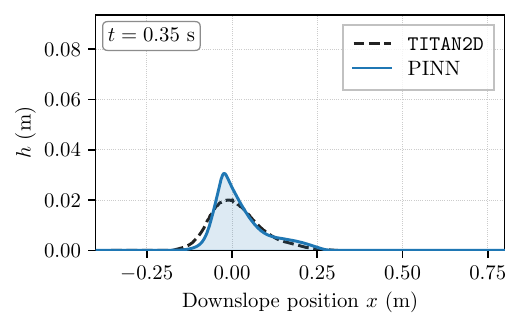}
\end{subfigure}
\hfill
\begin{subfigure}[b]{0.48\textwidth}
\includegraphics[width=\textwidth]{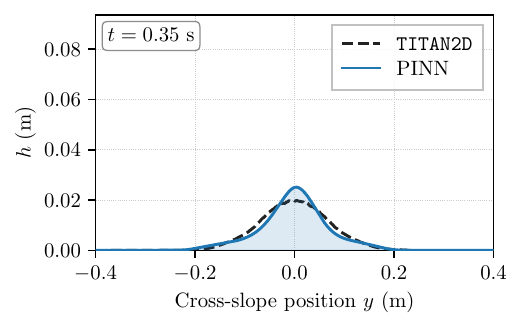}
\end{subfigure}
\vspace{-0.5em}
\begin{subfigure}[b]{0.48\textwidth}
\includegraphics[width=\textwidth]{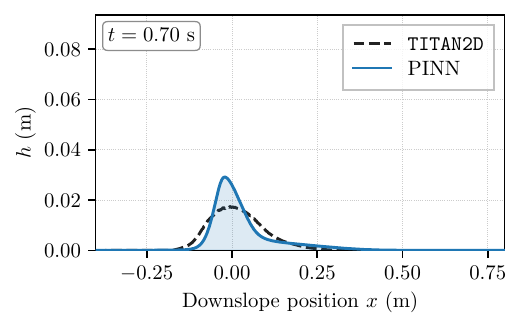}
\end{subfigure}
\hfill
\begin{subfigure}[b]{0.48\textwidth}
\includegraphics[width=\textwidth]{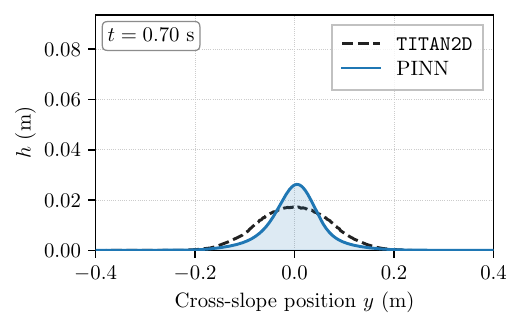}
\end{subfigure}
\caption{Case~C1 ($1\,\text{kg}$, $10^\circ$): centreline (left) and cross-slope (right) height profiles at $t = 0.00$, $0.18$, $0.35$, and $0.70\,\text{s}$. PINN (solid blue) and \texttt{TITAN2D} (dashed red).}
\label{fig:app_c1_profiles}
\end{figure}

At $10^\circ$, the deposit peak was recovered accurately at early times, and agreement with \texttt{TITAN2D} was closest at $t = 0.18$ and $0.35\,\text{s}$. By $t = 0.70\,\text{s}$, the PINN peak was slightly taller and shifted downslope relative to \texttt{TITAN2D}. The centreline and cross-slope profiles are shown in Figure~\ref{fig:app_c1_profiles}.

\begin{figure}[H]
\centering
\begin{subfigure}[b]{0.48\textwidth}
\includegraphics[width=\textwidth]{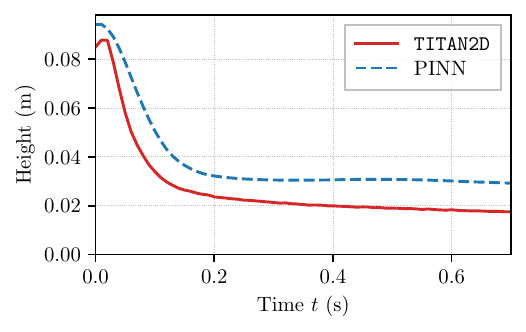}
\end{subfigure}
\hfill
\begin{subfigure}[b]{0.48\textwidth}
\includegraphics[width=\textwidth]{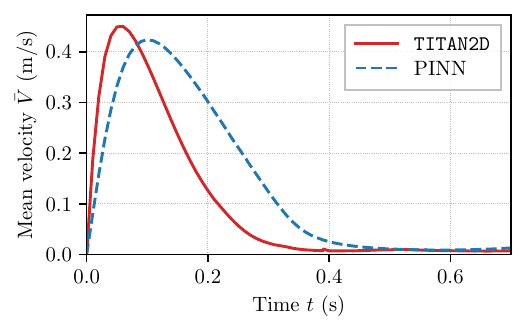}
\end{subfigure}
\vspace{-0.9em}
\begin{subfigure}[b]{0.48\textwidth}
\includegraphics[width=\textwidth]{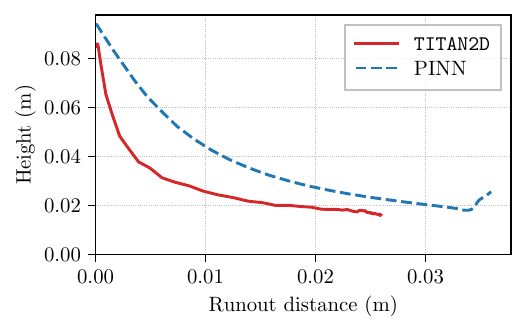}
\end{subfigure}
\hfill
\begin{subfigure}[b]{0.48\textwidth}
\includegraphics[width=\textwidth]{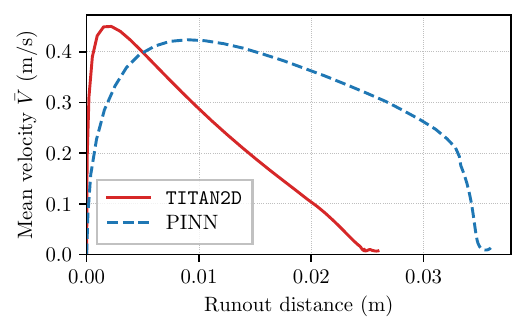}
\end{subfigure}
\caption{Integral flow diagnostics for Case~C1 ($1\,\text{kg}$,
$10^\circ$): $h_{\max}$ and $\bar{V}$ versus time (top row); centroid
displacement and $\bar{V}$ versus runout distance (bottom row).
\texttt{TITAN2D} (solid red); PINN (dashed blue).}
\label{fig:app_c1_diag}
\end{figure}

The integral diagnostics are shown in Figure~\ref{fig:app_c1_diag}.
The maximum height decayed monotonically from $H_0 = 85\,\text{mm}$, with
the PINN closely tracking \texttt{TITAN2D} throughout. The mean velocity
reached a lower peak than in C4, reflecting the weaker gravitational
drive at $10^\circ$.

\begin{figure}[H]
\centering
\includegraphics[width=0.55\textwidth]{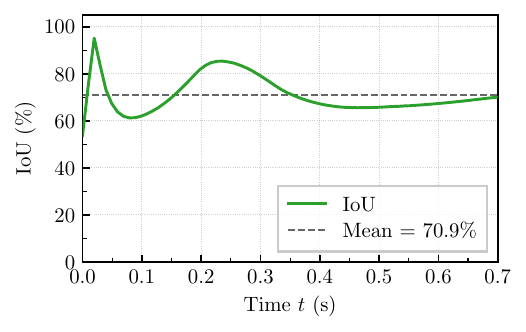}
\caption{Time-resolved IoU for Case~C1 ($1\,\text{kg}$, $10^\circ$).
Time-mean value: $70.9\,\%$.}
\label{fig:app_c1_iou}
\end{figure}

The IoU started near its maximum, as both solutions shared the same
initial cylindrical footprint, then decreased through the spreading phase
as the wavefronts diverged slightly, stabilising toward a time-mean of
$70.9\,\%$ (Figure~\ref{fig:app_c1_iou}).

\subsection{\texorpdfstring{Case C2: 1\,kg, $15^\circ$}
{Case C2: 1 kg, 15 degrees}}
\label{app:c2}

\begin{figure}[H]
\centering
\begin{subfigure}[b]{0.48\textwidth}
\includegraphics[width=\textwidth]{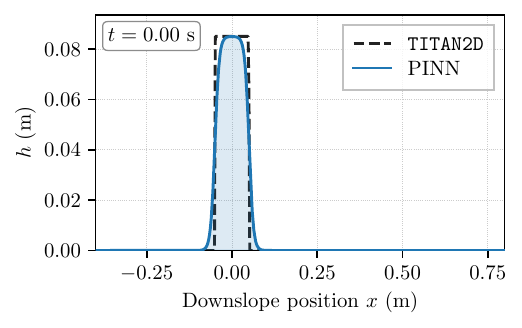}
\end{subfigure}
\hfill
\begin{subfigure}[b]{0.48\textwidth}
\includegraphics[width=\textwidth]{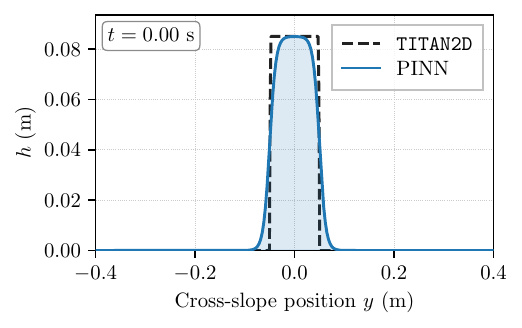}
\end{subfigure}
\vspace{-0.8em}
\begin{subfigure}[b]{0.48\textwidth}
\includegraphics[width=\textwidth]{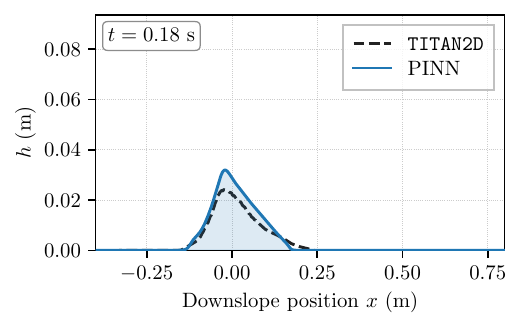}
\end{subfigure}
\hfill
\begin{subfigure}[b]{0.48\textwidth}
\includegraphics[width=\textwidth]{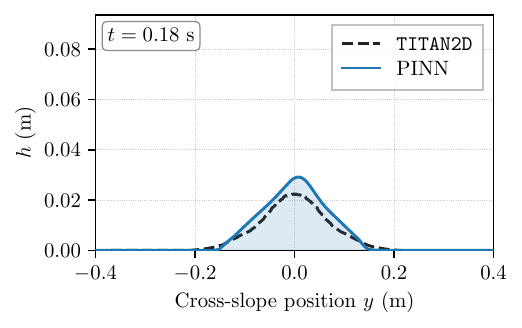}
\end{subfigure}
\vspace{-0.8em}
\begin{subfigure}[b]{0.48\textwidth}
\includegraphics[width=\textwidth]{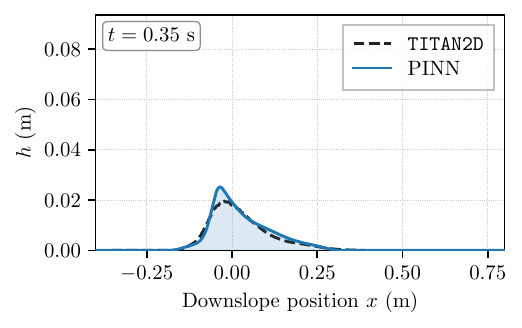}
\end{subfigure}
\hfill
\begin{subfigure}[b]{0.48\textwidth}
\includegraphics[width=\textwidth]{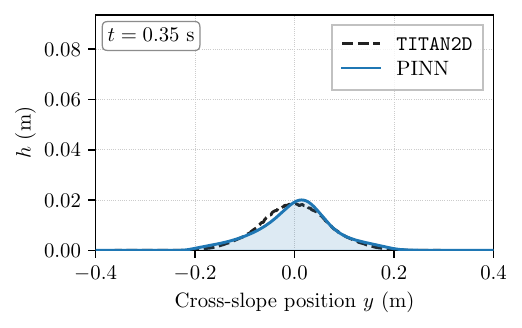}
\end{subfigure}
\vspace{-0.8em}
\begin{subfigure}[b]{0.48\textwidth}
\includegraphics[width=\textwidth]{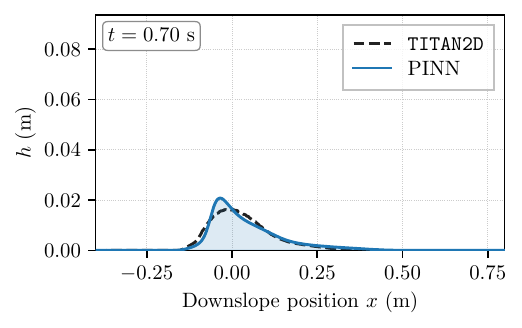}
\end{subfigure}
\hfill
\begin{subfigure}[b]{0.48\textwidth}
\includegraphics[width=\textwidth]{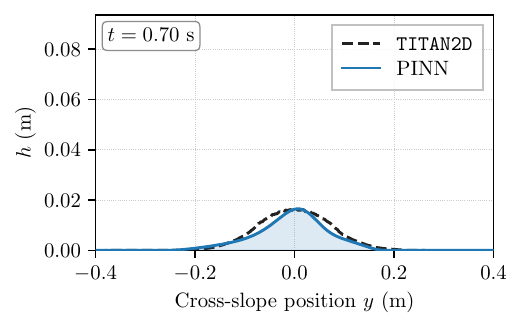}
\end{subfigure}
\caption{Case~C2 ($1\,\text{kg}$, $15^\circ$): centreline (left) and
cross-slope (right) height profiles at $t = 0.00$, $0.18$, $0.35$, and
$0.70\,\text{s}$. PINN (solid blue) and \texttt{TITAN2D} (dashed red).}
\label{fig:app_c2_profiles}
\end{figure}

Case~C2 shared the pile mass and geometry of Case~C1; only the slope angle
increased. The steeper incline drove a faster collapse and a more elongated
final deposit, and the PINN reproduced both features closely in both
profile directions (Figure~\ref{fig:app_c2_profiles}).

\begin{figure}[H]
\centering
\begin{subfigure}[b]{0.48\textwidth}
\includegraphics[width=\textwidth]{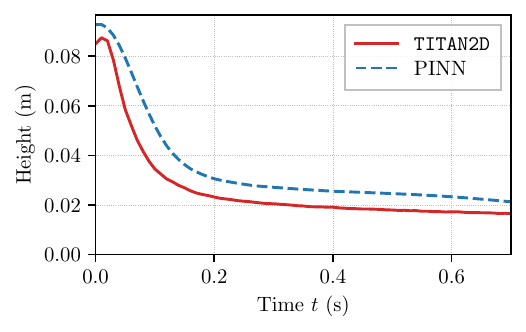}
\end{subfigure}
\hfill
\begin{subfigure}[b]{0.48\textwidth}
\includegraphics[width=\textwidth]{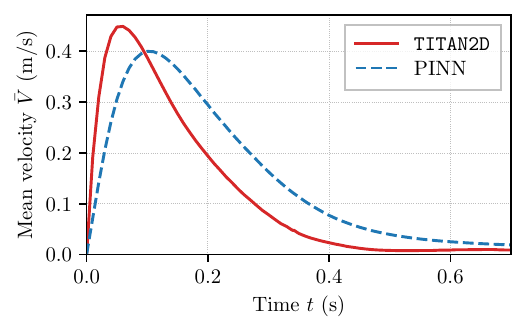}
\end{subfigure}
\vspace{-0.9em}
\begin{subfigure}[b]{0.48\textwidth}
\includegraphics[width=\textwidth]{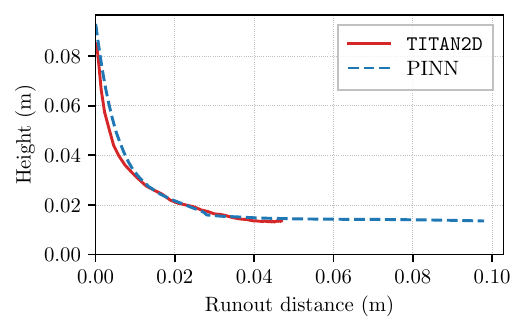}
\end{subfigure}
\hfill
\begin{subfigure}[b]{0.48\textwidth}
\includegraphics[width=\textwidth]{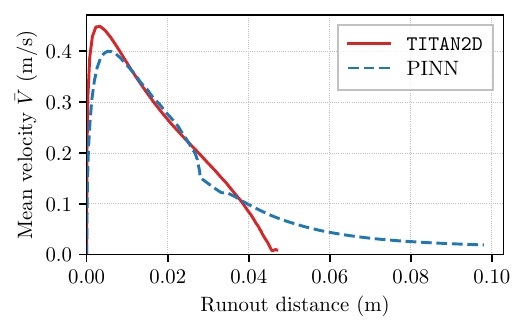}
\end{subfigure}
\caption{Integral flow diagnostics for Case~C2 ($1\,\text{kg}$,
$15^\circ$). Layout as in Figure~\ref{fig:app_c1_diag}.}
\label{fig:app_c2_diag}
\end{figure}

The integral diagnostics are shown in Figure~\ref{fig:app_c2_diag}. The maximum height decayed monotonically from $H_0 = 85\,\text{mm}$, with the PINN closely tracking \texttt{TITAN2D} throughout, and the mean velocity reached a higher peak than in C1, consistent with the stronger gravitational drive at $15^\circ$.

\begin{figure}[H]
\centering
\includegraphics[width=0.55\textwidth]{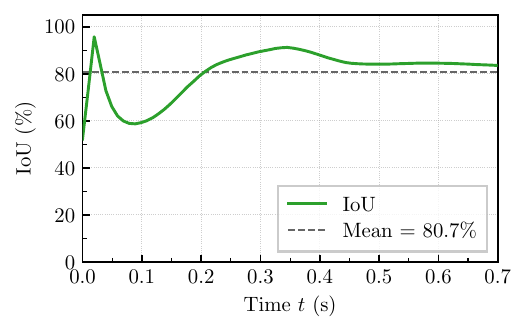}
\caption{Time-resolved IoU for Case~C2 ($1\,\text{kg}$, $15^\circ$).
Time-mean value: $80.7\,\%$.}
\label{fig:app_c2_iou}
\end{figure}

The time-mean IoU for Case~C2 was $80.7\,\%$, the highest value across all four cases (Figure~\ref{fig:app_c2_iou}).

\subsection{Case C3: 2.5 kg, \texorpdfstring{10$^\circ$}{10 degrees}}
\label{app:c3}

\begin{figure}[H]
\centering
\begin{subfigure}[b]{0.48\textwidth}
\includegraphics[width=\textwidth]{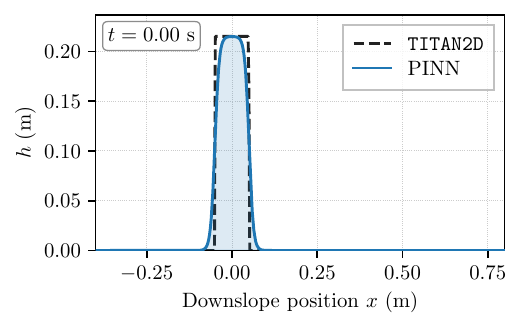}
\end{subfigure}
\hfill
\begin{subfigure}[b]{0.48\textwidth}
\includegraphics[width=\textwidth]{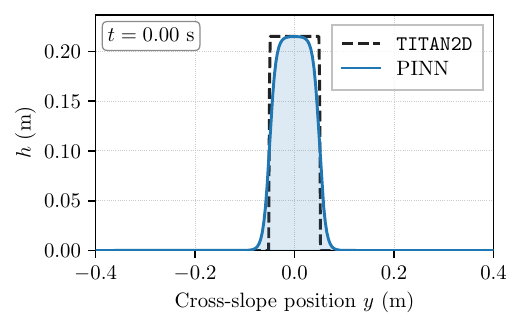}
\end{subfigure}
\vspace{-0.8em}
\begin{subfigure}[b]{0.48\textwidth}
\includegraphics[width=\textwidth]{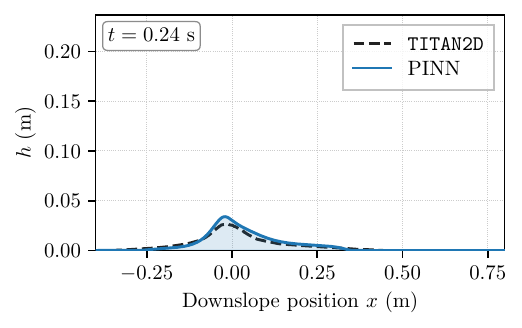}
\end{subfigure}
\hfill
\begin{subfigure}[b]{0.48\textwidth}
\includegraphics[width=\textwidth]{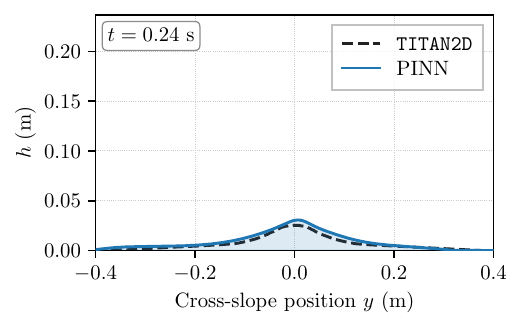}
\end{subfigure}
\vspace{-0.8em}
\begin{subfigure}[b]{0.48\textwidth}
\includegraphics[width=\textwidth]{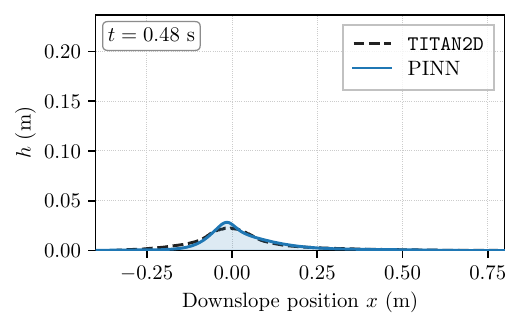}
\end{subfigure}
\hfill
\begin{subfigure}[b]{0.48\textwidth}
\includegraphics[width=\textwidth]{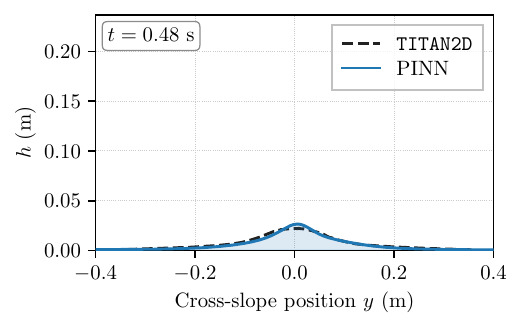}
\end{subfigure}
\vspace{-0.8em}
\begin{subfigure}[b]{0.48\textwidth}
\includegraphics[width=\textwidth]{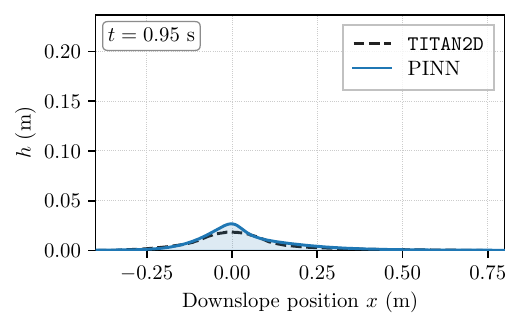}
\end{subfigure}
\hfill
\begin{subfigure}[b]{0.48\textwidth}
\includegraphics[width=\textwidth]{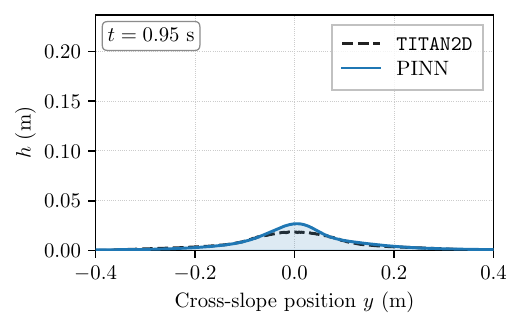}
\end{subfigure}
\caption{Case~C3 ($2.5\,\text{kg}$, $10^\circ$): centreline (left) and
cross-slope (right) height profiles at $t = 0.00$, $0.24$, $0.48$, and
$0.95\,\text{s}$. PINN (solid blue) and \texttt{TITAN2D} (dashed red).}
\label{fig:app_c3_profiles}
\end{figure}

Case~C3 paired the higher pile mass of C4 with the lower slope angle
of Case~C1. The larger initial volume produced steeper depth gradients
during the early collapse and a wider final deposit than either of the
$1\,\text{kg}$ cases, and agreement with \texttt{TITAN2D} was closest at
the final snapshot once the deposit had thinned and lateral gradients had
relaxed (Figure~\ref{fig:app_c3_profiles}).

\begin{figure}[H]
\centering
\begin{subfigure}[b]{0.48\textwidth}
\includegraphics[width=\textwidth]{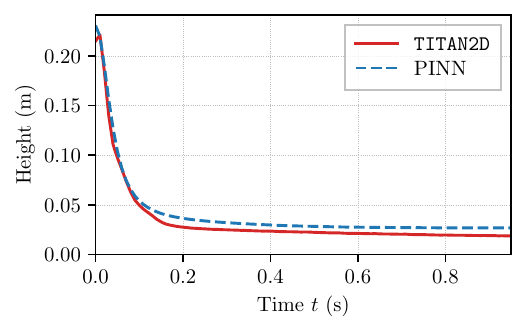}
\end{subfigure}
\hfill
\begin{subfigure}[b]{0.48\textwidth}
\includegraphics[width=\textwidth]{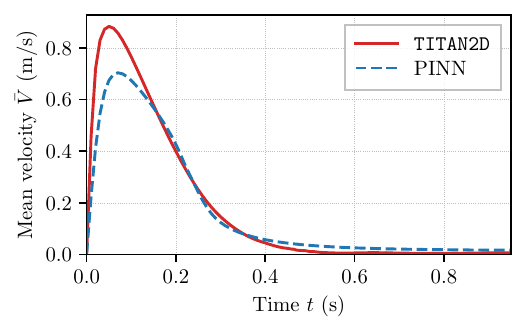}
\end{subfigure}
\vspace{-0.9em}
\begin{subfigure}[b]{0.48\textwidth}
\includegraphics[width=\textwidth]{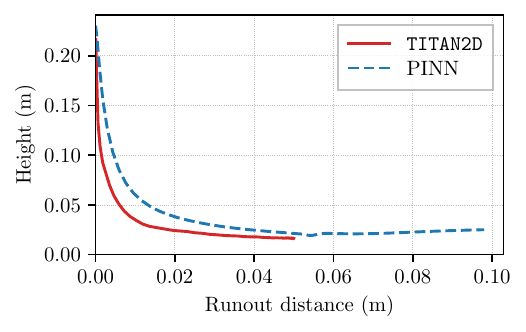}
\end{subfigure}
\hfill
\begin{subfigure}[b]{0.48\textwidth}
\includegraphics[width=\textwidth]{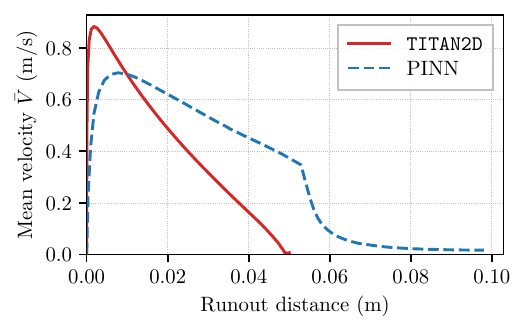}
\end{subfigure}
\caption{Integral flow diagnostics for Case~C3 ($2.5\,\text{kg}$,
$10^\circ$). Layout as in Figure~\ref{fig:app_c1_diag}.}
\label{fig:app_c3_diag}
\end{figure}

The integral diagnostics are shown in Figure~\ref{fig:app_c3_diag}. The maximum height decayed sharply within the first $0.1\,\text{s}$, with the PINN settling slightly above the \texttt{TITAN2D} reference at later times. The mean velocity reached a lower, later peak in the PINN prediction than in \texttt{TITAN2D}, and the velocity--distance curve shows that the PINN sustains motion over a longer runout distance than \texttt{TITAN2D}.

\begin{figure}[H]
\centering
\includegraphics[width=0.55\textwidth]{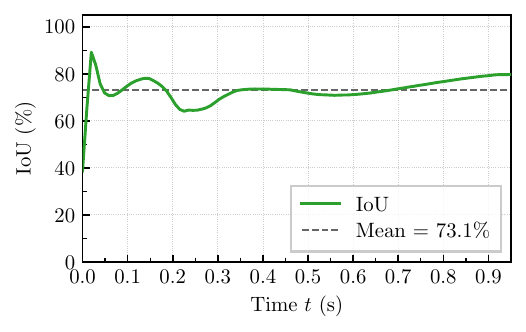}
\caption{Time-resolved IoU for Case~C3 ($2.5\,\text{kg}$, $10^\circ$).
Time-mean value: $73.1\,\%$.}
\label{fig:app_c3_iou}
\end{figure}

The time-resolved IoU for Case~C3 is shown in Figure~\ref{fig:app_c3_iou}.
The metric started near its maximum, dipped to a minimum during the early
spreading phase, and then recovered to stabilise near the time-mean of
$73.1\,\%$ for the remainder of the simulation.

\section{Repeatability Analysis}
\label{app:repeatability}
 
The results for Cases~C1--C4 in the main text each correspond to a single
training run. To assess how sensitive these results are to the choice of
random initialisation, three independent training runs were performed for
C4, each initiated from a different random seed. All other aspects of the setup, including the architecture, hyperparameters, collocation strategy, and observation set, were identical across runs, as described in the main text.
 
\subsection{Profile variability across runs}
\label{app:rep_profiles}
 
\begin{figure}[H]
\centering
\begin{subfigure}[b]{0.48\textwidth}
\includegraphics[width=\textwidth]{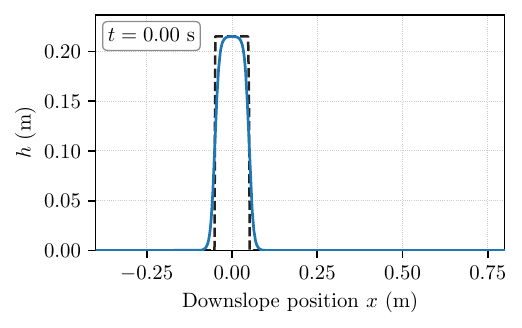}
\end{subfigure}
\hfill
\begin{subfigure}[b]{0.48\textwidth}
\includegraphics[width=\textwidth]{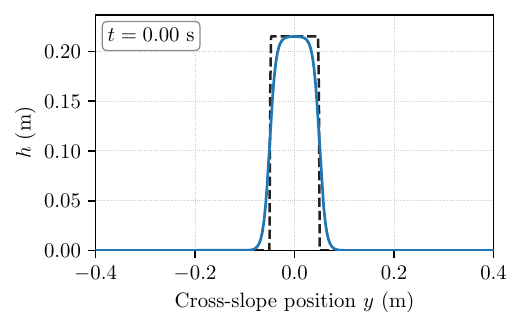}
\end{subfigure}
\vspace{-0.5em}
\begin{subfigure}[b]{0.48\textwidth}
\includegraphics[width=\textwidth]{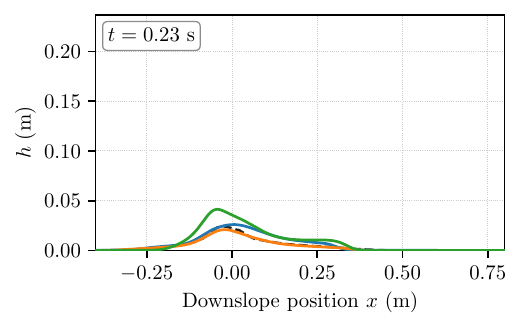}
\end{subfigure}
\hfill
\begin{subfigure}[b]{0.48\textwidth}
\includegraphics[width=\textwidth]{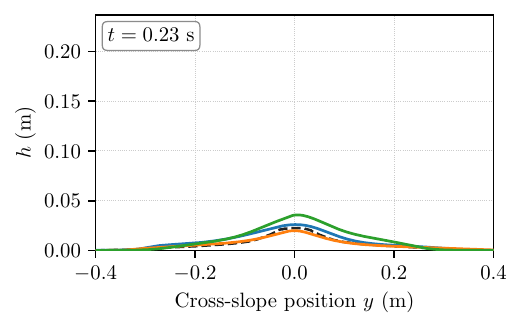}
\end{subfigure}
\vspace{-0.5em}
\begin{subfigure}[b]{0.48\textwidth}
\includegraphics[width=\textwidth]{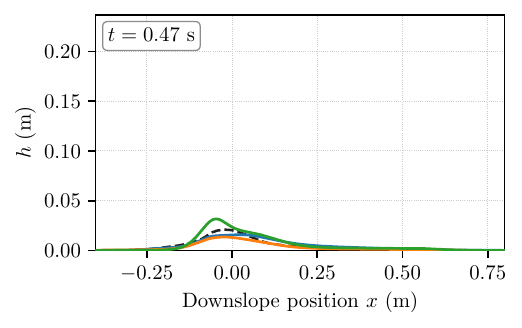}
\end{subfigure}
\hfill
\begin{subfigure}[b]{0.48\textwidth}
\includegraphics[width=\textwidth]{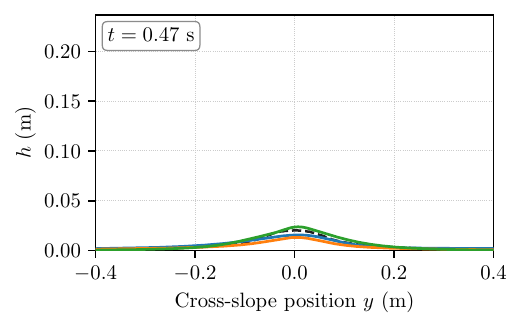}
\end{subfigure}
\hfill
\begin{subfigure}[b]{0.48\textwidth}
\includegraphics[width=\textwidth]{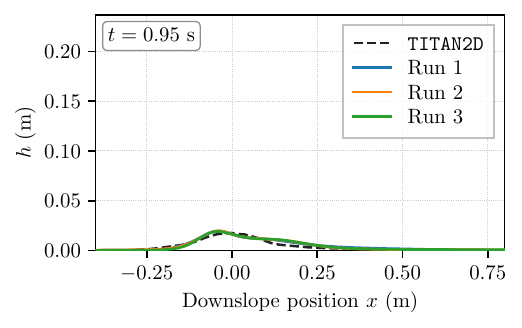}
\end{subfigure}
\hfill
\begin{subfigure}[b]{0.48\textwidth}
\includegraphics[width=\textwidth]{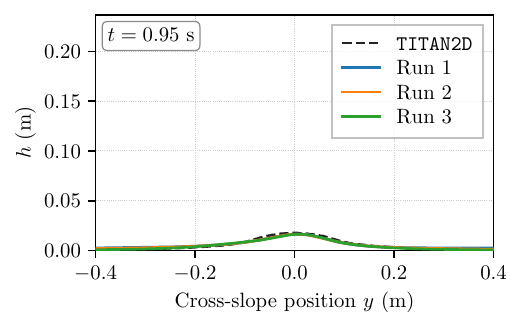}
\end{subfigure}
\caption{Repeatability study for C4 ($2.5\,\text{kg}$, $15^\circ$): centerline (left) and cross-slope (right) height profiles at $t = 0.00$, $0.23$, $0.47$, and $0.95\,\text{s}$. Each solid curve is one of three independent PINN runs. \texttt{TITAN2D} reference (dashed red).}
\label{fig:app_rep_profiles}
\end{figure}

Figure~\ref{fig:app_rep_profiles} overlays the centreline and cross-slope
height profiles from all three runs at four time instants alongside the
\texttt{TITAN2D} reference. The inter-run spread was small at every
snapshot and was comparable in magnitude to the single-run
PINN--\texttt{TITAN2D}. The three curves were nearly indistinguishable at $t = 0\,\text{s}$ and remained closely clustered through the final deposit.
 
\subsection{Final deposit comparison}
\label{app:rep_deposit}
 
\begin{figure}[H]
\centering
\begin{subfigure}[b]{0.48\textwidth}
\includegraphics[width=\textwidth]{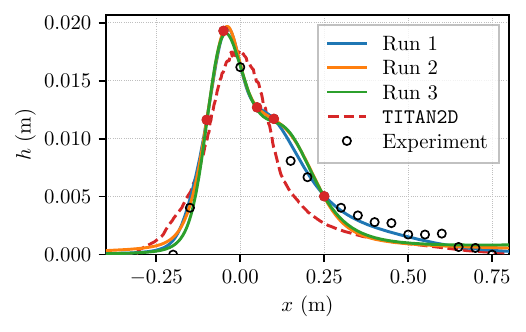}
\caption{Centreline ($y = 0$).}
\end{subfigure}
\hfill
\begin{subfigure}[b]{0.48\textwidth}
\includegraphics[width=\textwidth]{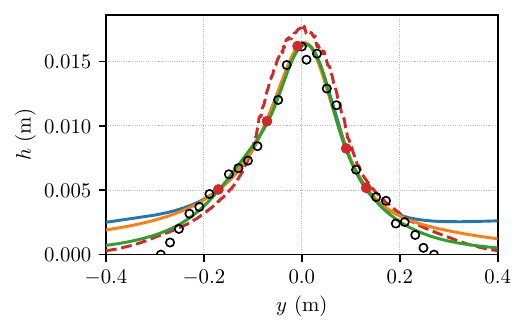}
\caption{Cross-slope ($x = 0$).}
\end{subfigure}
\caption{Final deposit profiles for C4 across three independent
training runs (solid lines). \texttt{TITAN2D} (dashed red) and
Maeno et al. (2013) experimental data (open circles) are shown
for reference.}
\label{fig:app_rep_deposit}
\end{figure}

The inter-run spread at the final deposit ($t = 0.95\,\text{s}$) was
substantially smaller than the gap between the PINN band and
\texttt{TITAN2D} in both the centreline and cross-slope directions
(Figure~\ref{fig:app_rep_deposit}), confirming that the discrepancies
reported in the main text reflect genuine model behaviour
rather than sensitivity to random initialisation.
 
\subsection{Error metrics across runs}
\label{app:rep_metrics}
 
\begin{figure}[H]
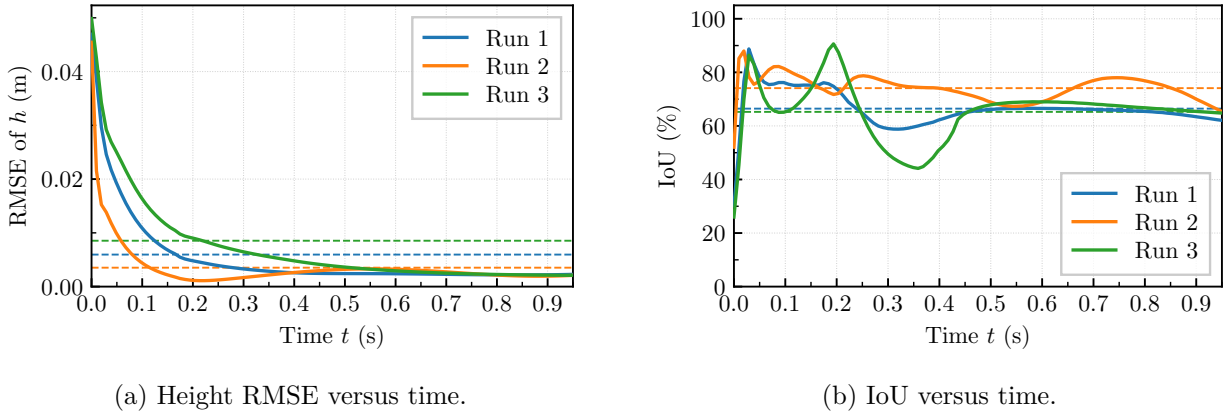

\centering
\begin{subfigure}[b]{0.48\textwidth}
\includegraphics[width=\textwidth]{fig_rmse_vs_time.pdf}
\caption{Height RMSE versus time.}
\end{subfigure}
\hfill
\begin{subfigure}[b]{0.48\textwidth}
\includegraphics[width=\textwidth]{fig_iou_vs_time.pdf}
\caption{IoU versus time.}
\end{subfigure}
\caption{Time-resolved error metrics across three independent training runs
for C4. Each curve corresponds to one run; the bandwidth quantifies
run-to-run variability.}
\label{fig:app_rep_metrics}
\end{figure}

The time-resolved RMSE and IoU for all three runs are shown in
Figure~\ref{fig:app_rep_metrics}. Both metrics followed the same
trajectory regardless of seed: RMSE peaked during the early collapse phase
and decayed as the deposit thinned. IoU was more variable across runs during the collapse and spreading phase (roughly $t < 0.5\,\text{s}$), with each run
showing a dip of differing depth and timing, before converging to a similar range by the final snapshot. The RMSE at the final snapshot lay within a narrow band across all three runs, and the time-mean IoU varied by only a few percentage points, confirming that a single training run is representative of the framework's predictive
capability for this problem.
 
\subsection{Training data point coordinates}
\label{app:data_points}

Table~\ref{tab:data_points} lists the coordinates of the ten training data points used for Case~C4, selected to sample the upstream face, deposit peak, and downstream tail along the centerline, as well as the full lateral extent of the deposit along the cross-slope profile.

\begin{table}[H]
\centering
\caption{Coordinates of the ten training data points used for 
C4. All values are in centimetres.}
\label{tab:data_points}
\begin{tabular}{clcc}
\toprule
\textbf{Profile} & \textbf{Point} & $x$ (cm) & $y$ (cm) \\
\midrule
\multirow{5}{*}{Centerline ($y = 0$)}   
    & P1 & $-10$ & $0$    \\
    & P2 & $-5$  & $0$    \\
    & P3 & $5$   & $0$    \\
    & P4 & $10$  & $0$    \\
    & P5 & $25$  & $0$    \\
\midrule
\multirow{5}{*}{Cross-slope ($x = 0$)}  
    & P6  & $0$ & $-17.0$ \\
    & P7  & $0$ & $-7.0$  \\
    & P8  & $0$ & $-1.0$  \\
    & P9  & $0$ & $9.0$   \\
    & P10 & $0$ & $13.0$  \\
\bottomrule
\end{tabular}
\end{table}

\end{document}